\documentclass{aa}  

\usepackage{graphicx}
\usepackage{enumerate}
\usepackage{epsfig}
\usepackage{color}
\usepackage{txfonts}
\usepackage{natbib}
\usepackage{multirow}
\usepackage{ulem} 
\usepackage{amssymb}
\usepackage[percent]{overpic}

\usepackage{movie15}

\usepackage{booktabs}
\usepackage{array}   

\usepackage{academicons}

\usepackage[export]{adjustbox}

\newcommand{\orcid}[1]{\href{https://orcid.org/#1}{\textcolor[HTML]{A6CE39}{\aiOrcid}}}
\usepackage{orcidlink}

\usepackage{hyperref}
\usepackage[dvipsnames]{xcolor} 
\hypersetup{
    colorlinks,
    citecolor=blue,
    filecolor=blue,
    linkcolor=blue,
    urlcolor=blue
}

\usepackage{xcolor}

\definecolor{mag}{rgb}{0.6, 0.4, 0.8}

\defcitealias{2025A&A...695A.220K}{Paper I}
\defcitealias{2025A&A...700A..89K}{Paper II}
\defcitealias{2025A&A...700A..90K}{Paper III}
\defcitealias{Mapping-extragalactic}{Paper IV}
\defcitealias{2026A&A...706A.103R}{Paper V}
\defcitealias{Mapping-migration}{Paper VI}

\definecolor{ao}{rgb}{0.0, 0.5, 0.0}

\newcommand{\kmps}{\rm km~s\ensuremath{^{-1} }\,}

\newcommand{\Msunyr}{M\ensuremath{_\odot}~yr\ensuremath{^{-1}}\,}

\newcommand{\Gaia}{{\it Gaia}\,}

\newcommand{\SKIRT}{\texttt{SKIRT}}
\newcommand{\ppxf}{\texttt{pPXF}}

\newcommand{\aMe}{\ensuremath{\rm [\alpha/M]}\,}
\newcommand{\aFe}{\ensuremath{\rm [\alpha/Fe]}\,}

\newcommand{\MgFe}{\ensuremath{\rm [Mg/Fe]}\,}
\newcommand{\FeH}{\ensuremath{\rm [Fe/H]}\,}

\begin{document}

\title{The Milky Way as a distant galaxy: an IFU and panchromatic view}

\titlerunning{IFU and panchromatic Milky Way}
\authorrunning{Khoperskov et al.}

\author{Sergey Khoperskov\thanks{sergey.khoperskov@gmail.com}\inst{1,2,3},
Alina Boecker\inst{2},
Nikolay Kacharov\inst{1}, \\
Davor Krajnović\inst{1}, 
Matthias Steinmetz\inst{1}, 
Glenn van de Ven\inst{2}, 
Jakob C. Walcher\inst{1}}

\institute{Leibniz-Institut für Astrophysik Potsdam (AIP),
              An der Sternwarte 16, 14482 Potsdam, Germany
              \and
              Department of Astrophysics, University of Vienna, T\"urkenschanzstrasse 17, A-1180 Vienna, Austria
              \and
              LIRA, Observatoire de Paris, Université PSL, Sorbonne Université, Université Paris Cité, CY Cergy Paris Université, CNRS, 92190 Meudon, France
              }
              
\abstract{
 Understanding the structure and evolution of galaxies increasingly benefits from placing the Milky Way~(MW), the best-studied stellar system, in an external-galaxy context. To analyse the MW as an extragalactic system, we construct an integrated-light view of it using two complementary approaches: composite stellar populations built from a selection-function-free orbit-superposition solution constrained by APOGEE DR17, and full \SKIRT\ radiative-transfer modelling of a hydrodynamical MW simulation initialised from the same solution and evolved into a non-equilibrium present-day state. 
 In both cases, the MW data are assembled into mock IFU datacubes and analysed with full spectral fitting~(\ppxf). We find, however, that the underlying LOSVD is more complex than can be captured by a Gauss--Hermite parametrisation, as is likely the case in MW-like late-type barred galaxies.
 
For the composite-population mocks, we recover the main large-scale kinematic structures, including the rotation and velocity-dispersion patterns associated with the thin and thick discs and the bar/bulge, together with the mean stellar-population maps. The known disc chemical bimodality in $\mathrm{[\alpha/M]}$--$\mathrm{[M/H]}$ cannot be recovered directly from the IFU data, regardless of the $\alpha$-resolution of the SSP templates, but its MW-like spatially varying double-sequence behaviour is recovered across the disc. The broad global star-formation history is reproduced, although artificial bursts, likely driven by the age--metallicity degeneracy, remain even after regularisation. For the radiative-transfer mocks, the main kinematic maps are recovered with moderate accuracy, whereas higher-order moments and stellar-population properties remain difficult to constrain, primarily because of the lower spectral resolution and lower statistical signal-to-noise ratio of the datacubes. These results indicate that forward panchromatic modelling remains challenging for high-resolution IFU-like observations. 

}

\keywords{}

\maketitle

\section{Introduction}

While understanding the Milky Way~(MW) in its own right constitutes a major and active field of research, its true value lies in serving as a uniquely detailed laboratory for galaxy evolution. The wealth of information now available on the formation of the thin and thick Galactic discs~\citep{2022MNRAS.514..689B, 2022Natur.603..599X}, the origin and present-day structure of the bar and the boxy/peanut bulge~\citep{2013MNRAS.435.1874W, 2015MNRAS.450.4050W,2017MNRAS.469.1587D}, the prevalence of non-equilibrium features~\citep{2018Natur.561..360A, 2018MNRAS.479L.108K}, and the Galaxy's merger history~\citep{2018Natur.563...85H, 2018MNRAS.478..611B}, including the contribution of accreted populations and various stellar halo substructures~\citep{2020MNRAS.494.3880B, 2020ARA&A..58..205H}, offers an unprecedented view of the processes operating in MW. The central challenge, however, is not just to recover the specific structural details of a single galaxy in the universe, but to distil the underlying physical mechanisms that govern galaxy evolution. 

Addressing how detailed near-field knowledge can be transferred to interpret and constrain the formation and evolution of other stellar systems is non-trivial, because MW and external-galaxy studies rely on fundamentally different methodologies. In the MW, our privileged vantage point allows individual stars to be resolved and stellar populations to be characterised in great detail, but at the cost of a coherent global view, even in the era of Gaia~\citep{2016A&A...595A...2G} and large spectroscopic surveys~(e.g., RAVE~\citep{2020AJ....160...82S}, LAMOST~\citep{2012RAA....12..723Z}, GALAH~\citep{2025PASA...42...51B}, SDSS~\citep{2025arXiv250707093S}). Conversely, high-resolution extragalactic observations provide integrated views of galaxies as complete systems, enabling direct measurements of global structural, photometric, and kinematic properties~(e.g., SAURON~\citep{2002MNRAS.329..513D}, ATLAS$^{3D}$~\citep{2011MNRAS.413..813C}, CALIFA~\citep{2012A&A...538A...8S}, MaNGA~\citep{2015ApJ...798....7B}, SAMI~\citep{2015MNRAS.447.2857B}, GECKOS~\citep{2024IAUS..377...27V}). These studies, however, necessarily rely on unresolved stellar populations and are limited by spatial resolution and projection effects, making inferred properties sensitive to distance and viewing angle. Recovering detailed star-formation histories and chemical enrichment from integrated spectra is therefore challenging, owing to degeneracies between age, metallicity, dust attenuation, and star-formation timescales. As a result, MW--external galaxy comparisons often rely on simplified or indirect diagnostics. Indeed, most studies identifying MW analogues~(MWAs) through selected properties, such as metallicity gradients~\citep{2023A&A...676A..57P,2025A&A...694A.113P}, stellar mass~\citep{2020ApJ...898..116K}, or morphology~\citep{2019ApJ...872..106K,2023MNRAS.521.5810Z}, have shown that galaxies closely matching the assumed properties of the MW are exceedingly rare; \citet{2019MNRAS.489.5030F}, for example, found that only $\sim 0.01$ per cent of SDSS DR7 galaxies satisfy commonly adopted combined MWA criteria.

For instance, \cite{2023MNRAS.521.5810Z} compared the radial structure of the MW with SDSS/MaNGA galaxies using semi-analytical spectral fitting, in which star-formation histories and chemical enrichment are modelled self-consistently. They identified MWAs that reproduce the MW radial profiles of stellar age, metallicity, and $\alpha$-enhancement in both the inner and outer disc. In these systems, the agreement is explained primarily through extended star formation in the outer disc and the inflow of chemically enriched gas toward the centre, thereby reproducing the observed gradients without invoking stellar radial migration. This interpretation contrasts sharply with much of the MW literature, where radial migration is regarded as a key process in reshaping the present-day thin and thick disc populations~\citep{2009MNRAS.396..203S, 2012A&A...548A.126M, 2013MNRAS.433..976R, 2016ApJ...818L...6L, 2018A&A...616A..86H} and contributing to the build-up of outer discs~\citep{2012MNRAS.426.2089R, 2012A&A...548A.126M}. Although the quantitative importance of migration in external galaxies remains debated~\citep{2017A&A...604A...4R}, this example exposes a broader methodological disconnect: similar observables may be assigned different physical origins in Galactic and extragalactic studies. This tension is further amplified by inconsistent definitions of concepts such as ``inner'' and ``outer'' disc, or ``thin'' and ``thick'' disc, which complicates any unified interpretation of how the processes resolved in the MW manifest themselves in external disc galaxies.

Several studies have attempted to place the MW in an extragalactic context by combining spectroscopic and photometric analyses~\citep{2015ApJ...809...96L, 2024MNRAS.533.4334T, 2020MNRAS.491.3672B, 2025arXiv251214664K}. For example, \citet{2021MNRAS.508.4459F} estimated the MW's photometric properties by constructing a full UV-to-IR SED as it would appear externally in a face-on configuration. They found that the MW lies in the green valley according to optical diagnostics, while standard UV and infrared indicators place it in the star-forming regime, typical of red spirals. Galaxies with MW-like SEDs were also shown to span a broad range of visual morphologies, highlighting the absence of a unique mapping between morphology and SED.

\citet{2025ApJ...990..203I} modelled a mock integrated MW spectrum using the classical selection-function correction~\citep{2012ApJ...753..148B} applied to APOGEE DR17. Their approach maps present-day mono-age-abundance populations across the Galactic disc, enabling estimates of the MW's stellar mass, integrated colours, mean metallicities, and their evolution with lookback time, but not of its kinematics. They concluded that the MW has likely resided in the green valley for several Gyr~\citep{2015ApJ...809...96L,2016A&A...589A..66H}. Using the same framework, \citet{2025ApJ...991...36Z} found evidence that the MW assembled its stellar mass earlier than typical MW analogues in TNG50~\citep[see also][]{2024MNRAS.533.3975K}.

Moving beyond a single integrated spectrum of the MW, \citet{2024MNRAS.534.1175W} generated an edge-on synthetic IFU datacube based on the E-GALAXIA model, which combines the kinematics, abundances, and ages of stellar populations in an axisymmetric disc extrapolated from APOGEE observations over $3-15$~kpc~\citep{2021MNRAS.507.5882S}. They then used full-spectrum fitting with \ppxf~\citep{2017MNRAS.466..798C} to test the recovery of line-of-sight kinematics and projected age and metallicity structure. The authors showed that biases in the recovered kinematics and stellar populations arise from the limited spectral resolution of the templates and from assuming a single line-of-sight velocity distribution~(LOSVD) for all stellar populations, an approximation that breaks down at the thin/thick-disc interface. Although they concluded that IFU analyses of MW analogues can recover the global observables of the MW reasonably well, their model does not include the observed chemical diversity of the MW or detailed information about the inner disc, bar, and bulge, which are often regarded as defining Galactic features.

To address the absence of the observed $\rm [\alpha/Fe]$ bimodality in recent attempts to analyse the MW as an external galaxy~(e.g. \citealt{2024MNRAS.534.1175W,2025ApJ...990..203I}), and to move beyond several idealised assumptions, we pursue two complementary approaches. First, we construct and analyse a mock IFU cube built from composite stellar populations~(CSP) based on a non-axisymmetric orbit superposition reconstruction of the MW~\citep{2025A&A...700A..89K}. Second, we develop a more complex evolutionary MW model that is locally out of equilibrium and includes a live dark matter halo, ISM dynamics, and self-consistent spiral structure. We then use radiative-transfer~(RT) calculations to generate mock IFU observations, thereby producing a more realistic external view of the present-day Galaxy.

This paper is organised as follows. Section~\ref{sec4::methods} describes the data and methodology, while Section~\ref{sec4::sos} summarises the orbit-superposition reconstruction of the MW. Sections~\ref{sec4::mock_vanila} and~\ref{sec4::mock_rt} introduce the CSP and RT mock IFU data, respectively. The full-spectrum fitting procedure is described in Section~\ref{sec4::ppxf}. Sections~\ref{sec4::CSP_results}-\ref{sec4::SFH} present the recovered kinematics, stellar populations, and star-formation histories of the extragalactic MW based on CSP; while the RT-based IFU analysis is given in Sect.~\ref{sec4::results_skirt}. Finally, Section~\ref{sec4::summary} discusses the results and summarises the main conclusions. Appendices~\ref{sec4::simulations_appendix}--\ref{sec4::kinematics_appendix} provide additional model descriptions and supplementary figures illustrating the methodology.

\begin{table*}
    \centering
    \caption{Parameters adopted to generate the mock IFU datacubes for the \texttt{eMW} models placed at a distance of 72~Mpc. The disc inclination is denoted by $i$. Two approaches are used to construct the datacubes: (i) a composite stellar population (CSP) method, in which spectra are built by combining sMILES SSP templates (see Section~\ref{sec4::mock_vanila}); and (ii) full radiative-transfer (RT) modelling, in which the panchromatic emission is computed with \SKIRT\ (see Section~\ref{sec4::mock_rt}). For each configuration, the table lists the total number of stellar particles within the IFU field of view of $60\times60$~arcsec, $N_{\mathrm{FOV}}$ is the number of star particles inside the FOV, and, for the RT runs, the number of photon packets used in the radiative-transfer calculations, $N_{\mathrm{phot}}$. The SSP column indicates the stellar population model used (sMILES~\citep{2023MNRAS.523.3450K} or BC03~\citep{2003MNRAS.344.1000B}), while the spectral resolution and sampling columns report the intrinsic SSP line-spread function (FWHM) and the adopted wavelength sampling relative to the native SSP grid, respectively.}
    \label{tab4:models}
    \begin{tabular}{@{} l c c c c c c c c @{}} 
    \toprule
    Model & $i$ & Bar & Setup & $N_{\rm FOV}$ & $N_{\rm phot}$ & SSP & Spectral res. & Sampling \\
          & [deg] & orientation &      & [$10^{6}$] & [$10^{9}$] & type & FWHM [\AA] & (SSP-native) \\
    \midrule
    \texttt{eMW.0070} & 70 & $27^\circ$  & CSP & 312.398412 & -- & sMILES & 2.51 & $\times 2$ \\
    \texttt{eMW.0071} & 70 & side-on     & CSP & 312.684196 & -- & sMILES & 2.51 & $\times 2$ \\
    \texttt{eMW.0072} & 70 & end-on      & CSP & 311.335200 & -- & sMILES & 2.51 & $\times 2$ \\
    \texttt{eMW.0073} & 90 & $27^\circ$  & CSP & 312.398412 & -- & sMILES & 2.51 & $\times 2$ \\
    \texttt{eMW.0074} & 90 & side-on     & CSP & 312.684196 & -- & sMILES & 2.51 & $\times 2$ \\
    \texttt{eMW.0075} & 90 & end-on      & CSP & 311.335200 & -- & sMILES & 2.51 & $\times 2$ \\
    \midrule
    \texttt{eMW.0170} & 70 & $27^\circ$  & RT  & 6.744199   & 26 & BC03   & 3.0  & $\times 1$ \\
    \texttt{eMW.0174} & 90 & side-on     & RT  & 6.714010   & 26 & BC03   & 3.0  & $\times 1$ \\
    \bottomrule
    \end{tabular}
    
\end{table*}

\section{Data and Methods}\label{sec4::methods}

\subsection{MW from orbit superposition}\label{sec4::sos}

The orbit superposition approach we use in this work and its application to the reconstruction of chrono-chemo-kinematics of the MW disc and the bulge are described in \cite{2025A&A...695A.220K}~(hereafter \citetalias{2025A&A...695A.220K}) and \cite{2025A&A...700A..89K, 2025A&A...700A..90K}~(hereafter \citetalias{2025A&A...700A..89K} and \citetalias{2025A&A...700A..90K}), respectively. In brief, we use the giant-star sample from APOGEE DR 17, selecting stars without problematic flags. We use radial velocities, atmospheric parameters and stellar abundances~(\FeH and \MgFe) from the APOGEE DR17~\citep{2022ApJS..259...35A}, which were complemented by the proper motions from the \Gaia DR3  catalogue~\citep{2023A&A...674A...1G}. We use only stars with radial velocity uncertainty $<2$~\kmps, distance error of $<20\%$, and proper motion errors better than $10\%$, as these are more critical for star orbits. In order to cover a larger area across the MW disc, we select giant stars with $\rm \log g < 2.2$. We further restrict the sample to $\rm ASPCAPFLAG=0$, excluding stars with ASPCAP pipeline warnings or bad parameter flags, and to $\rm EXTRATARG=0$, excluding objects tagged by the DR17 convenience targeting mask. For the final sample we adopted stellar ages from the \texttt{distmass} catalogue \citep{2024AJ....167...73S} with precision $\sigma_{\rm age} < 2\,\mathrm{Gyr}$. 

In this work, we use the output of the orbit superposition results obtained in the previous papers of the series~(\citetalias{2025A&A...700A..89K,2025A&A...700A..90K}). We adopt the 3D mass distribution of the MW, including its stellar component, from \cite{2022MNRAS.514L...1S}, which is an updated analytic model of the potential constructed by \cite{2017MNRAS.465.1621P}. This analytic potential, available in AGAMA~\citep{2019MNRAS.482.1525V}, shows the correct behaviour of the mass distribution outside the bar region and reproduces well the 3D density of the bar, including the X-shape structure of the bulge~\citep{2013MNRAS.435.1874W, 2015MNRAS.450.4050W}. We integrate orbits of the APOGEE stars, assuming a constant bar pattern speed of $\rm 37~km~s^{-1}~kpc^{-1}$. The weights of the orbits in the rotating rest frame were calculated by adjusting their total 3D density to the analytic solution for the stellar component from \cite{2024A&A...692A.216H}. In this work, we increased the cadence of the phase-space coordinates to 1000 along each orbit, and, as in the previous works, chemical abundances and ages for these data points were assigned using the uncertainties as a width of the distribution along orbit phase-space.

Following our previous work on the MW star formation history~\citep{2026A&A...706A.103R}, we applied specific corrections to the ages of two subsets of stars~(see a similar approach in \cite{Imig2025}): young high-$\alpha$ populations and stars with metallicities $\FeH < - 0.65$~dex. In the former case, these stars, although chemically consistent with the old high-$\alpha$ population, are often assigned younger ages typical of the low-$\alpha$ sequence. This mismatch likely results from mass-transfer or merger events, which rejuvenate stellar spectra and bias age estimates. The latter group, comprising metal-poor stars, is flagged as problematic in the \texttt{distmass} due to unreliable age assignments. These affected roughly 10\% of stellar mass across the age-metallicity plane, and the MW corrected age-metallicity relation derived from these adjustments is presented in~\citet{2026A&A...706A.103R}~(see their Fig. 1).

\subsection{Extragalactic MW: IFU based on composite stellar populations~(CSP)}\label{sec4::mock_vanila}
In this setup, we model each data point along the orbits as a simple stellar population~(SSP), whose masses remain constant along a given orbit but vary across different orbits. To represent the spectra of these SSPs, we used the semi-empirical sMILES stellar population models with variable $\aFe$ abundances~\citep{2023MNRAS.523.3450K, 2010MNRAS.404.1639V}, adopting a Kroupa initial mass function~\citep{2001MNRAS.322..231K}.

For each SSP, we derived an individual spectrum by interpolating from the eight nearest model spectra in age, $\mathrm{[M/H]}$, and $\mathrm{[\alpha/M]}$. In this setup, we adopt the iron abundance as a proxy for the total metallicity. While this is a simplification, it is motivated by the fact that the MILES stellar population templates are defined up to a maximum metallicity of $\rm [M/H]=0.4$, whereas APOGEE measurements of iron abundance extend to values as high as $\rm [Fe/H] \approx 0.55$ dex. This choice, therefore, reflects a practical limitation of the available templates rather than an implicit assumption of exact equivalence between iron abundance and total metallicity. In any case, we assume that stars with $\rm [Fe/H]>0.4$ are assigned a total metallicity of $\rm [M/H]=0.4$; this choice therefore limits the fraction of the stellar population affected by this approximation.

The individual spectra of each phase-space data point along orbits were then shifted relative to the rest frame by the line-of-sight velocity, as determined by the adopted extragalactic MW~(eMW) projection. Finally, the spectra were oversampled by a factor of two relative to the native MILES sampling; this improves numerical stability but does not alter the intrinsic spectral resolution.
For a given spaxel, we sum the stellar-mass-weighted SSP spectra on a spatial grid, producing spectra characteristic of unresolved stellar populations. This approach allowed us to generate realistic composite spectra for analysis, reflecting the cumulative properties of the stellar populations across eMW. 

To study the unresolved stellar populations of the eMW, we place the galaxy at a distance of 72 Mpc. At this distance, a $1'\times1'$ field of view~(FOV) corresponds to approximately $\rm 21~kpc\times 21~kpc$, providing extensive coverage of the galaxy. We adopt a spatial sampling of $0.2$~arcsec, yielding up to 90\,000 mock spectra across the full FOV. To assess the effect of viewing orientation on the recovered stellar populations, we generated six data cubes, adopting different orientations of the disc and bar. This setup is listed in Table~\ref{tab4:models} and marked as CSP.

\subsection{Extragalactic MW: IFU based on radiative transfer~(RT)}\label{sec4::mock_rt}
\subsubsection{MW beyond equilibrium: N-body/hydro simulation}\label{sec4::simulations}

While it remains challenging to find a simulation that is quantitatively identical to the MW in all aspects, especially the $\aMe$-bimodal trends, we can readily address this by using a simulation starting from conditions based on the real MW. By construction, our orbit superposition model does not capture spiral arms or other disequilibrium effects, and it also lacks the ISM. As mentioned in \citetalias{2025A&A...700A..89K}, one can use the orbit superposition solution as initial conditions~(ICs) for $N$-body/hydrodynamic simulation, where spiral arms naturally arise as a result of gravitational instability. 
To obtain self-consistent initial conditions for a simulation, we have constructed the orbit superposition solution for the DM halo, following the same approach we have developed for the APOGEE stars. In this case, the initial positions and velocities of DM particles were sampled from the NFW distribution; the particles' orbits were then integrated and weighted to match the DM 3D density distribution~(\citetalias{2025A&A...695A.220K}). 

Once the equilibrium orbit solution was obtained for both DM and stars, we sampled 6D phase-space information for each orbit with the number of particles proportional to the weight of the orbit. In this way, we arrive at the initial conditions for the MW galaxy, including 16.2~M DM particles and 7.2~M star particles. We also added a cold gas component distributed in a ring $5-20$~kpc from the Galactic centre. 

The initial conditions were used to run a simulation with SWIFT~\citep{2024MNRAS.530.2378S}, adopting star-formation and feedback models from EAGLE~\citep{2015MNRAS.450.1937C} and ISM cooling from \citep{2020MNRAS.497.4857P}, thereby enabling realistic treatment of the multi-phase ISM. 
Here, we focus on an isolated simulation evolved for 1.1 Gyr, mimicking the evolution of the MW over the same period. The stellar feedback parameters were chosen to match the present-day global star formation rate in the MW of about $1\Msunyr$. The evolution of the model shows nothing peculiar compared to many simulations of barred galaxies in this context~(see Appendix~\ref{sec4::simulations_appendix} for details). We emphasise that our orbit superposition approach enables the generation of an equilibrium for a non-axisymmetric system, and the simulation demonstrates that the bar preserves its properties well, at least on a timescale of several Gyr. Therefore, we have a representation of the MW galaxy, in which the stellar population parameters, such as abundances and ages, were derived from APOGEE and complemented by the \texttt{distmass} catalogue~\citep{2024AJ....167...73S}. Similarly to our CSP models~(see Sec.~\ref{sec4::sos}), the ages of star particles sampled from the orbits were adjusted to account for the metal-poor and young alpha-rich populations~(see~\cite{2026A&A...706A.103R} for details).

\begin{figure}[!h]
    \centering
    \includegraphics[width=1\linewidth]{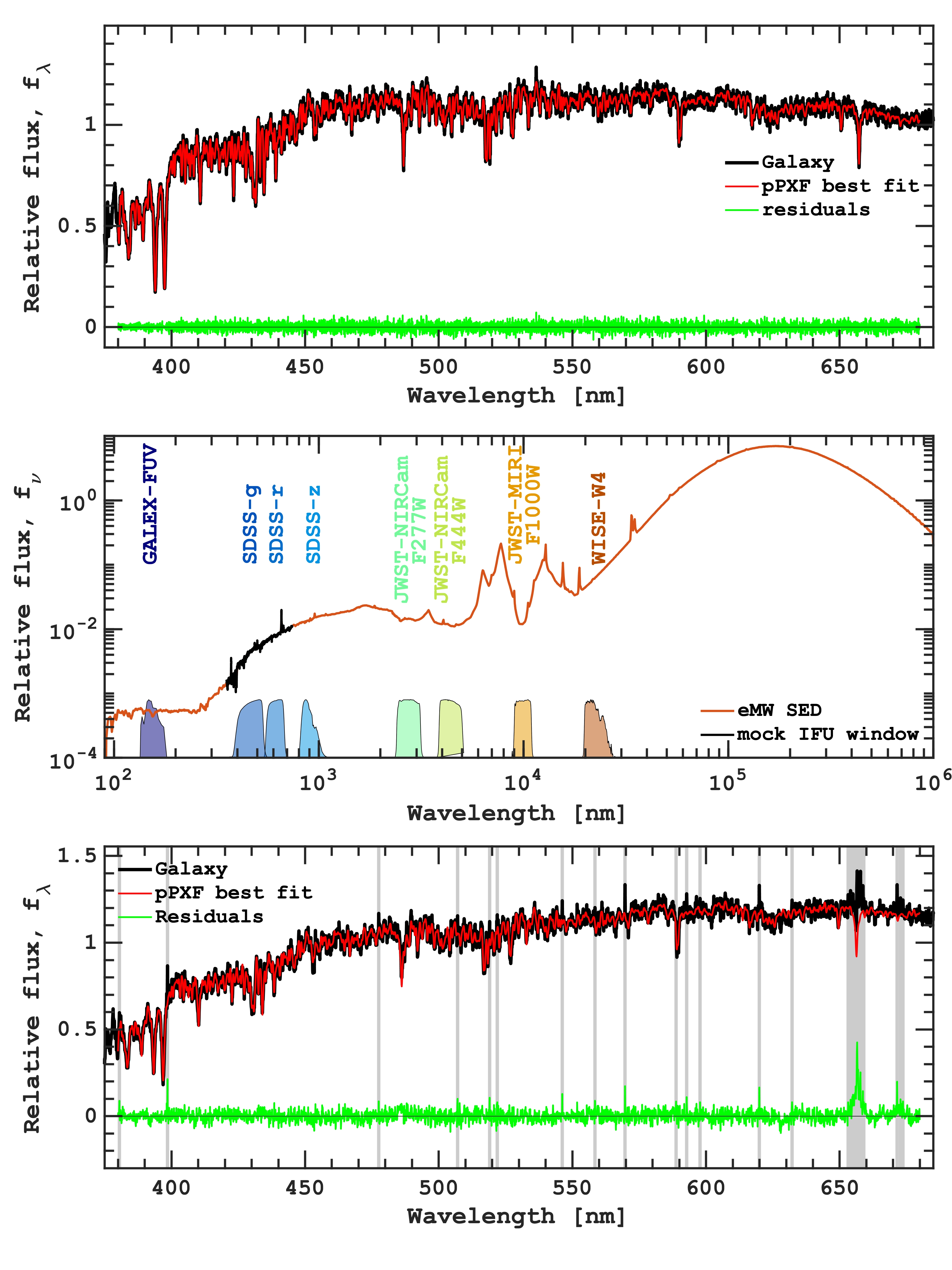}
    \caption{Examples of mock spectra generated using the composite stellar population approach and the radiative-transfer processing of the hydrodynamic simulation.
    {\it Top panel:} spectrum from a randomly selected Voronoi bin in \texttt{eMW.0074}~(CSP, Sect.~\ref{sec4::mock_vanila}), shown together with the corresponding \ppxf\ fit and residuals.
    {\it Middle panel:} full SED of the simulated MW obtained from the radiative-transfer calculation~(RT, Sect.~\ref{sec4::mock_real}); the wavelength range used for the mock IFU stellar-population analysis is highlighted in black. The filter response functions used to generate mock images (GALEX, SDSS, JWST–NIRCam, JWST–MIRI, and WISE) are displayed at the bottom of this panel.
    {\it Bottom panel:} zoom-in of the middle panel RT-spectrum extracted for the IFU analysis, shown together with the \ppxf\ fit; grey-shaded regions indicate masked emission-line wavelengths.}
    \label{fig04::spectr}
\end{figure}

\begin{figure*}
    \centering
      \adjincludegraphics[width=0.4\linewidth,
      clip,trim={{.0\height} {.0\height} {.0\height} {.0\height}}]{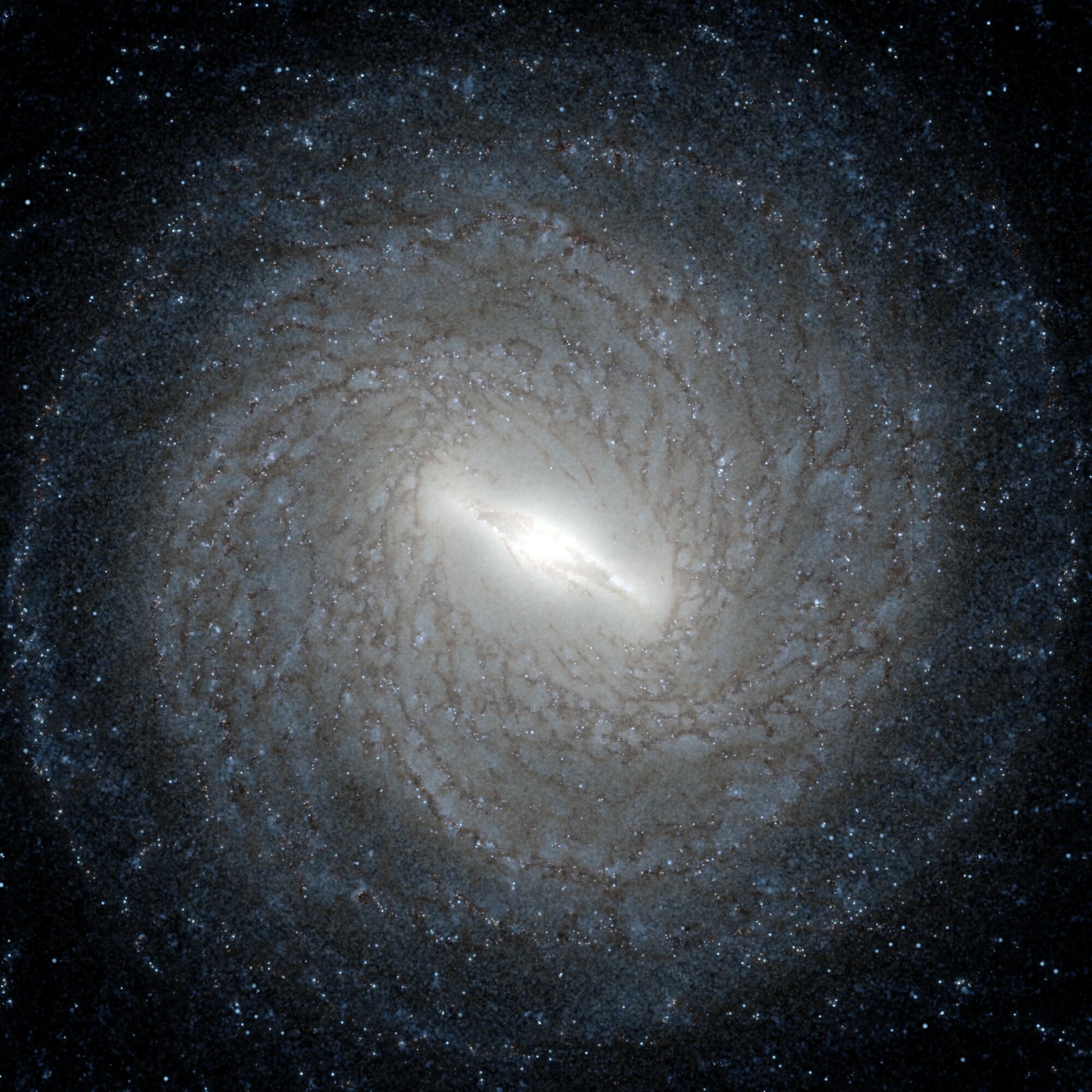}
          \adjincludegraphics[width=0.4\linewidth,
      clip,trim={{.0\height} {.0\height} {.0\height} {.0\height}}]{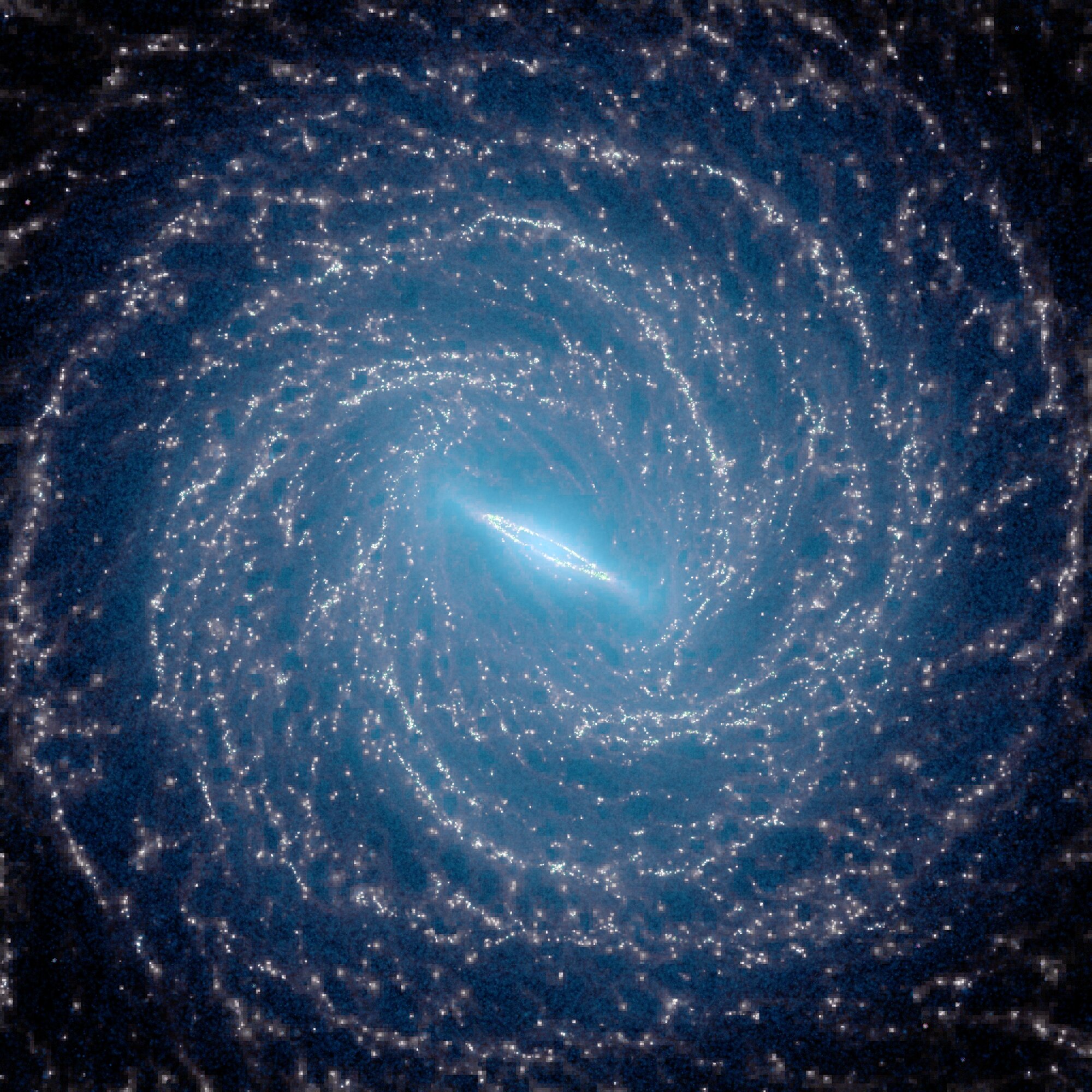} \par
      \adjincludegraphics[width=0.4\linewidth,
      clip,trim={{.0\height} {.15\height} {.0\height} {.15\height}}]{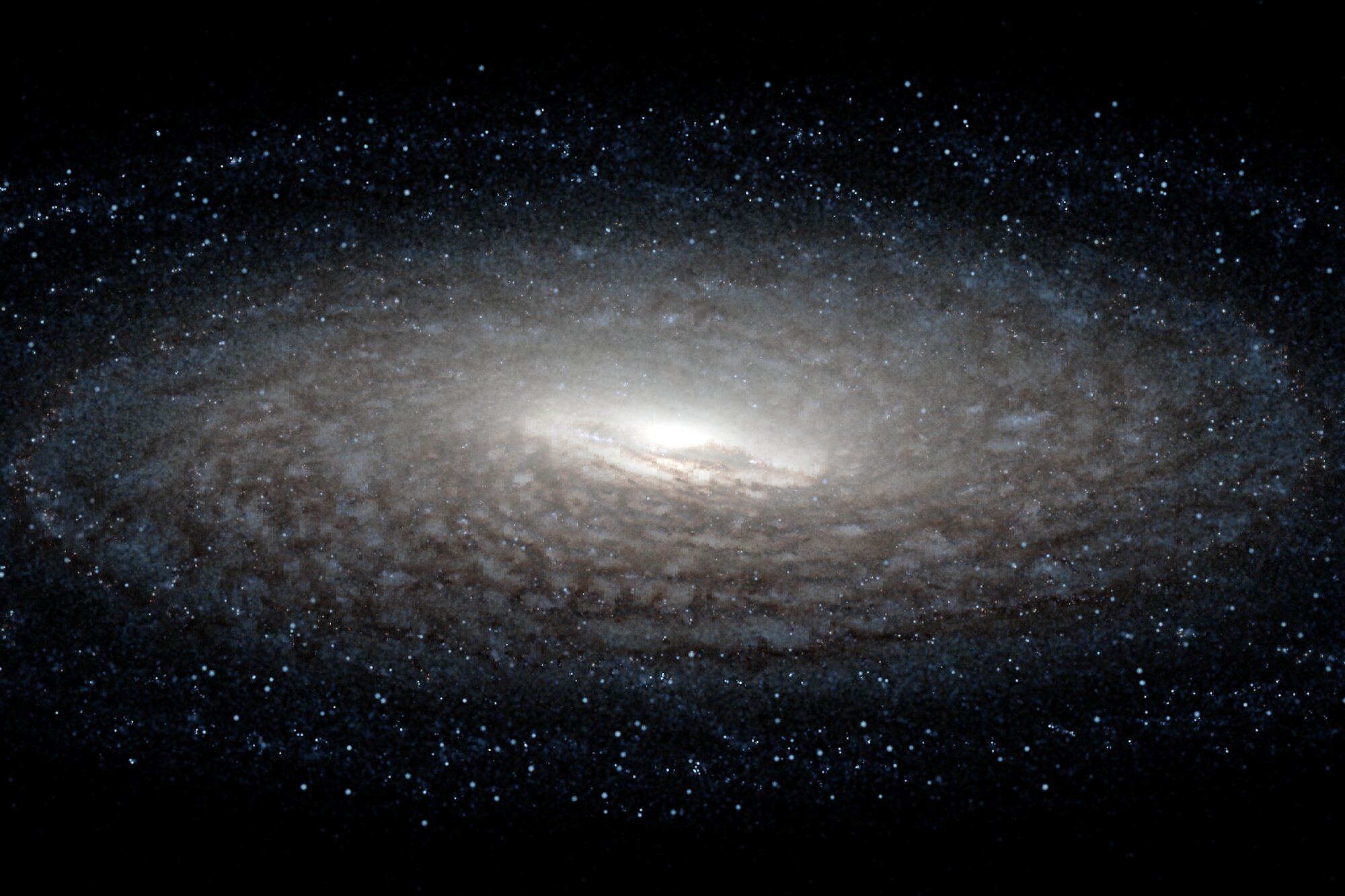}
          \adjincludegraphics[width=0.4\linewidth,
      clip,trim={{.0\height} {.15\height} {.0\height} {.15\height}}]{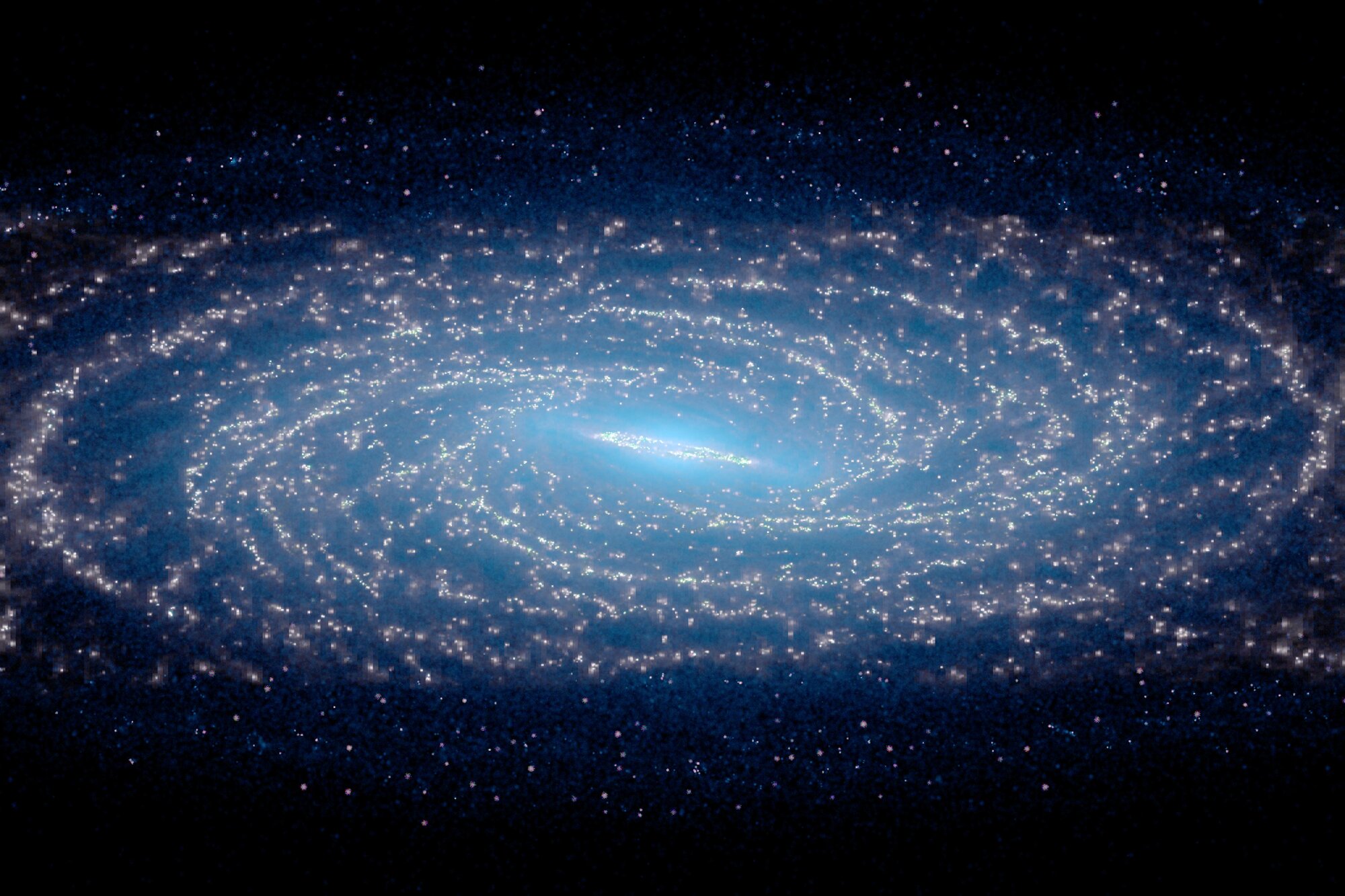} \par
          \adjincludegraphics[width=0.4\linewidth,
      clip,trim={{.0\height} {.0\height} {.0\height} {.0\height}}]{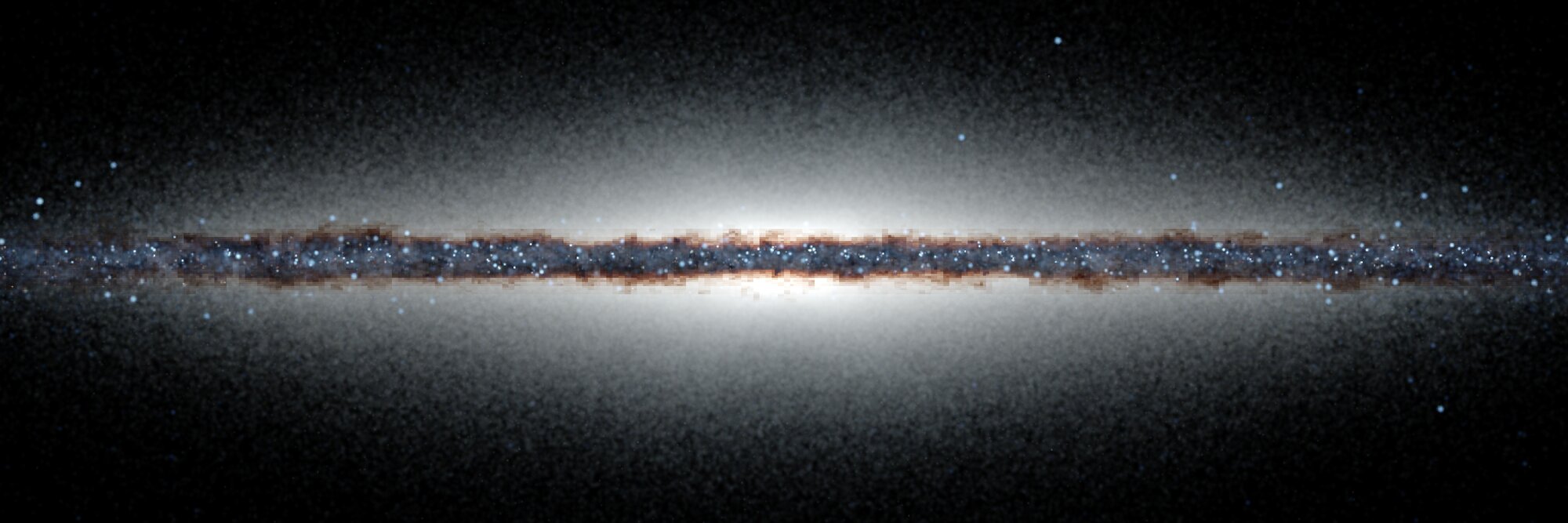}
          \adjincludegraphics[width=0.4\linewidth,
      clip,trim={{.0\height} {.0\height} {.0\height} {.0\height}}]{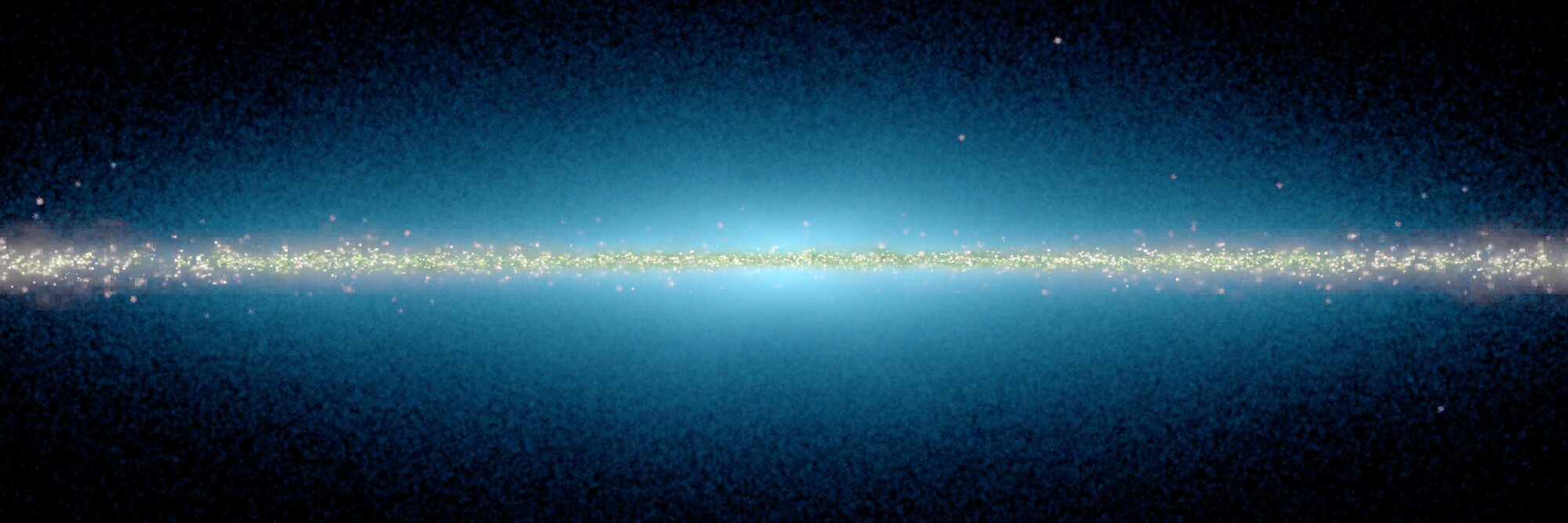}
    \caption{Synthetic multi-wavelength images of the MW galaxy obtained using \SKIRT\ radiative transfer post-processing of the hydrodynamic simulation started from the orbit superposition solution based on the APOGEE DR 17 data. {\it Left}: Composite image combining SDSS optical bands, GALEX far-ultraviolet, and WISE mid-infrared emission, highlighting the stellar continuum, ongoing star formation, and diffuse dust emission. {\it Right:} Synthetic observation in JWST-MIRI~(1000W) and JWST-NIRCam~(444W, 277W) bands, tracing the warm dust component and reprocessed starlight in the infrared regime. High-resolution images in other bands and projections are available at \url{https://cloud.aip.de/index.php/s/PHdB4qip7oooDeH}.}
   \label{fig04::images}
\end{figure*}

\begin{figure*}
    \centering
    \includegraphics[width=0.8\linewidth]{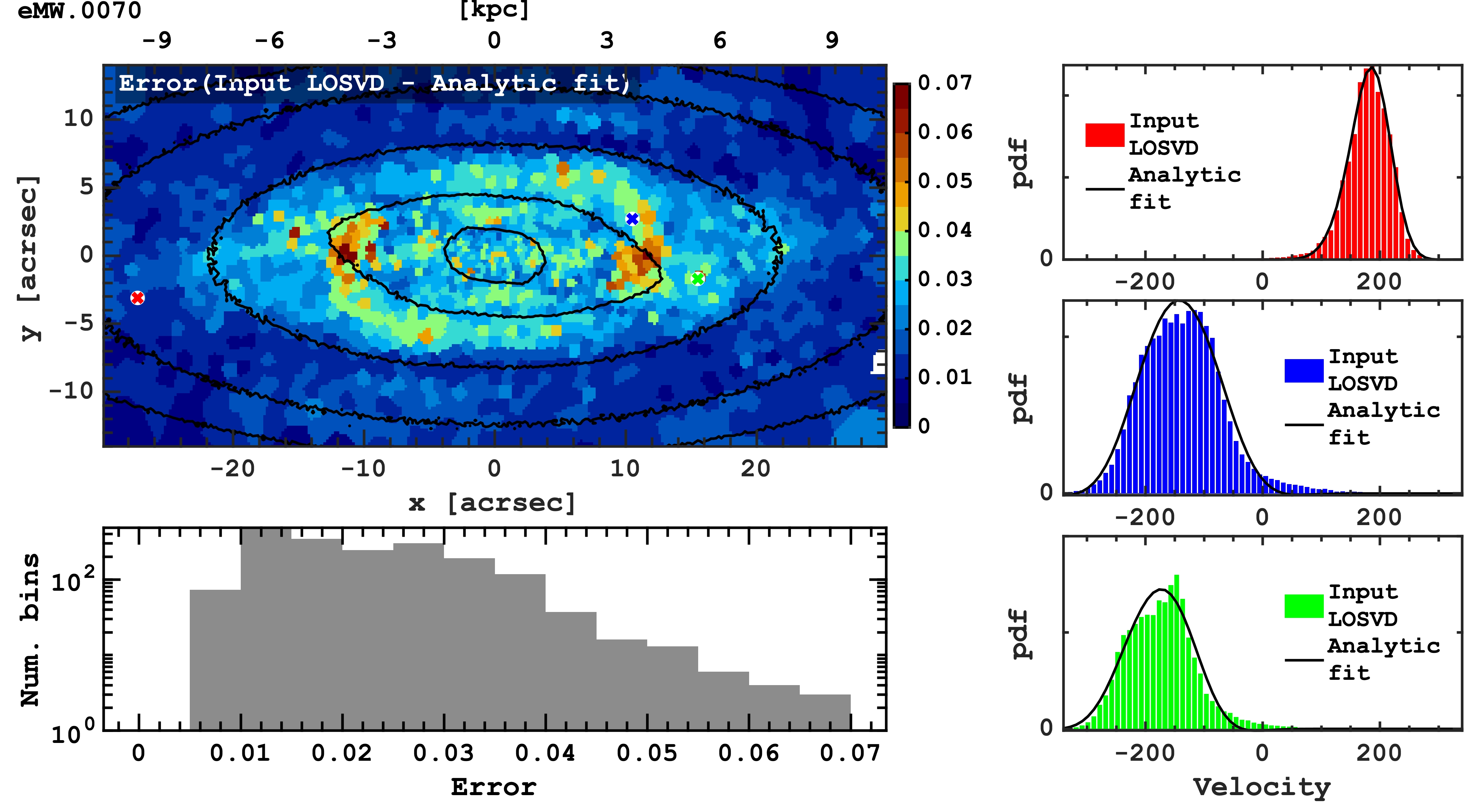}
    \caption{Deviation of the true LOSVD from the fourth-order G-H expansion fit in the \texttt{eMW.0070} model. {\it Top-left} panel shows the absolute difference between the input LOSVD and its G–H approximation in each Voronoi bin. {\it Bottom-left:} histogram indicating the number of Voronoi bins as a function of the computed deviation. {\it Right panels:} examples of the true LOSVD and the corresponding G-H fit for three randomly selected Voronoi bins, marked with crosses of the same colour in the top left panel. The figure illustrates a modest deviation of the input LOSVD from the G-H shape, which may give rise to artificial systematic patterns of the LOSVD moments obtained using parametric full-spectrum fitting approaches, which can be misinterpreted, as the ground truth is unknown in real systems. A comparison of the input LOSVD against the G–H fit for other CSP models with different disc and bar orientations is given in Fig.~\ref{fig4::losvd_error_appendix}.}
    \label{fig4::losvd_error}
\end{figure*}

\begin{figure*}
    \centering
    \includegraphics[width=1\linewidth]{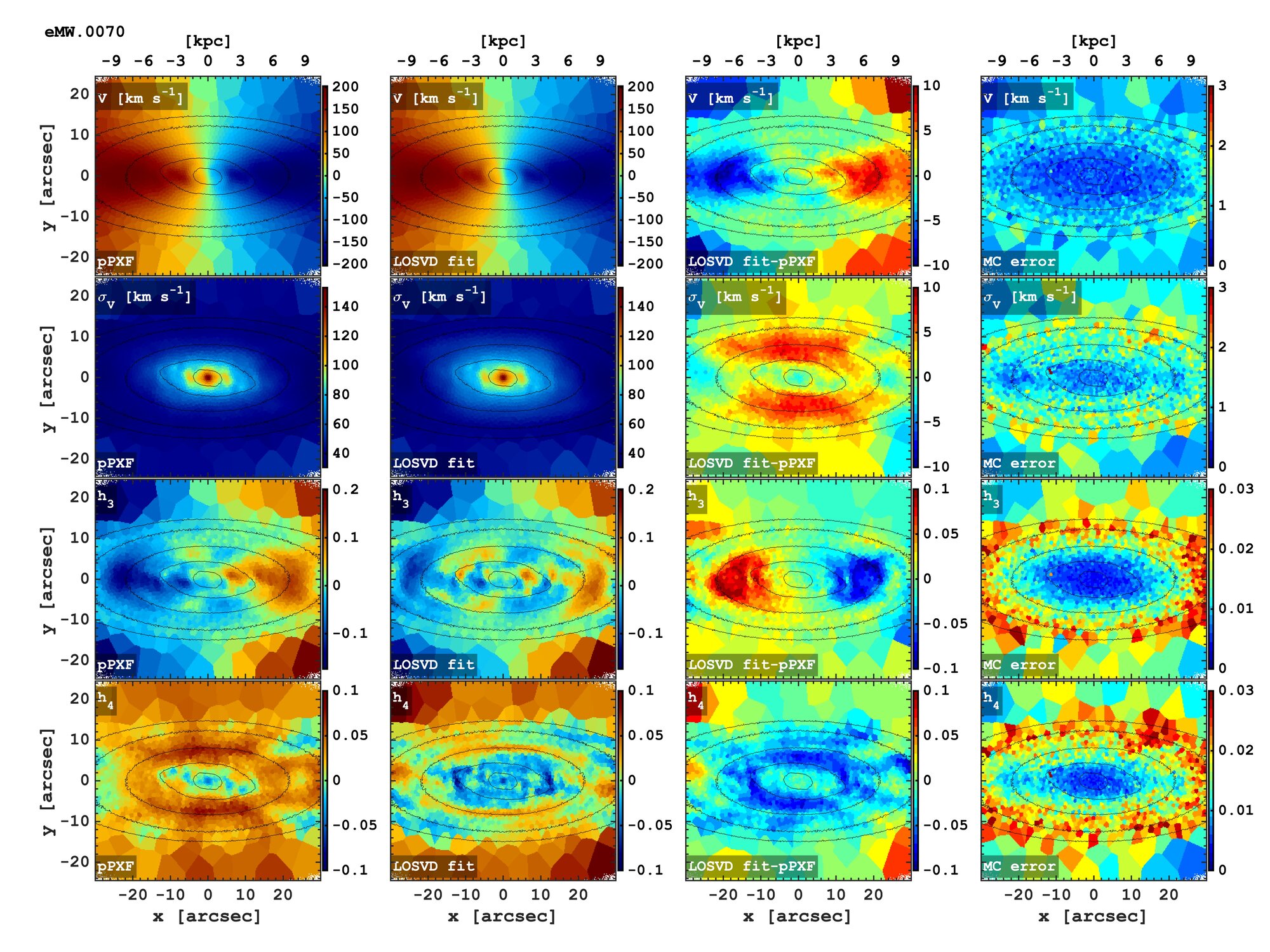}
    \caption{Comparison of the line-of-sight kinematics in the extragalactic MW model \texttt{eMW.0070}. The first column shows the \ppxf-derived kinematic maps, obtained by fitting the composite stellar populations spectra with a G–H parametrisation of the LOSVD, assuming no regularisation and no bias. The second column presents the corresponding kinematic moments extracted directly from the particle-based LOSVD using an analytic G–H fit. The third column displays the residuals between the \ppxf\ solution and the particle-based G–H moments. The fourth column shows the uncertainties estimated from the scatter among the Monte Carlo realisations.}
    \label{fig04:kinematics_70}
\end{figure*}

\subsubsection{Radiative transfer calculations}\label{sec4::mock_real}
To simulate the appearance of our Galaxy as an external system across a broad range of wavelengths, we employ the Monte Carlo radiative transfer code \SKIRT~\citep{2015A&C.....9...20C}. In these simulations, we trace photon packets emitted by stellar particles drawn from our hydrodynamic simulation, accounting for their ages and metallicities, which originate from APOGEE, and we propagate them along orbits used to sample the initial conditions for the hydrodynamic simulation. Compared to our CSP data cubes described in the previous section, we do not take into account $[\alpha/M]$ dependence because the SEDs are drawn from the~\citep{2003MNRAS.344.1000B} SED family~(BC03), under the assumption that each star particle follows a Chabrier IMF. \SKIRT\ treats SSP spectra as intrinsic input SEDs and does not modify their spectral resolution; consequently, the mock spectra inherit the native BC03 optical resolution ($\mathrm{FWHM}\simeq 3\,\AA$; \citealt{2003MNRAS.344.1000B,2011ApJS..196...22B, 2015A&C....12...33B,2015A&C.....9...20C}).

For stellar kinematics, the choice of SSP library is critical because the intrinsic spectral resolution directly limits the recoverable information content of the LOSVD. The BC03 models, based on the STELIB library, have a lower and less uniform optical resolution (FWHM $\simeq 3\,\AA$) than the MILES/sMILES models (FWHM $=2.51\,\AA$), which are therefore preferred for kinematic analyses, particularly when recovering higher-order moments ($h_3$, $h_4$; e.g. \citealt{2017MNRAS.466..798C,2008MNRAS.385.1998K}). Even after matching the line-spread functions, BC03-based spectra retain a reduced sensitivity to cold kinematic components and non-Gaussian LOSVD features.

Several studies have systematically compared BC03 SSP models, which are based on the STELIB stellar library, with Vazdekis/MILES models and other spectral libraries, and have reported significant differences in spectral features and inferred stellar population parameters (e.g. \citealt{2008MNRAS.385.1998K}, \citealt{2010A&A...515A.101C}, \citealt{2019MNRAS.485.1675G}). In particular, BC03-based models are known to exhibit systematic biases at non-solar metallicities, reflecting limitations of the underlying STELIB library. In this situation, we therefore expect more robust and internally consistent results from CSP modelling, where MILES-based templates are employed, than from more realistic radiative-transfer calculations, which necessarily rely on BC03 templates. This limitation should be borne in mind when interpreting differences between the two modelling approaches.

The RT modelling includes stochastic dust emission, with a dust-to-gas mass ratio of $0.01$. We adopt 15 grain sizes for both silicate and hydrocarbon dust species. To include radiation from star-forming regions, we adopted MAPPINGS III star-forming region template SEDs~\citep{2008ApJS..176..438G}, parametrised on metallicity, compactness, ISM pressure and photodissociation region covering factor, and scaled by star formation rate following the approach described by \cite{2023ApJ...957....7F}.

To generate realistic and detailed IFU data cubes, we performed 26 radiative-transfer simulations for a given galaxy projection, with a different random seed and using $6\times10^9$ photon packets each. This strategy allowed us to balance computational cost and memory requirements (with peak usage of 122 GB per run), while simultaneously providing multiple realisations from which we could estimate the spectral noise input for the full spectra fitting.

We adopted a broad wavelength range from 0.09 to $10^6\ \mu m$, sampled at 1500 points, to construct multi-band images spanning the UV to the far-infrared. For the IFU analysis, we further extracted a high-resolution segment of the SED covering the optical regime (350–740 nm), sampled with 4300 points, matching the native wavelength grid of the sMILES template library. This ensures that the \SKIRT\ output can be used directly for full spectra fitting without any additional resampling.

An example SED from one of the radiative-transfer realisations is shown in Figure~\ref{fig04::spectr}. The middle panel highlights the optical region, where the high spectral resolution reveals prominent emission lines and mid-infrared dust and nebular features. The same panel also marks the filter transmission curves used to generate the multi-band composite images of the MW displayed in Fig.~\ref{fig04::images}, including GALEX–FUV, several SDSS and WISE filters, which jointly produce a typical barred-spiral morphology and look remarkably similar to the SDSS images of NGC~5055 and NGC~3351~\citep[see Fig.10 in][]{2021MNRAS.508.4459F}. The right-hand panel shows a JWST–MIRI/NIRCam rendering, closely resembling the appearance of nearby spiral galaxies in the JWST–PHANGS survey~\citep{2023ApJ...944L..17L, 2024ApJS..273...13W}.

\subsection{Full spectra fitting}\label{sec4::ppxf}
To derive the kinematic and stellar population properties from the mock IFU observations of the eMW, we utilised the Penalised Pixel-Fitting (\ppxf) method \citep{2004PASP..116..138C, 2017MNRAS.466..798C}, fitting the line-of-sight velocity distribution directly at the pixel level to capture velocity information from the galaxy spectra precisely.

For the adaptive spatial binning of the two-dimensional data, we employed the Voronoi binning scheme of \citet{2003MNRAS.342..345C}. In the case of the CSP data cubes, we assumed that the mean signal-to-noise ratio scales with the square root of the number of star particles contributing to each spatial element. Given the very high particle numbers in our orbit superposition solution~(>300M), we imposed a minimum of 160,000 star particles per Voronoi bin. Noise was then added to the binned spectra by injecting Gaussian noise with a fixed amplitude corresponding to S/N = 80, independent of wavelength. This wavelength-independent treatment is, of course, a simplification, adopted here because no specific instrumental setup was assumed for the mock extragalactic observations of the MW. 

For the RT-generated datacubes, we adopted a strategy different from that used for the CSP models. For each eMW model, we performed 26 independent RT realisations with different random seeds. The mean of these realisations was used as the input IFU cube for the fitting, while the noise was estimated from the standard deviation of the individual RT runs about the mean. The resulting wavelength-dependent S/N was then used to define the Voronoi binning, ensuring a spatially consistent signal-to-noise ratio across the full field of view and a self-consistent coupling between the binning and the spectral-noise properties. Even with $6\times10^{9}$ photon packets emitted by $6\times10^{6}$ star particles and $6\times10^{6}$ gas particles, and after averaging over multiple independent datacube realisations, the RT-based IFU data still reach only relatively modest S/N. To preserve the spatial resolution of the mock observations and avoid excessively large Voronoi bins, we therefore adopted a target value of $\mathrm{S/N}=40$ for the binning. This choice represents a compromise between spectral quality and spatial fidelity: increasing the target S/N would require larger bins and hence stronger mixing of stellar populations with different kinematic properties in the plane of the sky. As a result, although the mean velocity and velocity-dispersion fields can still be recovered reasonably well, the degeneracy associated with population mixing becomes more severe for the higher-order LOSVD moments, contributing to the larger biases in $h_3$ and $h_4$. In addition, dust further degrades the effective data quality through wavelength-dependent absorption and re-emission, compounding the information loss and making detailed stellar-population constraints more uncertain.

Since \SKIRT\ generates RT spectra from BC03 templates, the resulting cube is inherently lower in resolution than that of MILES. The BC03 SSPs have an optical spectral resolution of $\mathrm{FWHM}\simeq 3~\text{\AA}$, lower and less uniform than the $2.51~\text{\AA}$ resolution of the MILES/sMILES models, which are therefore better suited for detailed stellar kinematics. Nevertheless, to fit BC03-based mock spectra with sMILES templates, we matched the line-spread functions by convolving the templates to the BC03 effective resolution (assumed $\mathrm{FWHM}\approx 3.0\,\AA$ in the optical), i.e. with a Gaussian kernel of $\mathrm{FWHM}_{\rm conv}=\sqrt{\mathrm{FWHM}_{\rm BC03}^2-\mathrm{FWHM}_{\rm MILES}^2}\approx 1.64\,\AA$. The convolution was performed on the linear wavelength grid prior to fitting. However, even with proper LSF matching, the RT-based cube is expected to have lower kinematic information content than a MILES-based CSP cube. 

We emphasise that the mock datacubes produced and analysed in this work are not intended to reproduce the full instrumental response of any specific IFU instrument. Instead, they should be regarded as idealised integrated-light datacubes designed to isolate the effects of unresolved stellar-population mixing and line-of-sight kinematics under controlled conditions. We tested the sensitivity of the spectral fitting to the adopted wavelength range, in particular by restricting the analysis to a wavelength interval within the MUSE spectral coverage~\citep[see also][]{2024MNRAS.534.1175W}, and found that, in this relatively idealised setup, the main recovered kinematic and stellar-population properties do not change significantly. We therefore use the full spectral range available in the MILES-based templates and do not impose a specific instrumental wavelength coverage, point-spread function, line-spread function, atmospheric transmission, or telluric residuals. This choice avoids introducing additional instrument-dependent systematics and ensures that any differences between the intrinsic and recovered stellar-population or kinematic quantities can be attributed primarily to the limitations of the spectral decomposition itself rather than to observational effects. Consequently, the results should not be interpreted as a direct performance forecast for MUSE or any other particular instrument, but rather as a controlled reference case against which more realistic, instrument-specific mock observations can be compared in future work.

Unless stated otherwise, the regularised solutions shown in this work adopt \texttt{regul = 5}. This value was chosen as a representative smoothing strength that suppresses noise-driven small-scale structure while preserving the main large-scale trends of the solution. We do not interpret it as a unique physically optimal value; rather, we use it to illustrate the sensitivity of the recovered distributions to regularisation.

\section{Results}\label{sec4::results_total}

\subsection{IFU based on Composite Stellar Populations}\label{sec4::CSP_results}

Before analysing the extragalactic eMW using \ppxf, we first examine the projected kinematics of the stellar populations in the \texttt{eMW.0070} model~(see other models in the Appendix~\ref{sec04::losvd_error_appendix}). Since \ppxf\ assumes that the line-of-sight velocity distribution~(LOSVD) can be adequately described by a Gauss–Hermite~(G–H) parameterisation, it is important to assess whether the true LOSVD of the model galaxy is well represented by this formalism. In Fig.~\ref{fig4::losvd_error}, we quantify the deviations between the intrinsic LOSVD and its best-fitting analytic G-H representation, showing both a spatial map of the fitting errors~(top left) and illustrative examples of the LOSVDs with their corresponding fits~(right). The largest discrepancies occur in the inner disc and near the edges of the bar, where stellar populations with distinct kinematic properties coexist within the same gravitational potential and where the relative contributions of the thin and thick discs become comparable. In these regions, the LOSVDs are often multi-modal, a structure that cannot be faithfully represented by a low-order G-H expansion \citep{2013ApJ...769..105K, 2021A&A...646A..31F, 2025A&A...701A..12R, 2025arXiv251103663J}. Increasing the expansion order does not necessarily remove this limitation, as higher-order G-H terms can instead reproduce multi-modality through oscillatory features, often at the cost of introducing unphysical structure in the LOSVD wings~\citep{1993MNRAS.265..213G,2022arXiv220603925H}.
 
These limitations motivate caution when applying the G-H formalism, originally developed and widely used for early-type galaxies~\citep{1993MNRAS.265..213G,2016ARA&A..54..597C}, to late-type spiral systems and, in particular, to MW analogues. More flexible LOSVD descriptions, including non-parametric or explicitly multi-component approaches, may be better suited to such cases~\citep[e.g.,][]{1993MNRAS.264..712K,2006MNRAS.365...46O,2013ApJ...769..105K}. In this work, however, we retain the standard G-H parametrisation to enable direct comparison with the existing literature and to use the conventional language of higher-order kinematic moments. Our goal is therefore not to optimise the LOSVD representation itself, but to assess how this commonly adopted assumption propagates into the recovered stellar-population properties. We expect the resulting biases to be most important in regions where several dynamically distinct populations contribute comparably to the light, such as the bar-disc interface, the thin-thick disc transition, and the transition from the thick disc to the stellar halo.

In the following sections, we focus on the LOS kinematics of the eMW, as recovered from mock IFU data cubes using \ppxf. We restrict the analysis to two representative models, \texttt{eMW.0070} and \texttt{eMW.0074}, which differ in bar orientation and viewing projection. Model \texttt{eMW.0070} is observed at an inclination of $i = 70^\circ$ with the bar oriented at $27^\circ$, while model \texttt{eMW.0074} is viewed edge-on ($i = 90^\circ$) with the bar seen side-on. Parameters of all other models are given in Table~\ref{tab4:models}, and their brief analysis is presented in the Appendix~\ref{sec4::kinematics_appendix}.

\begin{figure*}
    \centering
    \includegraphics[width=1\linewidth]{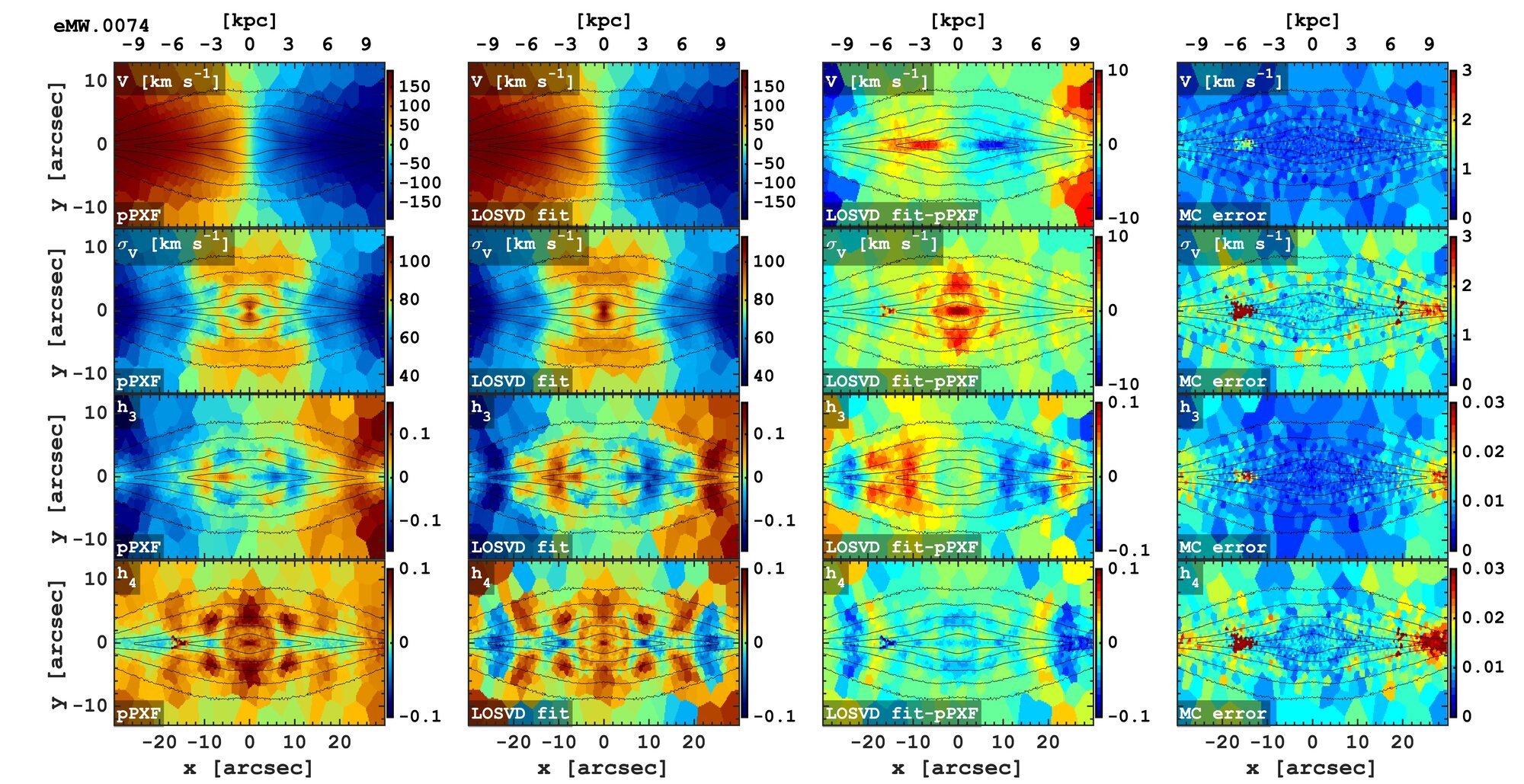}
    \caption{Same as in \ref{fig04:kinematics_70} but for model \texttt{eMW.0074}, revealing the edge-on projection with the side-on bar orientation.}
    \label{fig04:kinematics_74}
\end{figure*}

\begin{figure*}
    \centering
    \includegraphics[width=1\linewidth]{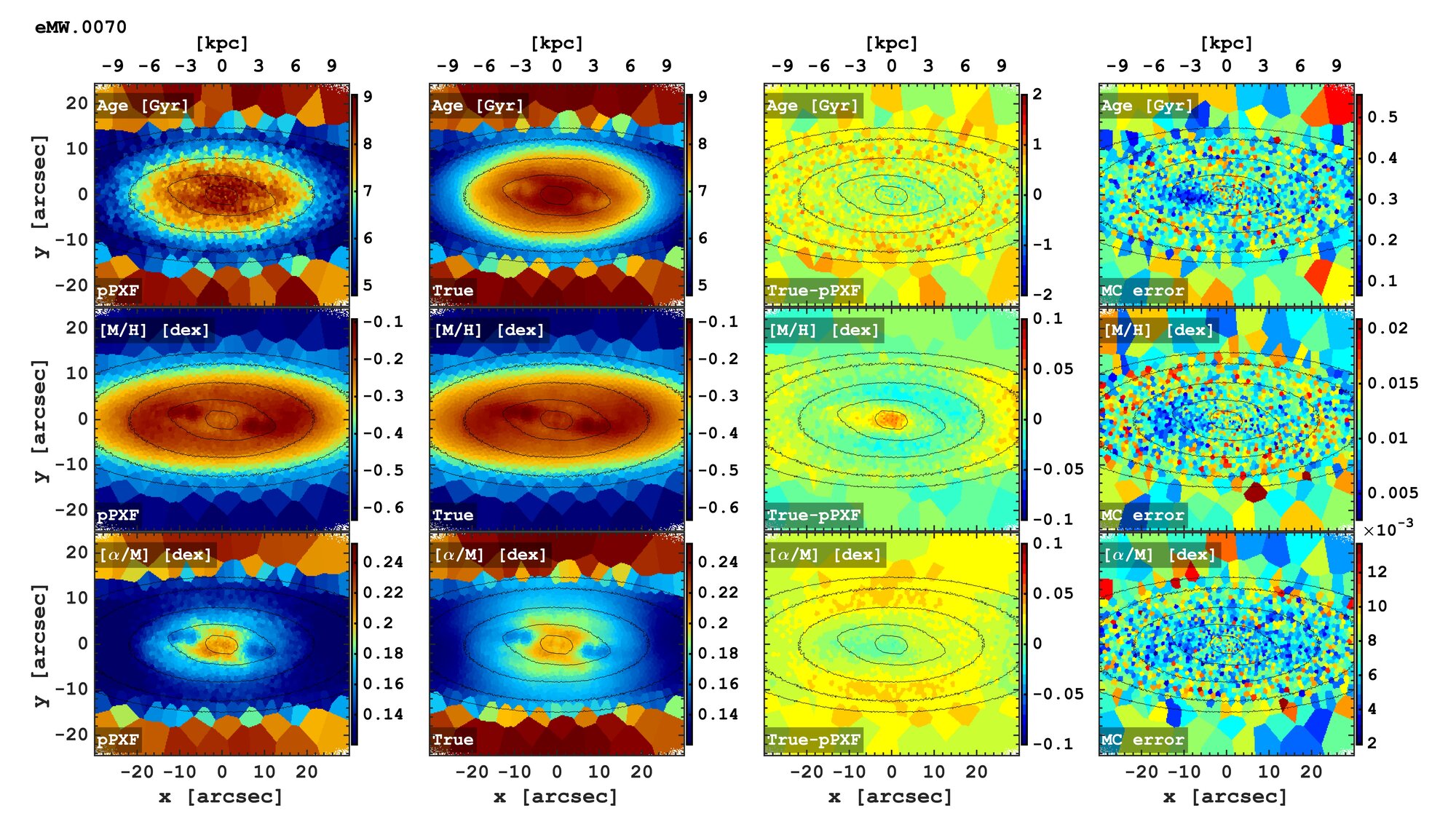}
    \caption{Comparison of the stellar populations parameters in the extragalactic MW model \texttt{eMW.0070}. The first column shows the \ppxf-derived age, $[M/H]$ and $[\alpha/M]$ maps, obtained by weighting template values by the weights of the solution, assuming no regularisation and no bias. The second column presents the corresponding mean value maps extracted directly from the particle data. The third column displays the residuals between the \ppxf\ solution and the particle-based values. The fourth column shows the uncertainties estimated from the scatter among the Monte Carlo realisations.}
    \label{fig04:sfh_70}
\end{figure*}

\subsection{Composite stellar populations models: eMW kinematics}\label{sec4::results_kinematics_csp}

Here, we present the line-of-sight kinematic analysis for two reference CSP-based models. For these models, the kinematics were recovered from 30 Monte Carlo \ppxf\ realisations, following a procedure similar to that adopted in \cite{2018MNRAS.480.1973K,2022A&A...659A.191E}. In each realisation, we perturbed the noiseless CSP spectra by adding an independent Gaussian noise realisation and then repeated the \ppxf\ fit. The remaining models, presented in the Appendix, were analysed using a single \ppxf\ fit in order to avoid unnecessary computational expense.

Figures~\ref{fig04:kinematics_70} and ~\ref{fig04:kinematics_74} present a comparison between the intrinsic and recovered LOSVD kinematics in the \texttt{eMW.0070} and \texttt{eMW.0074} models, respectively. From left to right, the columns show the kinematic moments recovered from the mock IFU data using \ppxf~(averaged over 30 Monte Carlo realisations), and the corresponding best-fitting G-H LOSVD representation based on the star-particle data. The third column highlights the residuals between the analytic G-H representation and the \ppxf-recovered kinematics. The fourth column shows the standard deviation of the kinematic moments obtained from independent Monte Carlo realisations (using different random seeds for the spectral noise) and therefore quantifies the statistical uncertainties in the recovery. The rows display, from top to bottom, the mean line-of-sight velocity $V$, the velocity dispersion $\sigma_V$, and the higher-order G-H moments $h_3$ and $h_4$. The \ppxf\ velocity field captures the large-scale non-axisymmetric rotation pattern induced by the bar, while the dispersion map shows a pronounced central peak associated with the dynamically hot bulge/bar component. The $h_3$ and $h_4$ maps exhibit coherent spatial structures that reflect asymmetries and variations in the LOSVD shape, driven by the interplay among thin and thick discs, X-shaped bulge and bar orbits. 

The line-of-sight kinematics of barred galaxies hosting boxy/peanut-shaped bulges have been extensively investigated in both theoretical~\citep[e.g.,][]{2005ApJ...626..159B, 2015MNRAS.450.2514I, 2016MNRAS.456..692M, 2020ApJ...889...39V, 2024MNRAS.527.3038Z} and observational studies~\citep[e.g.,][]{2016A&A...591A...7G,2017MNRAS.466L..93G,2020A&A...637A..56N,2025A&A...700A.237F}. We therefore do not aim to revisit or substantially extend these analyses here. A key distinction of our models is the explicit inclusion of two dynamically and chemically distinct thin and thick stellar discs, which is not always accounted for in the studies above and may affect the resulting kinematic signatures. Consequently, a direct one-to-one match with previous theoretical results is not expected, although the main qualitative trends are preserved. In the following, we briefly describe the line-of-sight kinematic features observed in Figs.~\ref{fig04:kinematics_70} and \ref{fig04:kinematics_74}.

\begin{itemize}
    \item The LOS velocity maps primarily reflect the regular rotation of the galaxy, with clear ``perturbations'' induced by the presence of the bar giving rise to non-circular motion in the inner region, seen also as a tilt of the line of nodes~(\texttt{eMW.0070}). In the edge-on projection, prominent kinematic arc-like features perpendicular to the midplane coincide spatially with the X-shaped lobes of the bulge~(\texttt{eMW.0074}).
    
    \item The LOS velocity dispersion exhibits a central peak in the $i = 70^\circ$ projection, elongated along the bar, and a secondary enhancement near the bar ends at $\sim 4$~kpc~(\texttt{eMW.0070}). In the edge-on view~(\texttt{eMW.0074}), the dispersion displays a more complex morphology, with a central peak extended perpendicular to the midplane and an additional oval-shaped enhancement at a larger distance from the centre, tracing the X-shaped bulge structure further out.
    
    \item The $h_3$ moment is largely anti-correlated with the mean velocity in the $i = 70^\circ$ projection, providing a clear signature of the disc rotation. The bar causes $h_3-V$ more correlation while also affected by the presence of the X-shaped bulge~(\texttt{eMW.0070}). In the edge-on configuration~(\texttt{eMW.0074}), $h_3$ shows a positive correlation with velocity in the innermost regions, transitions to an anti-correlation associated with the X-shaped bulge, and alternates again between correlated and anti-correlated behaviour at larger radii.
    
    \item The recovery of the $h_4$ moment is observationally challenging; models suggest that in the case of a strong bar, the $h_4$ profile exhibits a pronounced and relatively flat central minimum, followed by a sharp increase and a more gradual decline at intermediate radii. In agreement, the $h_4$ moment decreases to negative values along the bar for the $i = 70^\circ$ projection~(\texttt{eMW.0070}), with some additional minor features, and rises to positive values outside the bar without a clear large-scale pattern. In the edge-on case~(\texttt{eMW.0074}), the $h_4$ map broadly mirrors features seen in the velocity dispersion, remaining mostly positive but becoming weakly negative near the midplane.
\end{itemize}

\begin{figure*}
    \centering
    \includegraphics[width=1\linewidth]{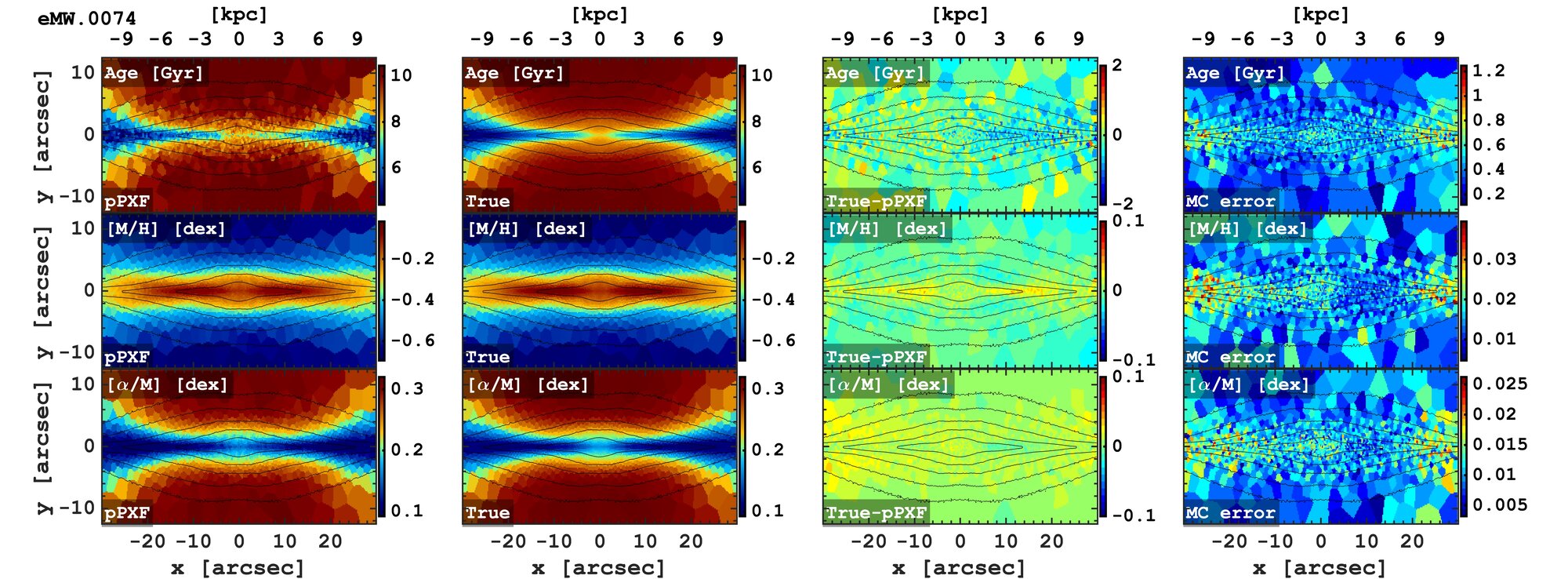}
    \caption{Same as in \ref{fig04:sfh_70} but for model \texttt{eMW.0074}, revealing the edge-on projection with the side-on bar orientation.}
    \label{fig04:sfh_74}
\end{figure*}

Our primary focus is on assessing how reliably various kinematic features can be recovered through full-spectrum fitting of unresolved stellar populations in the case of the MW or more broadly MWAs. Overall, we find good agreement between the analytic G-H LOSVD representation and the kinematics recovered with \texttt{pPXF} in both models. The variations in the kinematic parameters are small across all LOSVD moments, remaining below $\sim 3\,\mathrm{km\,s^{-1}}$ for the mean velocity and velocity dispersion, and below $\sim 0.03$ for the higher-order moments $h_3$ and $h_4$. Therefore, systematic residuals are not attributable to fitting uncertainties.

\begin{itemize}
    \item The residuals in the line-of-sight velocity and velocity dispersion do not exceed $\sim 5-10$\,\kmps across the field of view; while \citet{2024MNRAS.534.1175W} reported higher values~(up to 17 \kmps) in a similar analysis, which, however, can be decreased by increasing spectral resolution. However, similarly to \citet{2024MNRAS.534.1175W}, we identify systematic patterns in these residuals. Both \texttt{eMW.0070} and \texttt{eMW.0074} exhibit mismatches in the LOS velocity along the disc midplane, as expected \citep{2025arXiv251103663J}, where the contribution of dynamically cold thin-disc orbits cannot be fully captured in the presence of a broader LOSVD component associated with the thick disc. In addition, the velocity dispersion is slightly underestimated in the ''off-plane'' regions for the \texttt{eMW.0070} model.
    
    \item Although the $h_3$ and $h_4$ moments qualitatively reproduce the main features and the global pattern of the analytic G-H LOSVD fits, while no kinematic bias in the \ppxf\ fitting was used, systematic offsets in their amplitudes are present in both models considered here. One likely origin of these differences is that the mock spectra combine stellar populations with different ages, metallicities, and potentially different intrinsic kinematics. In the analytic LOSVD analysis, the kinematic distribution is fitted directly and independently of the stellar-population mix, whereas in the \ppxf\ fits all templates are effectively assigned a common LOSVD. As a result, \ppxf\ may partly compensate for population-dependent kinematic differences through the template mixture and continuum/spectral-feature matching, leading to biased higher-order G-H moments. Similar effects, arising from the coupling between stellar-population variations and LOSVD recovery in full-spectrum fitting were discussed in ~\cite{2024MNRAS.534.1175W}.

\end{itemize}

Using \ppxf\ to fit mock MW CSP data, we find that the mean line-of-sight velocity and velocity dispersion are recovered robustly across most of the field, including the complex inner regions. This supports the use of integrated-light fitting to study the large-scale kinematics of the bar, bulge, and thin-thick disc transition in MW analogues. The interpretation of higher-order G-H moments is more subtle. In regions where the intrinsic LOSVD is multi-component or strongly non-Gaussian, the best-fitting G-H expansion is not necessarily a faithful representation of the full LOSVD. Thus, the systematic offsets between the \ppxf\ maps and the direct G--H fits to the intrinsic LOSVD do not only measure limitations of the spectral fitting, but also the limitations of the G-H parametrisation itself. Fitting higher-order terms, for example up to $h_6$ in the highest-S/N bins, may help determine whether part of the apparent $h_4$ offset is redistributed into higher-order moments. Nevertheless, even if such fits improve the formal agreement, the resulting moments may still be difficult to interpret physically in regions where several dynamical components overlap. 

\subsubsection{Composite stellar populations models: projected age and abundance information}\label{sec4::age_abundance_csp}

Figures~\ref{fig04:sfh_70} and~\ref{fig04:sfh_74} present the projected maps of mean stellar age, metallicity, and $\alpha$-enhancement in separate rows for models \texttt{eMW.0070} and \texttt{eMW.0074}, respectively. For each quantity, the first column shows the \ppxf\ solutions averaged over the Monte Carlo realisations, the second column displays the corresponding mass-weighted values computed directly from the input particle data, and the third column presents the residuals between the recovered and input maps. The fourth column reports the uncertainties estimated from the scatter among the Monte Carlo realisations.

In both models, the mean stellar population properties are recovered with good precision. In the \texttt{eMW.0070} model, the most prominent bar-related signatures are well reproduced, manifesting as slightly younger ages and lower $\alpha$-enhancement, accompanied by enhanced metallicity. Similar patterns have been detected both in the MW and in external barred galaxies~\citep{2020A&A...637A..56N, 2024MNRAS.534.2438N}, and are commonly interpreted as the result of kinematic separation during bar formation, which traps metal-rich stellar populations onto more elongated orbits \citep{2013A&A...553A.102D, 2017MNRAS.469.1587D}. In the edge-on projection (\texttt{eMW.0074}), a clear stratification of stellar populations is visible, with younger, more metal-rich, and $\alpha$-poor stars dominating near the midplane, contrasted against an older, more metal-poor, and relatively $\alpha$-enhanced thick disc component. This vertical population separation is well established in the MW and has also been observed in edge-on external disc galaxies \citep[see, e.g.,][]{2019A&A...623A..19P, 2019A&A...625A..95P, 2025A&A...705A...1F}.

Nevertheless, we detect a weak but systematic offset between the recovered and true projected mean stellar population properties, with typical differences of $\sim 0.5$--$1$~Gyr in age and $\sim 0.03$--$0.05$~dex in metallicity and $\alpha$-enhancement. These offsets are not accompanied by any pronounced spatial structure in the uncertainties derived from the Monte Carlo realisations, indicating that they are systematic rather than noise-driven~\citep[see also][who attributed this to insufficient spectral resolution]{2024MNRAS.534.1175W}.

The projected abundance trends in the eMWs, presented here, do not deviate from other studies of the MW resolved stellar populations~\citep[e.g.,][]{2023A&A...674A..38G, 2023ApJ...954..124I}, which we have discussed intensively in \cite{2025A&A...700A..89K}. Therefore, the stellar population maps shown in Figs.~\ref{fig04:sfh_70} and~\ref{fig04:sfh_74} once obtained from the unresolved spectra can serve as a reference for identifying MWAs. Deviations from these templates, when observed, can then be used to constrain the diversity of evolutionary pathways followed by MW–like galaxies.

\begin{figure*}
    \centering
    \includegraphics[width=0.8\linewidth]{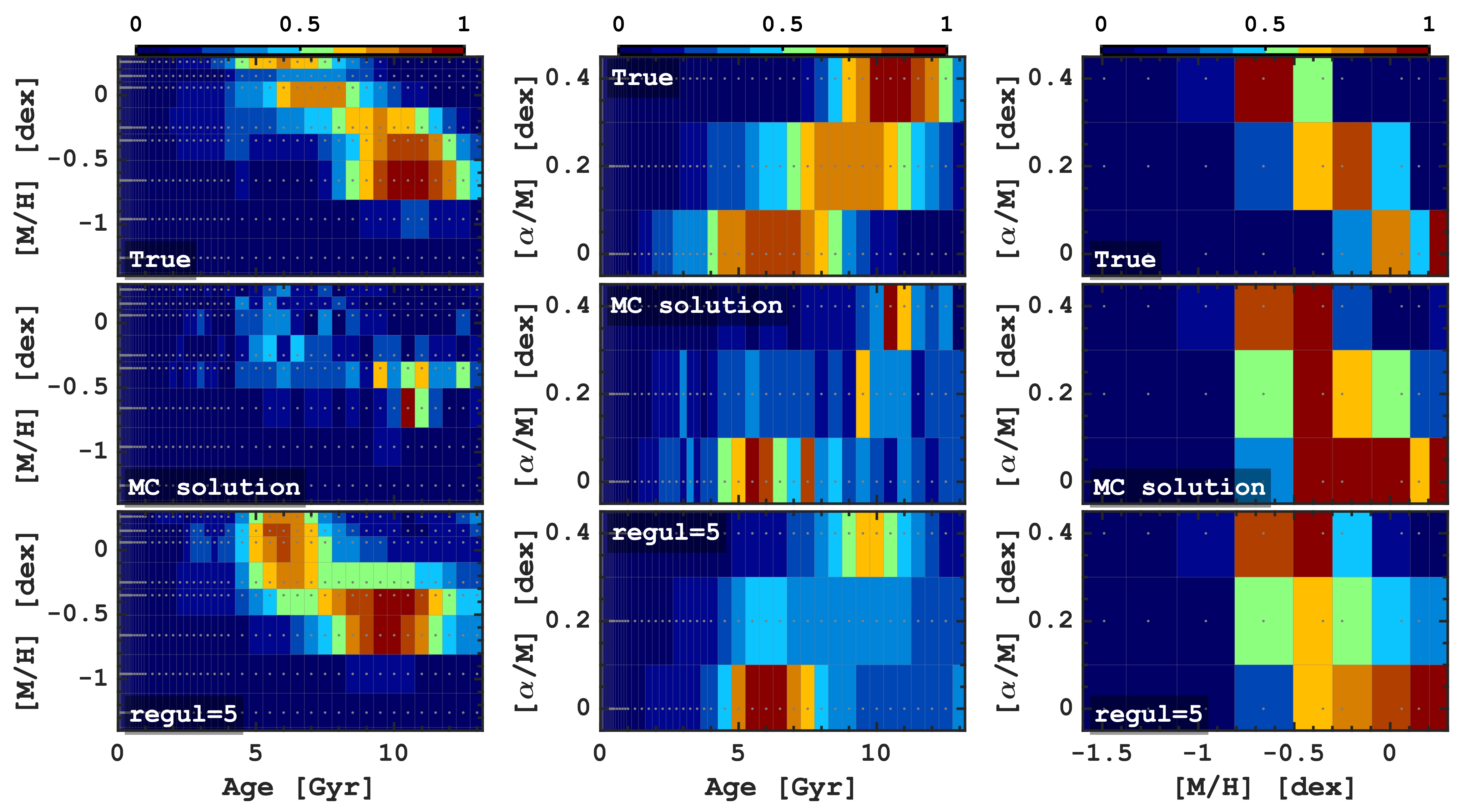}
    \caption{Comparison of the recovered stellar-population distributions in the CSP models. From left to right, the panels show the age--$\rm [M/H]$, age--\aMe, and \aMe--$\rm [M/H]$ planes. The top row presents the orbit-superposition solution projected onto the sMILES SSP template grid, that is, the true distributions. The middle row shows the \ppxf\ results from the Monte Carlo runs without regularisation. The bottom row shows the corresponding \ppxf\ results obtained with regularisation set to 5.}
    \label{fig04::age_abundance}
\end{figure*}

\subsubsection{Composite stellar populations models: age-abundance relations}\label{sec4::age_abundance_relations_csp}

We emphasise that the primary goal of this and the following section is not to quantify in detail how accurately full-spectrum fitting recovers the true age–abundance relations of the MW. Instead, we treat the \ppxf-based solutions as a projection of the MW data into an extragalactic observational framework, designed to mimic how such relations would be inferred when the Galaxy or an MWA is analysed as an unresolved external system.

Figure~\ref{fig04::age_abundance} shows the distribution of weights in the age--[M/H] (left), age--$\alpha$-enhancement (middle), and $\rm [\alpha/M]-[M/H]$ (right) planes. The rows correspond, from top to bottom, to the true mass distribution binned using the adopted sMILES SSP templates, the distribution recovered from \ppxf\ solutions averaged over the Monte Carlo realisations, and a single \ppxf\ solution obtained using a representative regularisation strength (\texttt{regul = 5}), all shown for edge-on model~\texttt{eMW.0074}. The Monte Carlo averaging is used to suppress noise-driven fluctuations and to highlight robust features of the recovered distributions, while the maximally regularised solution illustrates the smoothest age-abundance relations compatible with the data. We restrict this analysis to a single \texttt{eMW.0074} model, as the recovered age-abundance relations depend only weakly on viewing geometry and bar orientation. This model can serve as a template for the GECKOS survey~\citep{2024IAUS..377...27V,2025A&A...700A.237F} to contrast the MWAs with other systems.

\textbf{Age-metallicity relation~(AMR).} Naturally, the detailed shape of the recovered age-abundance relations in full-spectrum fitting depends to some extent on the adopted SSP template grid. To enable a fair comparison, we therefore projected the input resolved stellar-population data onto the same sMILES grid, with equidistant boundaries midway between adjacent template centre values. In this representation, the MW age-metallicity relation is dominated by the old AMR-sequence \citep{2020A&A...640A..81N}, while the younger sequence associated with the outer disc~\citep{2018A&A...618A..78H,2025A&A...700A..89K} appears significantly fainter. This effect is partly intrinsic, as the young MW stellar component is less massive, but is further amplified by the narrower age bins at young ages, which dilute the signal and make the sequence appear even less prominent. We also recall that our FOV is limited to $\sim 10$ kpc from the centre in the horizontal direction, which further reduces the contribution from the young AMR sequence.

\begin{figure*}
    \sidecaption
    \adjincludegraphics[width=0.70\linewidth]{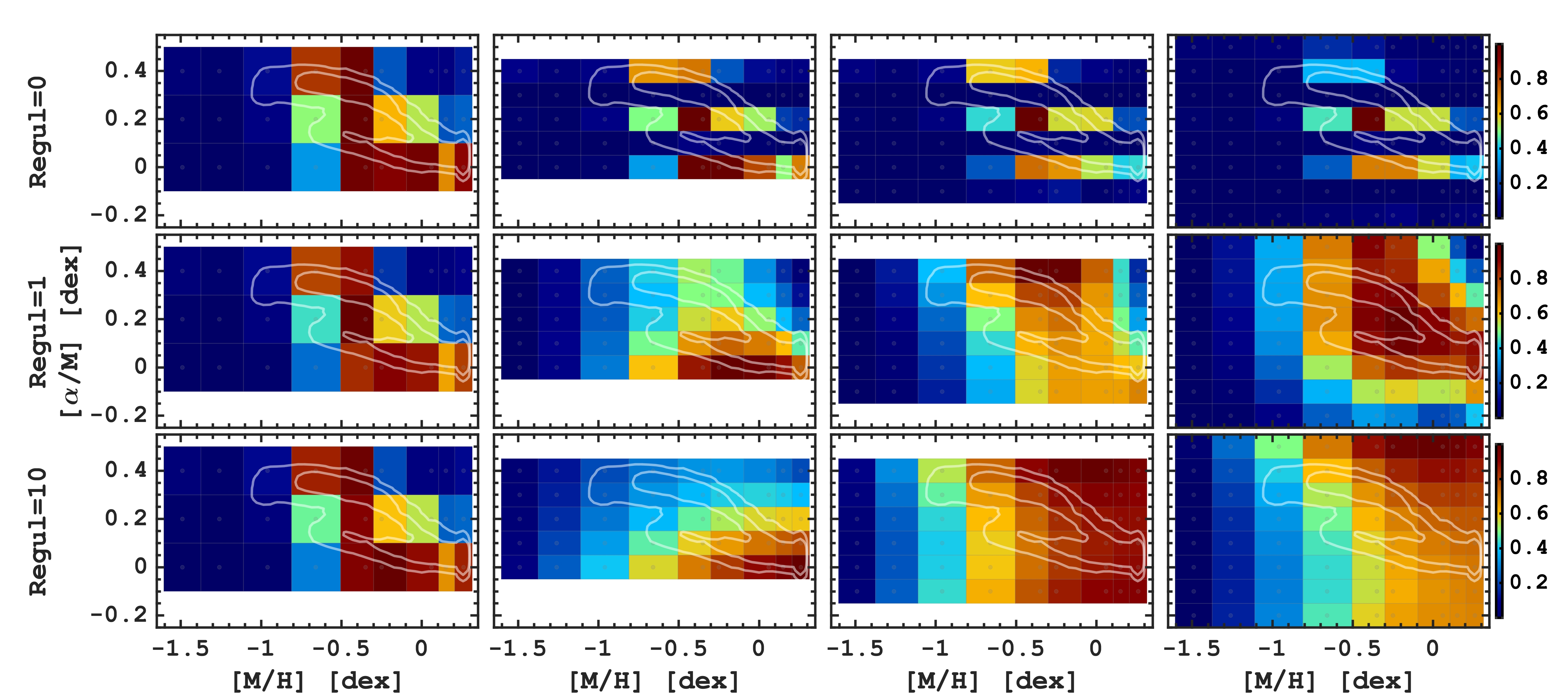} 
    \caption{Recovery of the $\rm [\alpha/M] -[M/H]$ relation in model \texttt{eMW.0070} as a function of both regularisation strength and the number of $\rm  [\alpha/M]$ templates. The rows correspond to increasing \ppxf\ regularisation strength from 0~(top), 1~(middle) and 10~(bottom), while the columns show increasing numbers of $\rm  [\alpha/M]$ bins, and hence higher chemical abundance resolution, from left to right. The white contours correspond to the resolved selection-function-corrected APOGEE measurements, limited to the same FOV as the IFU mock data, and are provided for reference. \\ }\label{fig04::bimodality_resolution}
\end{figure*}

The non-regularised \ppxf\ solution averaged over the Monte Carlo realisations (middle left panel of Fig.~\ref{fig04::age_abundance}), while still affected by noise from individual fits, reproduces the overall structure of the true MW AMR reasonably well. However, the metal-rich component, largely associated with the innermost regions of the Galaxy~\citep{2024ApJ...975..293R}, appears less pronounced, while the old population dominates at ages of $\sim 10-11$~Gyr. This behaviour naturally explains the largest metallicity residuals observed in the central regions of the \texttt{eMW.0070} model in Fig.~\ref{fig04:sfh_74}. The \texttt{regul=5} solution partially rebalances the relative contributions of metal-rich and metal-poor populations, but at the cost of introducing a systematic bias toward younger ages across the field. This likely stems from a fixed level of regularisation as stellar populations properties vary strongly across the FOV and might require a more adaptive regularisation strategy. We suggest that such subtleties are readily apparent in our controlled setup, where the true distribution is known; in the absence of this reference, the AMRs recovered by \ppxf\ would otherwise appear trustworthy.

\textbf{Age-\aMe relation.} The ability, or failure, to recover the correct relative prominence of these $\alpha$-enhanced populations is particularly important, as the age-$\alpha$ relation encodes key information about star-formation timescales and chemical enrichment pathways, and is therefore central to interpreting disc formation scenarios in the MW and in external galaxies. When the true MW stellar population is rebinned onto the SSP template grid, the age-$\alpha$-enhancement relation exhibits the expected bimodality, with two dominant sequences~(now seen as blobs) centred at $[\alpha/\mathrm{M}]\simeq 0.4$ and $[\alpha/\mathrm{M}]\simeq 0.0$~(centre row in Fig.~\ref{fig04::age_abundance}). The intermediate-$\alpha$ population, although present in the intrinsic distribution, appears significantly suppressed after the binning, reflecting both its lower intrinsic weight and the limited age-$\alpha$ resolution imposed by the template grid. This behaviour is largely reproduced by both the Monte Carlo–averaged and the regularised \ppxf\ solutions, indicating that the bimodal structure itself is robustly recovered. However, the Monte Carlo solution still exhibits additional small-scale features not present in the input distribution, highlighting the susceptibility of non-regularised fits to noise-driven artefacts.

\textbf{$\rm \aMe-[M/H]$ relation.} The rightmost column shows the $[\alpha/\mathrm{M}]$--[M/H] plane, where the well-known chemical disc bimodality of the MW is commonly discussed. This parameter space is particularly sensitive to the discretisation, coverage, and effective resolution of the SSP template grid, making it unrealistic to expect a faithful recovery of the detailed intrinsic distribution from full-spectrum fitting of unresolved data alone \citep{2021ApJ...913L..11S,2024MNRAS.534.1175W}. Consistent with these limitations, the binned true distribution does not display a clear thin-thick disc bimodality at fixed metallicity. However, it instead reveals an $\alpha$-bimodality primarily associated with the inner disc and largely independent of metallicity~(``bulge'' bimodality). In this regime, the \ppxf-based solutions reasonably reproduce the global structure of the distribution, as expected, though not guaranteed, given the good agreement observed in the projected abundance maps discussed above.

This comparison highlights an important caveat for Galactic archaeology and its extension to external galaxies: even when global chemical trends are robustly recovered, subtle population separations such as the classical thin/thick disc dichotomy may be blurred, reshaped, or reinterpreted by the inversion process. In an extragalactic context, where only integrated-light measurements are available, such effects become particularly critical, as apparent chemical bimodalities (or their absence) may reflect methodological limitations rather than fundamental differences in galaxy formation pathways. We discuss the recovery of the \aMe-bimodality in more detail in the following section.

\begin{figure*}
    \centering
    \includegraphics[width=0.9\linewidth]{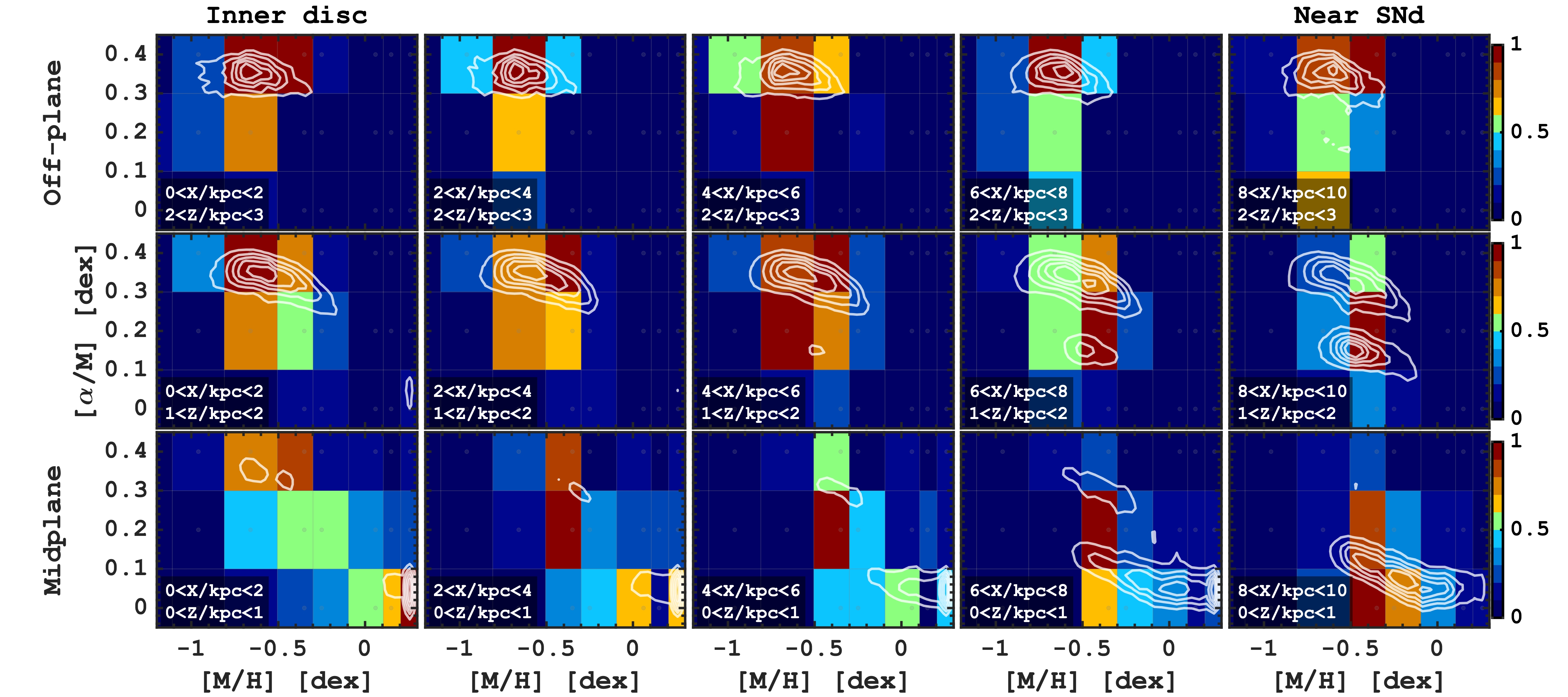}
    \caption{Spatial variation of the $\rm \aMe-[M/H]$ relation in an edge-on projection of the \texttt{eMW.0074} mock galaxy, obtained from Monte Carlo \ppxf\ fits without regularisation. From left to right, the panels correspond to increasing projected galactocentric distance, out to the solar neighbourhood, SNd~(8-10~kpc). The bottom row samples the midplane, while the top row shows off-plane regions. In each panel, the density is normalised by the maximum template weight.
    The white contours in each panel correspond to the resolved selection-function corrected APOGEE measurements limited by the same as in the IFU FOV. Despite being derived from projected IFU data, the recovered distributions reproduce the main qualitative trends observed in the MW, including the dominance of low-$\alpha$ populations near the midplane and the coexistence of high- and low-$\alpha$ populations in the inner regions, resulting in $\alpha$-bimodality across metallicities. However, the simultaneous presence of both populations at a fixed metallicity cannot be disentangled for the adopted set of sMILES templates.}
    \label{fig04::bimodality_spatial}
\end{figure*}

\subsubsection{Composite stellar populations models: extragalactic view on $[\alpha/M]$--$[M/H]$ bimodality}\label{sec4::bimodality_csp}

As we mentioned above, one of the most intriguing features of MW stellar populations in chemical-abundance space is the so-called $\rm [\alpha/M]-[M/H]$ bimodality. This characteristic structure has attracted considerable attention, as it encodes the imprint of multiple physical processes, including star-formation timescales, gas accretion, and chemical enrichment pathways. Recovering this bimodality is therefore crucial for understanding the enrichment history of the MW and for assessing whether similar chemical patterns are present in other galaxies. Therefore, once discovered outside the Galaxy, this can reinforce efforts to understand the origin of the bimodality in the MW. 

In this section, we assess the detectability of the $\alpha$-bimodality using several complementary approaches. As shown in Fig.~\ref{fig04::age_abundance}, the recovery of a bimodal $\alpha$-abundance distribution from IFU-based full-spectrum fitting is not guaranteed, as it depends sensitively on the effective $\alpha$-resolution of the adopted SSP template grid \citep{2021ApJ...913L..11S,2024MNRAS.534.1175W}. Our analysis is based on the $\alpha$-dependent empirical sMILES templates \citep{2023MNRAS.523.3450K}, which span a range of $[\alpha/\mathrm{M}]$ values from $-0.2$ to $0.6$. For the MW, this full range is not required; therefore, our default setup includes only three $\rm [\alpha/M]$ values (0.0, 0.2, and 0.4), which encompass the known $\alpha$-abundance range of Galactic stellar populations.

To explore whether a higher chemical resolution improves the detectability of the $\alpha$-bimodality, we additionally construct an extended template set by interpolating intermediate $[\alpha/\mathrm{M}]$ values between the original sMILES templates. This effectively increases the resolution in $\alpha$-abundance space and allows us to test the impact of template sampling on the recovered distributions. However, this increased resolution comes at the cost of substantially higher computational expense, particularly when regularisation is applied, reflecting the intrinsic trade-off between chemical fidelity and computational feasibility in full-spectrum fitting analyses. 

Figure~\ref{fig04::bimodality_resolution} illustrates the qualitative sensitivity of the recovered distributions to regularisation. It compares the \ppxf\ recovered template-weight distributions, collapsed along the age axis and shown in the $\aMe-[M/H]$ plane, for different numbers of templates along the \aMe\ axis (increasing from left to right) and for different regularisation strengths (\texttt{regul = 0, 1, 10} from top to bottom). We find that, for our default set of SSP templates, the application of regularisation does not significantly alter the recovered weight distributions: the two dominant low- and high-$\alpha$ populations remain clearly identifiable, consistent with the results discussed in the previous section. Increasing the effective $\alpha$-resolution by introducing interpolated templates leads to a more puzzling outcome. In the absence of regularisation, \ppxf\ continues to assign weight primarily to the original template nodes, leaving gaps between them and effectively producing a tri-modal distribution.

This behaviour is straightforward to understand. The design matrix constructed from the $\alpha$-dependent templates contains columns that are linear combinations of one another. Since \ppxf\ solves a linear least-squares problem for the template weights, any model that assigns weight to interpolated templates at $\rm [\alpha/M]=0.1$ or $0.3$ can be equivalently expressed using only the original sMILES templates at $0.0$, $0.2$, and $0.4$. When non-negativity constraints and/or no regularisation are applied, \ppxf\ naturally favours sparse solutions with fewer non-zero coefficients and therefore tends to discard redundant basis vectors. As a result, the solution remains concentrated almost exclusively on the original grid points, making little or no use of the interpolated templates. In this sense, adding such templates does not introduce genuinely new degrees of freedom; rather, it increases the redundancy of the basis and provides no incentive for \ppxf\ to populate them.

One might nonetheless expect that combining additional templates with regularisation could yield smoother or more visually appealing solutions, perhaps closer to the abundance distributions inferred from resolved MW data. However, Fig.~\ref{fig04::bimodality_resolution} demonstrates that this is not the case. While regularisation suppresses the somewhat counter-intuitive yet internally consistent tri-modal structure, the resulting distributions deviate even further from the true underlying distribution without adding physical interpretability. This highlights a key limitation of attempting to recover fine chemical structure through increased template sampling alone: unless the templates represent genuinely independent spectral information, higher resolution in parameter space does not translate into more meaningful constraints~\citep{2020ApJ...896...13B}.

One of the key advantages of the MW chemical dichotomy is its well-established spatial variation with both Galactocentric radius and distance from the midplane \citep{2015ApJ...808..132H}. While it remains unclear how ubiquitous this behaviour is among disc galaxies, a similar pattern has been reported by \citet{2021ApJ...913L..11S}, who showed, using stellar population synthesis modelling with only two \aMe values (0.0 and 0.2), that the external disc galaxy UGC~10738 hosts distinct $\alpha$-rich and $\alpha$-poor stellar populations with spatial distributions closely resembling those of the MW. Motivated by this result, we apply an analogous approach to our data, aiming to test the recoverability of such a spatially dependent chemical structure in external MWAs.

In Fig.~\ref{fig04::bimodality_spatial}, we present the spatial variation of SSP template weights in the $\rm [\alpha/M]-[M/H]$ plane for the edge-on \texttt{eMW.0074} configuration. Following a classical Galactic archaeology perspective, Voronoi bins are selected at increasing galactocentric distances along the midplane (from left to right) and at different heights above and below the midplane. This representation allows us to directly trace how the relative contributions of $\alpha$-rich and $\alpha$-poor populations vary both radially and vertically, providing a controlled framework for assessing the detectability of spatially dependent chemical bimodality in unresolved MW analogues. 

We find that, while the exact fractional contributions are not fully recovered, the overall trends in the $\rm [\alpha/M]-[M/H]$ plane broadly follow expectations based on resolved MW data. High-$\alpha$ templates dominate in the inner eMW and at larger distances from the midplane, whereas low-$\alpha$ templates contribute more strongly near the midplane. However, the separation in $[\alpha/\mathrm{M}]$ between the two sequences cannot be resolved with the current sMILES template grid (see the rightmost columns in Fig.~\ref{fig04::bimodality_spatial}), and, as we showed above, artificially increasing the resolution by interpolating the template grid does not alleviate this limitation. As a result, the intrinsically bimodal chemical structure of the MW is effectively blended into a single sequence in the full-spectrum fitting analysis.

Nevertheless, the spatial variation of the high- and low-$\alpha$ populations appears to be the most promising avenue for assessing whether external galaxies exhibit a MW-like chemical bimodality. A key open question is whether these two sequences are spatially segregated, as observed in the MW, with only limited overlap in the solar neighbourhood~\citep{2025A&A...700A..89K}, or whether they have similar spatial extents. In the latter case, our analysis suggests that detecting bimodality would be particularly challenging with standard SSP templates. This limitation is not only a matter of spatial resolution: even at high angular resolution, integrated-light spectra measure a luminosity-weighted average over many stars, rather than the individual stars from which the MW bimodality is defined. As a result, chemical separation can be washed out whenever distinct populations contribute comparably to the same spectrum. This is further complicated by the fact that individual $\alpha$-elements do not necessarily trace one another, whereas SSP templates usually encode abundance variations through a single global $\rm [\alpha/Fe]$ parameter~\citep[see, e.g.][]{2024A&A...687L..14P}. At present, simulations cannot fully resolve this ambiguity from the theoretical side, because they still struggle to reproduce the detailed chemical-abundance patterns observed in the MW~\citep{2026MNRAS.545f1551O,2025MNRAS.537.1571P}. They therefore provide useful physical guidance, but not yet a definitive prediction for how common, spatially extended, or observationally detectable MW-like bimodality should be in external discs.

\subsection{SFH recovery}\label{sec4::SFH}
In recent years, a growing number of observational studies have reported evidence for apparently distinct episodes of enhanced star formation across the MW disc~\citep{2019A&A...624L...1M, 2020NatAs...4..965R, 2022A&A...666A.130S, 2024MNRAS.527..583M}. These SF bursts are typically inferred to be short-lived ($<0.5-1$~Gyr) yet sufficiently pronounced to remain detectable despite observational selection effects, systematic uncertainties, and modelling assumptions. Intriguingly, a number of IFU studies showed that galaxies characterised by a ``bursty'' or ``bumpy'' rather than smooth SFHs, exhibiting recurrent peaks at similar ages across different methods and datasets~\citep{2013A&A...557A..86C, 2017A&A...607A.128G, 2019MNRAS.482.1557S, 2023MNRAS.526.3273C}. On the one hand, the apparent similarity between the evolutionary histories inferred for the MW and for external disc galaxies could point to a largely stochastic nature of star-formation evolution, operating in a broadly universal manner across different systems and on galactic scales. Such Stochasticity would be expected to imprint substantial scatter on the Schmidt–Kennicutt law and Larson’s relations; however, these relations are observed to be remarkably tight. For example, increasing the star-formation rate by a factor of two generally requires a comparable increase in the molecular gas reservoir. While such an enhancement is plausible locally, e.g. in galactic nuclei \citep{2015MNRAS.448.1107M} or in regions associated with spiral structure \citep{2017MNRAS.468.4189P}, it is difficult to envision mechanisms that would produce a similar, coherent increase in molecular gas across an entire disc, and hence in the global star-formation rate, in low-redshift disc galaxies, unless the entire system moves from the canonical Schmidt–Kennicutt relation of star-forming disks to the regime of starbursts~\citep{2019A&A...625A..65R}.

\begin{figure}[!h]
    \centering
    \includegraphics[width=1\linewidth]{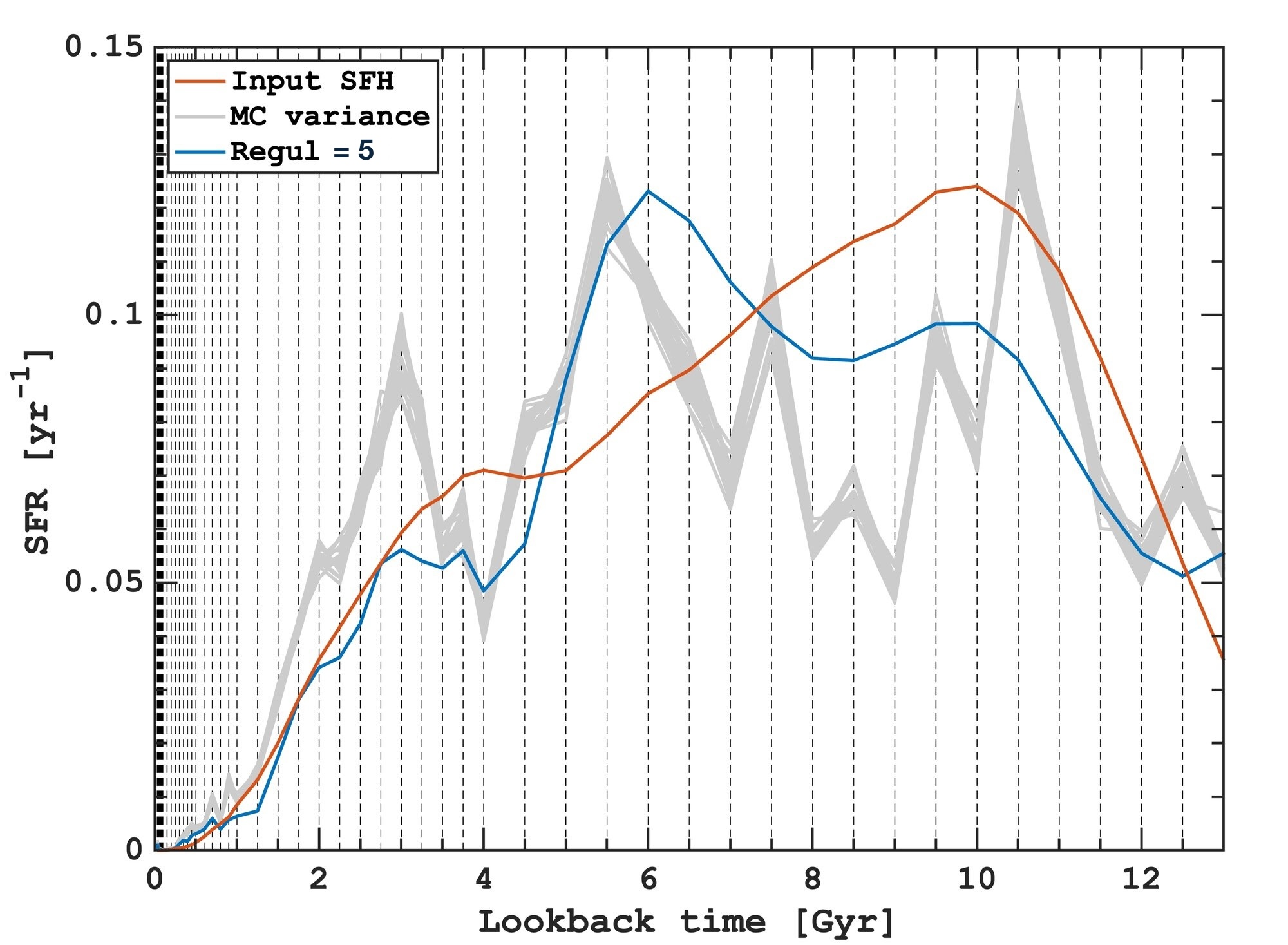}
    \caption{Comparison of star formation histories recovered from the mock IFU data using \ppxf\ with the input star formation history (red), derived from the orbit-superposition solution for the resolved stellar populations of the MW~\citep{2026A&A...706A.103R}. The grey-shaded region shows the range of star formation histories obtained from multiple Monte Carlo \ppxf\ realisations without regularisation. The blue curve corresponds to a single \ppxf\ fit performed with the regularisation strength set to 5.}
    \label{fig04:SFR}
\end{figure}

\begin{figure*}
    \centering
    \includegraphics[width=0.99\linewidth]{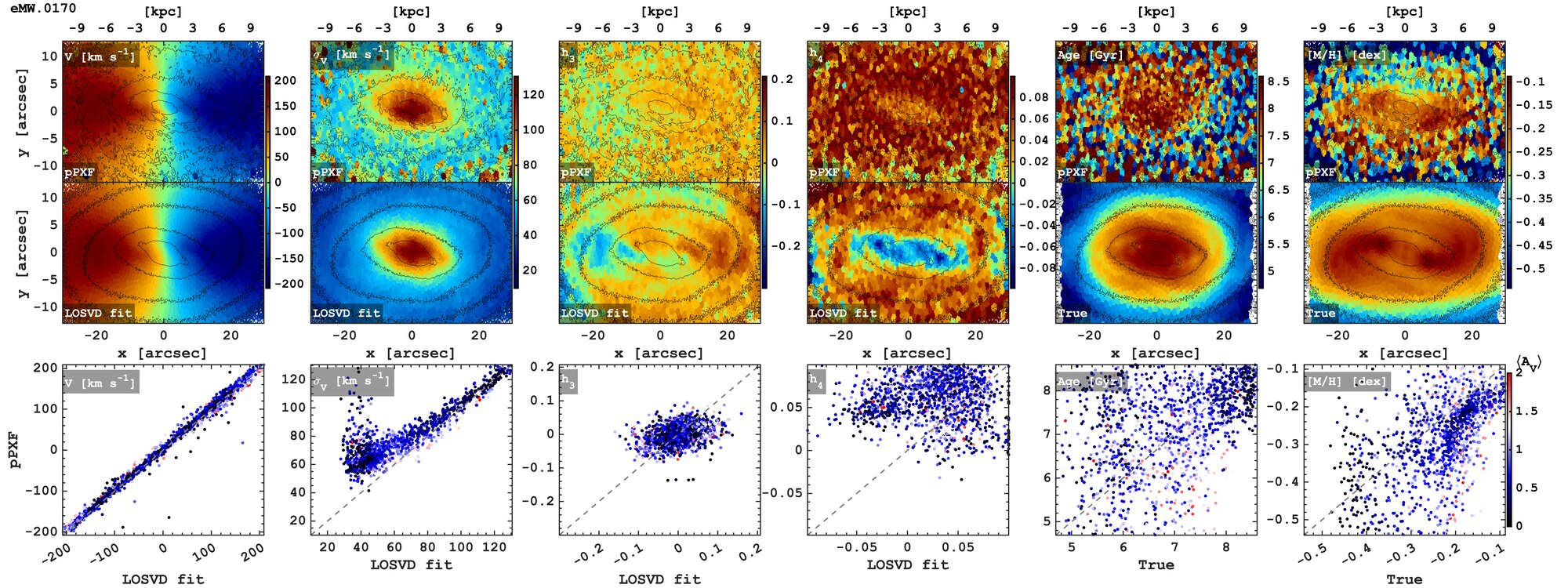} \par
    \includegraphics[width=0.99\linewidth]{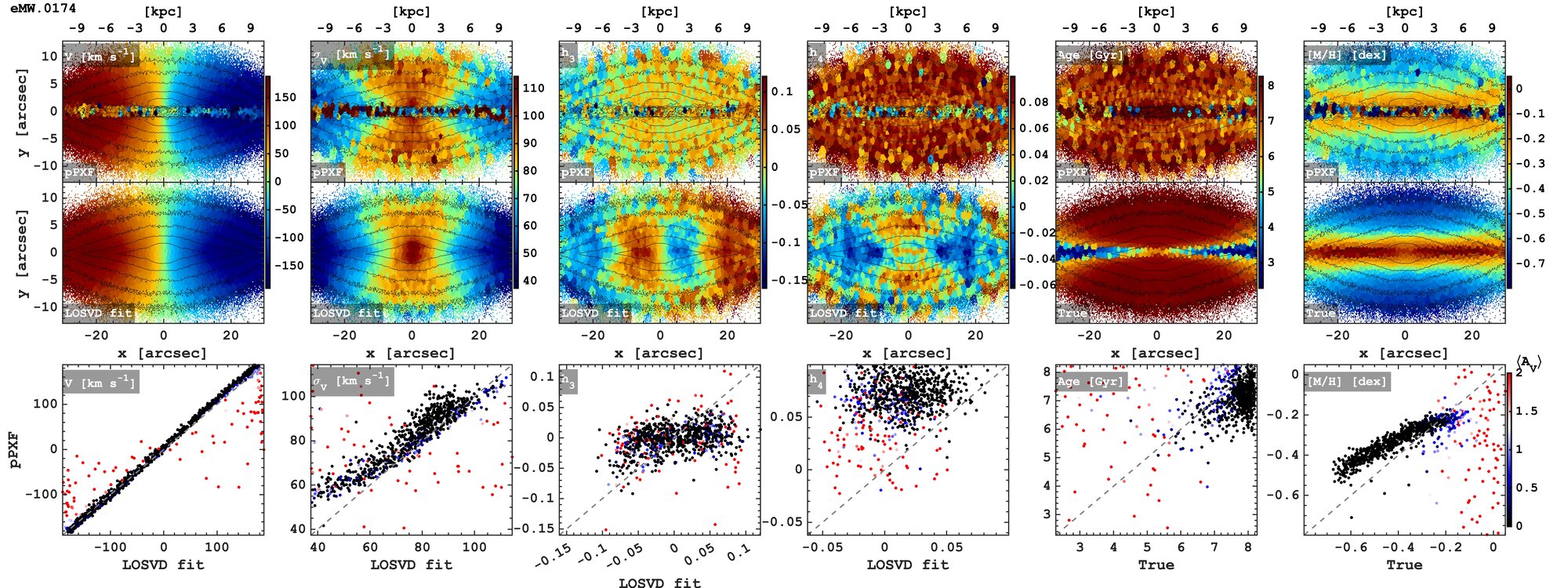}
    \caption{Stellar kinematics and stellar population properties recovered with \ppxf\ from the RT mock IFU datacubes of the extragalactic MW. The two main blocks of panels correspond to models \texttt{eMW.0170} (top) and \texttt{eMW.0174} (bottom). The model parameters are given in Table~\ref{tab4:models}. From left to right, the maps show the line-of-sight mean velocity, velocity dispersion, G-H moments $h_3$ and $h_4$, and the mean projected stellar age and metallicity. In each block, the upper row presents the \ppxf\ results, while the row below shows the corresponding reference values based on the analytic LOSVD fits (for kinematics) or the input particle properties (for stellar populations). The bottom row of each block displays a bin-by-bin comparison between the \ppxf\ measurements and the reference values for individual Voronoi bins; points are colour-coded by the mean visual extinction, $A_V$~(see Appendix~\ref{sec4::dust_appendix} for details).}
    \label{fig04::ppxf_skirt_results}
\end{figure*}

From a theoretical point of view, the most attractive mechanism potentially responsible for the bursts of SF in the MW is interactions with the Sgr dwarf galaxy~\citep{2020NatAs...4..965R}, while this has not confirmed theoretically \citep{2024MNRAS.527.2426A,2025arXiv250800690S}. However, the MW departed the blue star-forming sequence 6-8 Gyr ago and has been located in the green valley since then~\citep{Imig2025,2025ApJ...991...36Z}. In general, the appearance of the recurrent SF modes across various systems may suggest that at least part of this phenomenon may arise from methodological biases rather than genuine star-formation variability~\citep{2024MNRAS.528.2790Z}. Indeed, recovering an SFH from an integrated spectrum is a classic ill-conditioned inverse problem, where the weights encode the SFH. Many combinations of weights produce spectra that are statistically indistinguishable in the presence of noise. As a result, smooth input SFHs do not map uniquely to smooth solutions, noise and template degeneracies are amplified in the inversion. This interpretation is further supported by the comparison of resolved and integrated-light analyses in M54 by \cite{2020ApJ...896...13B}, where a continuous age distribution is compressed into a narrow peak in the integrated SFH. 

Motivated by this controversy, we analyse the star-formation history of the MW inferred from the mock IFU analysis and compare it with both the input SFH and the present-day, mass-weighted age distribution obtained from the orbit-superposition approach. The latter has been explored in detail in \cite{2026A&A...706A.103R} and, as described in Section~\ref{sec4::sos}, relies on stellar ages derived for the APOGEE sample. Consequently, although we apply corrections for several known biases, the analysis remains subject to the intrinsic limitations of the adopted stellar age estimates. In addition, age uncertainties are propagated along orbits, effectively smoothing the recovered age distribution over a characteristic window of approximately 1.5 Gyr, which may suppress genuine short-timescale SFR variations, even if such variations are present in the MW. Despite these caveats, we treat the orbit-superposition age distribution as the reference solution and quantify how the SFHs recovered in our mock IFU analysis deviate from this input.

In Fig.~\ref{fig04:SFR} we present the star-formation histories obtained by collapsing the AMRs along the metallicity axis. The grey curves correspond to multiple Monte Carlo realisations of \ppxf\ fits performed without regularisation, whereas the blue curve shows the result of a representative run with regularisation. For reference, the red curve denotes the ``input'' or resolved SFH of the MW, derived from the stellar populations within the field of view of the \texttt{eMW.0070} model. As expected, the unregularized solutions exhibit a series of pronounced peaks that could be naively interpreted as star-formation bursts. The Monte Carlo realisations are mutually consistent, showing only $\approx 10\%$ variations in amplitude, which indicates that these features are robust to noise but not necessarily physical. Such artificial burstiness is a well-known outcome of IFU-based spectral fitting~\citep{2017MNRAS.466..798C}: when multiple SSP templates provide comparably good fits to the same spectral features, \ppxf\ tends to select a single ``winner'' age bin, thereby concentrating weight into discrete ages. Consequently, these peaks often trace ages where the SSP templates are most spectrally distinct, rather than epochs of genuinely enhanced star formation, as discussed above. Although the regularised solution is substantially smoother, it still departs significantly from the input SFH, with a dominant peak at $\sim 10$~Gyr followed by a gradual decline and a secondary feature around $\sim 4$~Gyr, which is associated with the emergence of the outer MW disc \citep{2025A&A...700A..89K}.

\subsection{Analysis of IFU data based on Radiative Transfer}\label{sec4::results_skirt}

Up to this point, we have analysed results of full-spectrum fitting applied to CSP-based IFU data, in which the extragalactic MW spectra are constructed from superposed sMILES SSP spectra. This ``vanilla'' setup omits several sources of physical and observational complexity, including dust attenuation, spiral structure, and strongly irregular kinematics; only Gaussian noise is added to generate high-SNR spectra. At the same time, it provides a controlled test case, since the mock and fitting SSPs share the same spectral resolution and sampling, and the LOSVDs are very well sampled by the large number of particles contributing to each Voronoi bin.

In this section, we turn to an IFU datacube produced with RT calculations of an MW simulation whose initial conditions were adapted from the orbit-superposition model~(see Sect.~\ref{sec4::mock_rt}). The goal is not to reproduce the level of detail achieved in the CSP case, but to assess which kinematic and stellar-population signatures remain detectable once dust attenuation, finite photon statistics, and lower signal-to-noise are included. Given these additional complications, we restrict the stellar-population analysis of the RT cubes to global and large-scale trends.

In Fig.~\ref{fig04::ppxf_skirt_results}, we present the analysis of kinematics and stellar population properties for two eMW configurations, analogous to \texttt{eMW.0070} and \texttt{eMW.0074}, but now generated using full RT calculations. For each configuration, the panels from top to bottom show the \ppxf\ solutions, averaged over 30 realisations, for the LOSVD parameters, followed by the mean stellar age and metallicity. The bottom row provides a Voronoi-bin–by–bin comparison between the true input values and the \ppxf-recovered quantities, colour-coded by the mean visual extinction (see Appendix~\ref{sec4::dust_appendix} for details).

In both projections, the mean LOS velocity and velocity dispersion are recovered robustly, except near the mid-plane in the edge-on case (\texttt{eMW.0174}). In the lower-resolution configuration (\texttt{eMW.0170}), the velocity dispersion is systematically overestimated outside the bar region, owing to the limited intrinsic resolution of the BC03 templates. This is evident in the bottom panels, where deviations from the one-to-one relation appear for true dispersions of $\sim 70 -80\ \mathrm{km\ s^{-1}}$. Strong dust attenuation, unless treated explicitly, further hampers the recovery of the main LOSVD moments near the mid-plane.

Although the amplitudes of $h_3$ are strongly suppressed, their spatial patterns and weak correlations with the true values are partly preserved. This suggests that the underlying non-Gaussian LOSVD structure remains detectable even in the RT-based cubes, and that higher signal-to-noise and/or improved photon statistics may allow more quantitative constraints in future applications. By contrast, $h_4$ cannot be reliably recovered in the present setup.

The recovery of age–metallicity trends is also adversely affected in the RT-based models. In the $70^\circ$ inclination configurations, a negative radial age gradient is still discernible; however, in the outer disc, this trend becomes dominated by uncertainties, as illustrated by the increased scatter in the corresponding comparison panels. The metallicity fits are more robust in specific regions, successfully recovering the enhanced metallicity at the bar edges, whereas other large-scale trends appear substantially blurred.

In the edge-on projection, the situation deteriorates further. The age gradient perpendicular to the midplane becomes effectively inaccessible, as the younger thin-disc component is largely masked along the line of sight. In contrast, the metallicity distribution remains qualitatively reasonable, although the true versus \ppxf\ comparison in the bottom panels reveals a correlated yet biased recovery, with systematically reduced metallicity variations. The quality of the fits, however, should be sufficient to recover the metallicity gradients or large-scale trends.

In light of these limitations, we remain conservative in the further interpretation of higher-order kinematic moments and detailed stellar population properties. More broadly, this highlights an important limitation for the analysis of numerical galaxy-formation simulations: while radiative-transfer post-processing enables the generation of visually realistic and physically motivated mock observations (Fig.~\ref{fig04::images}), its substantial computational cost and the associated degradation in recoverable IFU-level information ultimately reduce the predictive power of such models when robust stellar population and kinematic constraints are required.

In conclusion, the reduced performance of full-spectrum fitting in the RT-based analysis can be attributed to the combined effects of lower effective signal-to-noise ratio, dust attenuation, and the intrinsically limited spectral resolution of the BC03 templates, rather than to shortcomings of the fitting methodology itself. Although the RT treatment does not permit a detailed reconstruction of higher-order kinematic moments or fine stellar population substructure, it nevertheless demonstrates that the principal dynamical and population signatures identified in the CSP analysis remain qualitatively detectable under more realistic observational conditions.

\section{Summary}\label{sec4::summary}
In this work, we investigated the stellar populations of the MW as they would be observed in a distant galaxy with an IFU instrument. We considered two complementary approaches for constructing external views of the Galaxy. The first, referred to as the CSP method, uses an idealised setup in which unresolved spectra are generated by stacking SSPs with their associated line-of-sight kinematics across the Galaxy obtained from our orbit-superposition reconstruction of the MW. This controlled framework allows us to isolate the intrinsic effects of Galactic structure, including the bar and X-shaped bulge, without additional observational complications.

The second approach is based on full RT calculations with \SKIRT, applied to a dynamical MW simulation whose initial conditions are derived from an orbit-superposition model. The subsequent evolution allows spiral structure to develop, provides a more realistic sampling of the youngest stellar populations~($<1$~Gyr), and self-consistently includes dust attenuation and emission-line contamination in the resulting IFU data cubes. 

Applying full-spectrum fitting with \ppxf\ to these extragalactic MW realisations, we draw the following conclusions:

\begin{itemize}

\item Prior to spectral fitting, we quantified the degree to which the LOSVD departs from a G–H representation. While this parametrisation performs well over most of the Galaxy, we identify significant deviations in transitional regions, such as the interface between the inner thick and outer discs, within the lobes of the X-shaped bulge, and near the ends of the bar, where the LOSVD is difficult to describe using the G-H parameterisation. In these regions, kinematic fits, particularly higher-order moments, are substantially less reliable and, more importantly, may ultimately be less informative, even if they match the analytic G-H fit.

\item For the idealised CSP data, or in cases of high signal-to-noise ratio and negligible dust attenuation and emission-line contamination, the detailed recovery of bar–bulge kinematics is feasible, including constraints on higher-order kinematic moments. The detailed reconstruction of stellar population properties depends sensitively on the adopted regularisation scheme in the \ppxf\ optimisation. Nevertheless, the recovered age–\aMe–metallicity relations, although not identical to their Galactic counterparts, broadly reproduce similar trends. In particular, the \aMe-bimodality is robustly detected and exhibits its characteristic spatial dependence. We show that, when considered as a purely global relation, the appearance of the $\alpha$-bimodality may vary with the adopted template grid and regularisation strength; however, tracing the spatial distributions of the low- and high-$\alpha$ populations across the discs provides a more robust avenue for identifying MW analogues.

\item The recovery of the SFH remains challenging. As commonly found in extragalactic studies, full-spectrum fitting tends to introduce artificial burst-like features into an otherwise smooth MW SFH. The physical origin of this effect remains incompletely understood, particularly in the MW context, where independent techniques are available to infer the SFH. Further investigation is required to assess whether bursty SFHs are genuinely ubiquitous among disc galaxies and whether such features can be robustly recovered across different observational regimes and analysis methods.

\item The reduced performance of full-spectrum fitting in the RT-based analysis is primarily driven by the combined effects of lower effective signal-to-noise ratio, dust attenuation, and the limited spectral resolution of the available templates, rather than by intrinsic shortcomings of the fitting methodology. Despite these challenges, the large-scale kinematic structure, most notably the mean velocity and velocity dispersion, is robustly recovered, and several broad age and metallicity gradients remain discernible. While higher-order kinematic moments and fine stellar population substructure cannot be reliably constrained, the principal dynamical and population signatures identified in the CSP analysis remain qualitatively detectable under realistic observational conditions.

\end{itemize}
Viewing the MW as an external galaxy provides a practical bridge between resolved Galactic archaeology and integrated-light studies of galaxy populations~\citep{2021NatAs...5..879V}. The key challenge is to determine which processes inferred in the MW, including bar-driven evolution, disc heating, radial redistribution, chemical bimodality, and inside-out growth, remain identifiable once galaxies are observed only through projected, unresolved light. The MW therefore offers a unique benchmark for calibrating how galactic-scale physics appears in spatially resolved integrated-light diagnostics, particularly in the era of current and future IFU facilities such as MUSE~\citep{2010SPIE.7735E..08B}, WEAVE~\citep{2024MNRAS.530.2688J}, WST/IFS~\citep{2024arXiv240305398M}, and BlueMUSE~\citep{2019arXiv190601657R}. 

An important next step will be to extend the present analysis using alternative stellar template libraries and associated SSP models beyond those adopted here (MILES and BC03), such as XSL~\citep{2022A&A...660A..34V} and FASTAR~\citep{2026arXiv260524093M}, together with future template sets offering different abundance coverage, in order to assess how strongly the current conclusions depend on the choice of spectral library. An especially promising direction is the construction and analysis of mock IFU observations tailored to future blue-sensitive instruments, with particular interest in BlueMUSE and in fully RT-based mock data cubes, where dust, geometry, emission, and instrumental effects can be treated in a more realistic way. Such comparisons are essential for developing a common physical language between near-field and far-field studies, and for assessing whether external MW analogues selected from global or spatially resolved observables truly share similar evolutionary histories.

\begin{acknowledgements}

S.K. acknowledges the hospitality of the University of Vienna through the Ida Pfeiffer Visiting Professorship and the Paris Observatory, during which the initial and final parts of this work were carried out, respectively. S.K. acknowledges support by the Deutsche Forschungsgemeinschaft under the grant KH~500/2-1. La Silla Paranal Observatory under programme IDs 110.24AS and 179.A-2010. The authors wish to thank Prashin Jethwa for useful conversations related to this work. \\

The simulations presented in this paper were performed using the open-source SWIFT simulation code (\url{http://www.swiftsim.com}; \citealt{2024MNRAS.530.2378S}).\\

This work presents results from the European Space Agency (ESA) space mission Gaia. Gaia data are being processed by the Gaia Data Processing and Analysis Consortium (DPAC). Funding for the DPAC is provided by national institutions, in particular the institutions participating in the Gaia Multi-Lateral Agreement (MLA). The Gaia mission website is \url{https://www.cosmos.esa.int/gaia}. The Gaia Archive website is \url{http://archives.esac.esa.int/gaia}. \\

Funding for the Sloan Digital Sky Survey IV has been provided by the Alfred P. Sloan Foundation, the U.S. Department of Energy Office of Science, and the Participating Institutions. SDSS acknowledges support and resources from the Center for High-Performance Computing at the University of Utah. The SDSS web site is \url{www.sdss4.org}. \\

SDSS is managed by the Astrophysical Research Consortium for the Participating Institutions of the SDSS Collaboration including the Brazilian Participation Group, the Carnegie Institution for Science, Carnegie Mellon University, Center for Astrophysics | Harvard \& Smithsonian (CfA), the Chilean Participation Group, the French Participation Group, Instituto de Astrofísica de Canarias, The Johns Hopkins University, Kavli Institute for the Physics and Mathematics of the Universe (IPMU) / University of Tokyo, the Korean Participation Group, Lawrence Berkeley National Laboratory, Leibniz Institut für Astrophysik Potsdam (AIP), Max-Planck-Institut für Astronomie (MPIA Heidelberg), Max-Planck-Institut für Astrophysik (MPA Garching), Max-Planck-Institut für Extraterrestrische Physik (MPE), National Astronomical Observatories of China, New Mexico State University, New York University, University of Notre Dame, Observatório Nacional / MCTI, The Ohio State University, Pennsylvania State University, Shanghai Astronomical Observatory, United Kingdom Participation Group, Universidad Nacional Autónoma de México, University of Arizona, University of Colorado Boulder, University of Oxford, University of Portsmouth, University of Utah, University of Virginia, University of Washington, University of Wisconsin, Vanderbilt University, and Yale University.

\end{acknowledgements}

\bibliographystyle{aa}
\bibliography{refs}

\begin{thebibliography}{137}
\expandafter\ifx\csname natexlab\endcsname\relax\def\natexlab#1{#1}\fi

\bibitem[{{Abdurro'uf} {et~al.}(2022){Abdurro'uf}, {Accetta}, {Aerts}, {Silva Aguirre}, {Ahumada}, {Ajgaonkar}, {Filiz Ak}, {Alam}, {Allende Prieto}, {Almeida}, {Anders}, {Anderson}, {Andrews}, {Anguiano}, {Aquino-Ort{\'\i}z}, {Arag{\'o}n-Salamanca}, {Argudo-Fern{\'a}ndez}, {Ata}, {Aubert}, {Avila-Reese}, {Badenes}, {Barb{\'a}}, {Barger}, {Barrera-Ballesteros}, {Beaton}, {Beers}, {Belfiore}, {Bender}, {Bernardi}, {Bershady}, {Beutler}, {Bidin}, {Bird}, {Bizyaev}, {Blanc}, {Blanton}, {Boardman}, {Bolton}, {Boquien}, {Borissova}, {Bovy}, {Brandt}, {Brown}, {Brownstein}, {Brusa}, {Buchner}, {Bundy}, {Burchett}, {Bureau}, {Burgasser}, {Cabang}, {Campbell}, {Cappellari}, {Carlberg}, {Wanderley}, {Carrera}, {Cash}, {Chen}, {Chen}, {Cherinka}, {Chiappini}, {Choi}, {Chojnowski}, {Chung}, {Clerc}, {Cohen}, {Comerford}, {Comparat}, {da Costa}, {Covey}, {Crane}, {Cruz-Gonzalez}, {Culhane}, {Cunha}, {Dai}, {Damke}, {Darling}, {Davidson}, {Davies}, {Dawson}, {De Lee}, {Diamond-Stanic}, {Cano-D{\'\i}az}, {S{\'a}nchez},
  {Donor}, {Duckworth}, {Dwelly}, {Eisenstein}, {Elsworth}, {Emsellem}, {Eracleous}, {Escoffier}, {Fan}, {Farr}, {Feng}, {Fern{\'a}ndez-Trincado}, {Feuillet}, {Filipp}, {Fillingham}, {Frinchaboy}, {Fromenteau}, {Galbany}, {Garc{\'\i}a}, {Garc{\'\i}a-Hern{\'a}ndez}, {Ge}, {Geisler}, {Gelfand}, {G{\'e}ron}, {Gibson}, {Goddy}, {Godoy-Rivera}, {Grabowski}, {Green}, {Greener}, {Grier}, {Griffith}, {Guo}, {Guy}, {Hadjara}, {Harding}, {Hasselquist}, {Hayes}, {Hearty}, {Hern{\'a}ndez}, {Hill}, {Hogg}, {Holtzman}, {Horta}, {Hsieh}, {Hsu}, {Hsu}, {Huber}, {Huertas-Company}, {Hutchinson}, {Hwang}, {Ibarra-Medel}, {Chitham}, {Ilha}, {Imig}, {Jaekle}, {Jayasinghe}, {Ji}, {Johnson}, {Jones}, {J{\"o}nsson}, {Katkov}, {Khalatyan}, {Kinemuchi}, {Kisku}, {Knapen}, {Kneib}, {Kollmeier}, {Kong}, {Kounkel}, {Kreckel}, {Krishnarao}, {Lacerna}, {Lane}, {Langgin}, {Lavender}, {Law}, {Lazarz}, {Leung}, {Leung}, {Lewis}, {Li}, {Li}, {Lian}, {Liang}, {Lin}, {Lin}, {Lin}, {Lintott}, {Long}, {Longa-Pe{\~n}a}, {L{\'o}pez-Cob{\'a}}, {Lu},
  {Lundgren}, {Luo}, {Mackereth}, {de la Macorra}, {Mahadevan}, {Majewski}, {Manchado}, {Mandeville}, {Maraston}, {Margalef-Bentabol}, {Masseron}, {Masters}, {Mathur}, {McDermid}, {Mckay}, {Merloni}, {Merrifield}, {Meszaros}, {Miglio}, {Di Mille}, {Minniti}, {Minsley}, {Monachesi}, {Moon}, {Mosser}, {Mulchaey}, {Muna}, {Mu{\~n}oz}, {Myers}, {Myers}, {Nadathur}, {Nair}, {Nandra}, {Neumann}, {Newman}, {Nidever}, {Nikakhtar}, {Nitschelm}, {O'Connell}, {Garma-Oehmichen}, {Luan Souza de Oliveira}, {Olney}, {Oravetz}, {Ortigoza-Urdaneta}, {Osorio}, {Otter}, {Pace}, {Padilla}, {Pan}, {Pan}, {Parikh}, {Parker}, {Peirani}, {Pe{\~n}a Ram{\'\i}rez}, {Penny}, {Percival}, {Perez-Fournon}, {Pinsonneault}, {Poidevin}, {Poovelil}, {Price-Whelan}, {B{\'a}rbara de Andrade Queiroz}, {Raddick}, {Ray}, {Rembold}, {Riddle}, {Riffel}, {Riffel}, {Rix}, {Robin}, {Rodr{\'\i}guez-Puebla}, {Roman-Lopes}, {Rom{\'a}n-Z{\'u}{\~n}iga}, {Rose}, {Ross}, {Rossi}, {Rubin}, {Salvato}, {S{\'a}nchez}, {S{\'a}nchez-Gallego}, {Sanderson}, {Santana
  Rojas}, {Sarceno}, {Sarmiento}, {Sayres}, {Sazonova}, {Schaefer}, {Schiavon}, {Schlegel}, {Schneider}, {Schultheis}, {Schwope}, {Serenelli}, {Serna}, {Shao}, {Shapiro}, {Sharma}, {Shen}, {Shetrone}, {Shu}, {Simon}, {Skrutskie}, {Smethurst}, {Smith}, {Sobeck}, {Spoo}, {Sprague}, {Stark}, {Stassun}, {Steinmetz}, {Stello}, {Stone-Martinez}, {Storchi-Bergmann}, {Stringfellow}, {Stutz}, {Su}, {Taghizadeh-Popp}, {Talbot}, {Tayar}, {Telles}, {Teske}, {Thakar}, {Theissen}, {Tkachenko}, {Thomas}, {Tojeiro}, {Hernandez Toledo}, {Troup}, {Trump}, {Trussler}, {Turner}, {Tuttle}, {Unda-Sanzana}, {V{\'a}zquez-Mata}, {Valentini}, {Valenzuela}, {Vargas-Gonz{\'a}lez}, {Vargas-Maga{\~n}a}, {Alfaro}, {Villanova}, {Vincenzo}, {Wake}, {Warfield}, {Washington}, {Weaver}, {Weijmans}, {Weinberg}, {Weiss}, {Westfall}, {Wild}, {Wilde}, {Wilson}, {Wilson}, {Wilson}, {Wolf}, {Wood-Vasey}, {Yan}, {Zamora}, {Zasowski}, {Zhang}, {Zhao}, {Zheng}, {Zheng}, \& {Zhu}}]{2022ApJS..259...35A}
{Abdurro'uf}, {Accetta}, K., {Aerts}, C., {et~al.} 2022, \href{http://dx.doi.org/10.3847/1538-4365/ac4414}{\color{magenta}\apjs}, \href{https://ui.adsabs.harvard.edu/abs/2022ApJS..259...35A}{259, 35}

\bibitem[{{Annem} \& {Khoperskov}(2024)}]{2024MNRAS.527.2426A}
{Annem}, B. \& {Khoperskov}, S. 2024, \href{http://dx.doi.org/10.1093/mnras/stad3244}{\color{magenta}\mnras}, \href{https://ui.adsabs.harvard.edu/abs/2024MNRAS.527.2426A}{527, 2426}

\bibitem[{{Antoja} {et~al.}(2018){Antoja}, {Helmi}, {Romero-G{\'o}mez}, {Katz}, {Babusiaux}, {Drimmel}, {Evans}, {Figueras}, {Poggio}, {Reyl{\'e}}, {Robin}, {Seabroke}, \& {Soubiran}}]{2018Natur.561..360A}
{Antoja}, T., {Helmi}, A., {Romero-G{\'o}mez}, M., {et~al.} 2018, \href{http://dx.doi.org/10.1038/s41586-018-0510-7}{\color{magenta}\nat}, \href{https://ui.adsabs.harvard.edu/abs/2018Natur.561..360A}{561, 360}

\bibitem[{{Bacon} {et~al.}(2010){Bacon}, {Accardo}, {Adjali}, {Anwand}, {Bauer}, {Biswas}, {Blaizot}, {Boudon}, {Brau-Nogue}, {Brinchmann}, {Caillier}, {Capoani}, {Carollo}, {Contini}, {Couderc}, {Daguis{\'e}}, {Deiries}, {Delabre}, {Dreizler}, {Dubois}, {Dupieux}, {Dupuy}, {Emsellem}, {Fechner}, {Fleischmann}, {Fran{\c{c}}ois}, {Gallou}, {Gharsa}, {Glindemann}, {Gojak}, {Guiderdoni}, {Hansali}, {Hahn}, {Jarno}, {Kelz}, {Koehler}, {Kosmalski}, {Laurent}, {Le Floch}, {Lilly}, {Lizon}, {Loupias}, {Manescau}, {Monstein}, {Nicklas}, {Olaya}, {Pares}, {Pasquini}, {P{\'e}contal-Rousset}, {Pell{\'o}}, {Petit}, {Popow}, {Reiss}, {Remillieux}, {Renault}, {Roth}, {Rupprecht}, {Serre}, {Schaye}, {Soucail}, {Steinmetz}, {Streicher}, {Stuik}, {Valentin}, {Vernet}, {Weilbacher}, {Wisotzki}, \& {Yerle}}]{2010SPIE.7735E..08B}
{Bacon}, R., {Accardo}, M., {Adjali}, L., {et~al.} 2010, in Society of Photo-Optical Instrumentation Engineers (SPIE) Conference Series, Vol. 7735, Ground-based and Airborne Instrumentation for Astronomy III, ed. {McLean}, I.~S., {Ramsay}, S.~K., \& {Takami}, H., \href{https://ui.adsabs.harvard.edu/abs/2010SPIE.7735E..08B}{773508}

\bibitem[{{Baes} \& {Camps}(2015)}]{2015A&C....12...33B}
{Baes}, M. \& {Camps}, P. 2015, \href{http://dx.doi.org/10.1016/j.ascom.2015.05.006}{\color{magenta}Astronomy and Computing}, \href{https://ui.adsabs.harvard.edu/abs/2015A&C....12...33B}{12, 33}

\bibitem[{{Baes} {et~al.}(2011){Baes}, {Verstappen}, {De Looze}, {Fritz}, {Saftly}, {Vidal P{\'e}rez}, {Stalevski}, \& {Valcke}}]{2011ApJS..196...22B}
{Baes}, M., {Verstappen}, J., {De Looze}, I., {et~al.} 2011, \href{http://dx.doi.org/10.1088/0067-0049/196/2/22}{\color{magenta}\apjs}, \href{https://ui.adsabs.harvard.edu/abs/2011ApJS..196...22B}{196, 22}

\bibitem[{{Belokurov} {et~al.}(2018){Belokurov}, {Erkal}, {Evans}, {Koposov}, \& {Deason}}]{2018MNRAS.478..611B}
{Belokurov}, V., {Erkal}, D., {Evans}, N.~W., {Koposov}, S.~E., \& {Deason}, A.~J. 2018, \href{http://dx.doi.org/10.1093/mnras/sty982}{\color{magenta}\mnras}, \href{https://ui.adsabs.harvard.edu/abs/2018MNRAS.478..611B}{478, 611}

\bibitem[{{Belokurov} \& {Kravtsov}(2022)}]{2022MNRAS.514..689B}
{Belokurov}, V. \& {Kravtsov}, A. 2022, \href{http://dx.doi.org/10.1093/mnras/stac1267}{\color{magenta}\mnras}, \href{https://ui.adsabs.harvard.edu/abs/2022MNRAS.514..689B}{514, 689}

\bibitem[{{Belokurov} {et~al.}(2020){Belokurov}, {Sanders}, {Fattahi}, {Smith}, {Deason}, {Evans}, \& {Grand}}]{2020MNRAS.494.3880B}
{Belokurov}, V., {Sanders}, J.~L., {Fattahi}, A., {et~al.} 2020, \href{http://dx.doi.org/10.1093/mnras/staa876}{\color{magenta}\mnras}, \href{https://ui.adsabs.harvard.edu/abs/2020MNRAS.494.3880B}{494, 3880}

\bibitem[{{Boardman} {et~al.}(2020){Boardman}, {Zasowski}, {Seth}, {Newman}, {Andrews}, {Bershady}, {Bird}, {Chiappini}, {Fielder}, {Fraser-McKelvie}, {Jones}, {Licquia}, {Masters}, {Minchev}, {Schiavon}, {Brownstein}, {Drory}, \& {Lane}}]{2020MNRAS.491.3672B}
{Boardman}, N., {Zasowski}, G., {Seth}, A., {et~al.} 2020, \href{http://dx.doi.org/10.1093/mnras/stz3126}{\color{magenta}\mnras}, \href{https://ui.adsabs.harvard.edu/abs/2020MNRAS.491.3672B}{491, 3672}

\bibitem[{{Boecker} {et~al.}(2020){Boecker}, {Alfaro-Cuello}, {Neumayer}, {Mart{\'\i}n-Navarro}, \& {Leaman}}]{2020ApJ...896...13B}
{Boecker}, A., {Alfaro-Cuello}, M., {Neumayer}, N., {Mart{\'\i}n-Navarro}, I., \& {Leaman}, R. 2020, \href{http://dx.doi.org/10.3847/1538-4357/ab919d}{\color{magenta}\apj}, \href{https://ui.adsabs.harvard.edu/abs/2020ApJ...896...13B}{896, 13}

\bibitem[{{Bovy} {et~al.}(2012){Bovy}, {Rix}, {Liu}, {Hogg}, {Beers}, \& {Lee}}]{2012ApJ...753..148B}
{Bovy}, J., {Rix}, H.-W., {Liu}, C., {et~al.} 2012, \href{http://dx.doi.org/10.1088/0004-637X/753/2/148}{\color{magenta}\apj}, \href{https://ui.adsabs.harvard.edu/abs/2012ApJ...753..148B}{753, 148}

\bibitem[{{Bruzual} \& {Charlot}(2003)}]{2003MNRAS.344.1000B}
{Bruzual}, G. \& {Charlot}, S. 2003, \href{http://dx.doi.org/10.1046/j.1365-8711.2003.06897.x}{\color{magenta}\mnras}, \href{https://ui.adsabs.harvard.edu/abs/2003MNRAS.344.1000B}{344, 1000}

\bibitem[{{Bryant} {et~al.}(2015){Bryant}, {Owers}, {Robotham}, {Croom}, {Driver}, {Drinkwater}, {Lorente}, {Cortese}, {Scott}, {Colless}, {Schaefer}, {Taylor}, {Konstantopoulos}, {Allen}, {Baldry}, {Barnes}, {Bauer}, {Bland-Hawthorn}, {Bloom}, {Brooks}, {Brough}, {Cecil}, {Couch}, {Croton}, {Davies}, {Ellis}, {Fogarty}, {Foster}, {Glazebrook}, {Goodwin}, {Green}, {Gunawardhana}, {Hampton}, {Ho}, {Hopkins}, {Kewley}, {Lawrence}, {Leon-Saval}, {Leslie}, {McElroy}, {Lewis}, {Liske}, {L{\'o}pez-S{\'a}nchez}, {Mahajan}, {Medling}, {Metcalfe}, {Meyer}, {Mould}, {Obreschkow}, {O'Toole}, {Pracy}, {Richards}, {Shanks}, {Sharp}, {Sweet}, {Thomas}, {Tonini}, \& {Walcher}}]{2015MNRAS.447.2857B}
{Bryant}, J.~J., {Owers}, M.~S., {Robotham}, A.~S.~G., {et~al.} 2015, \href{http://dx.doi.org/10.1093/mnras/stu2635}{\color{magenta}\mnras}, \href{https://ui.adsabs.harvard.edu/abs/2015MNRAS.447.2857B}{447, 2857}

\bibitem[{{Buder} {et~al.}(2025){Buder}, {Kos}, {Wang}, {McKenzie}, {Howell}, {Martell}, {Hayden}, {Zucker}, {Nordlander}, {Montet}, {Traven}, {Bland-Hawthorn}, {de Silva}, {Freeman}, {Lewis}, {Lind}, {Sharma}, {Simpson}, {Stello}, {Zwitter}, {Amarsi}, {Armstrong}, {Banks}, {Beavis}, {Beeson}, {Chen}, {Ciuc{\u{a}}}, {da Costa}, {de Grijs}, {Martin}, {Nataf}, {Ness}, {Rains}, {Scarr}, {Vogrin{\v{c}}i{\v{c}}}, {Wang}, {Wittenmyer}, {Xie}, \& {The Galah Collaboration}}]{2025PASA...42...51B}
{Buder}, S., {Kos}, J., {Wang}, X.~E., {et~al.} 2025, \href{http://dx.doi.org/10.1017/pasa.2025.26}{\color{magenta}\pasa}, \href{https://ui.adsabs.harvard.edu/abs/2025PASA...42...51B}{42, e051}

\bibitem[{{Bundy} {et~al.}(2015){Bundy}, {Bershady}, {Law}, {Yan}, {Drory}, {MacDonald}, {Wake}, {Cherinka}, {S{\'a}nchez-Gallego}, {Weijmans}, {Thomas}, {Tremonti}, {Masters}, {Coccato}, {Diamond-Stanic}, {Arag{\'o}n-Salamanca}, {Avila-Reese}, {Badenes}, {Falc{\'o}n-Barroso}, {Belfiore}, {Bizyaev}, {Blanc}, {Bland-Hawthorn}, {Blanton}, {Brownstein}, {Byler}, {Cappellari}, {Conroy}, {Dutton}, {Emsellem}, {Etherington}, {Frinchaboy}, {Fu}, {Gunn}, {Harding}, {Johnston}, {Kauffmann}, {Kinemuchi}, {Klaene}, {Knapen}, {Leauthaud}, {Li}, {Lin}, {Maiolino}, {Malanushenko}, {Malanushenko}, {Mao}, {Maraston}, {McDermid}, {Merrifield}, {Nichol}, {Oravetz}, {Pan}, {Parejko}, {Sanchez}, {Schlegel}, {Simmons}, {Steele}, {Steinmetz}, {Thanjavur}, {Thompson}, {Tinker}, {van den Bosch}, {Westfall}, {Wilkinson}, {Wright}, {Xiao}, \& {Zhang}}]{2015ApJ...798....7B}
{Bundy}, K., {Bershady}, M.~A., {Law}, D.~R., {et~al.} 2015, \href{http://dx.doi.org/10.1088/0004-637X/798/1/7}{\color{magenta}\apj}, \href{https://ui.adsabs.harvard.edu/abs/2015ApJ...798....7B}{798, 7}

\bibitem[{{Bureau} \& {Athanassoula}(2005)}]{2005ApJ...626..159B}
{Bureau}, M. \& {Athanassoula}, E. 2005, \href{http://dx.doi.org/10.1086/430056}{\color{magenta}\apj}, \href{https://ui.adsabs.harvard.edu/abs/2005ApJ...626..159B}{626, 159}

\bibitem[{{Calzetti} {et~al.}(2000){Calzetti}, {Armus}, {Bohlin}, {Kinney}, {Koornneef}, \& {Storchi-Bergmann}}]{2000ApJ...533..682C}
{Calzetti}, D., {Armus}, L., {Bohlin}, R.~C., {et~al.} 2000, \href{http://dx.doi.org/10.1086/308692}{\color{magenta}\apj}, \href{https://ui.adsabs.harvard.edu/abs/2000ApJ...533..682C}{533, 682}

\bibitem[{{Camps} \& {Baes}(2015)}]{2015A&C.....9...20C}
{Camps}, P. \& {Baes}, M. 2015, \href{http://dx.doi.org/10.1016/j.ascom.2014.10.004}{\color{magenta}Astronomy and Computing}, \href{https://ui.adsabs.harvard.edu/abs/2015A&C.....9...20C}{9, 20}

\bibitem[{{Cappellari}(2016)}]{2016ARA&A..54..597C}
{Cappellari}, M. 2016, \href{http://dx.doi.org/10.1146/annurev-astro-082214-122432}{\color{magenta}\araa}, \href{https://ui.adsabs.harvard.edu/abs/2016ARA&A..54..597C}{54, 597}

\bibitem[{{Cappellari}(2017)}]{2017MNRAS.466..798C}
{Cappellari}, M. 2017, \href{http://dx.doi.org/10.1093/mnras/stw3020}{\color{magenta}\mnras}, \href{https://ui.adsabs.harvard.edu/abs/2017MNRAS.466..798C}{466, 798}

\bibitem[{{Cappellari}(2023)}]{2023MNRAS.526.3273C}
{Cappellari}, M. 2023, \href{http://dx.doi.org/10.1093/mnras/stad2597}{\color{magenta}\mnras}, \href{https://ui.adsabs.harvard.edu/abs/2023MNRAS.526.3273C}{526, 3273}

\bibitem[{{Cappellari} \& {Copin}(2003)}]{2003MNRAS.342..345C}
{Cappellari}, M. \& {Copin}, Y. 2003, \href{http://dx.doi.org/10.1046/j.1365-8711.2003.06541.x}{\color{magenta}\mnras}, \href{https://ui.adsabs.harvard.edu/abs/2003MNRAS.342..345C}{342, 345}

\bibitem[{{Cappellari} \& {Emsellem}(2004)}]{2004PASP..116..138C}
{Cappellari}, M. \& {Emsellem}, E. 2004, \href{http://dx.doi.org/10.1086/381875}{\color{magenta}\pasp}, \href{https://ui.adsabs.harvard.edu/abs/2004PASP..116..138C}{116, 138}

\bibitem[{{Cappellari} {et~al.}(2011){Cappellari}, {Emsellem}, {Krajnovi{\'c}}, {McDermid}, {Scott}, {Verdoes Kleijn}, {Young}, {Alatalo}, {Bacon}, {Blitz}, {Bois}, {Bournaud}, {Bureau}, {Davies}, {Davis}, {de Zeeuw}, {Duc}, {Khochfar}, {Kuntschner}, {Lablanche}, {Morganti}, {Naab}, {Oosterloo}, {Sarzi}, {Serra}, \& {Weijmans}}]{2011MNRAS.413..813C}
{Cappellari}, M., {Emsellem}, E., {Krajnovi{\'c}}, D., {et~al.} 2011, \href{http://dx.doi.org/10.1111/j.1365-2966.2010.18174.x}{\color{magenta}\mnras}, \href{https://ui.adsabs.harvard.edu/abs/2011MNRAS.413..813C}{413, 813}

\bibitem[{{Chen} {et~al.}(2010){Chen}, {Liang}, {Hammer}, {Prugniel}, {Zhong}, {Rodrigues}, {Zhao}, \& {Flores}}]{2010A&A...515A.101C}
{Chen}, X.~Y., {Liang}, Y.~C., {Hammer}, F., {et~al.} 2010, \href{http://dx.doi.org/10.1051/0004-6361/200913894}{\color{magenta}\aap}, \href{https://ui.adsabs.harvard.edu/abs/2010A&A...515A.101C}{515, A101}

\bibitem[{{Cid Fernandes} {et~al.}(2013){Cid Fernandes}, {P{\'e}rez}, {Garc{\'\i}a Benito}, {Gonz{\'a}lez Delgado}, {de Amorim}, {S{\'a}nchez}, {Husemann}, {Falc{\'o}n Barroso}, {S{\'a}nchez-Bl{\'a}zquez}, {Walcher}, \& {Mast}}]{2013A&A...557A..86C}
{Cid Fernandes}, R., {P{\'e}rez}, E., {Garc{\'\i}a Benito}, R., {et~al.} 2013, \href{http://dx.doi.org/10.1051/0004-6361/201220616}{\color{magenta}\aap}, \href{https://ui.adsabs.harvard.edu/abs/2013A&A...557A..86C}{557, A86}

\bibitem[{{Crain} {et~al.}(2015){Crain}, {Schaye}, {Bower}, {Furlong}, {Schaller}, {Theuns}, {Dalla Vecchia}, {Frenk}, {McCarthy}, {Helly}, {Jenkins}, {Rosas-Guevara}, {White}, \& {Trayford}}]{2015MNRAS.450.1937C}
{Crain}, R.~A., {Schaye}, J., {Bower}, R.~G., {et~al.} 2015, \href{http://dx.doi.org/10.1093/mnras/stv725}{\color{magenta}\mnras}, \href{https://ui.adsabs.harvard.edu/abs/2015MNRAS.450.1937C}{450, 1937}

\bibitem[{{de Zeeuw} {et~al.}(2002){de Zeeuw}, {Bureau}, {Emsellem}, {Bacon}, {Carollo}, {Copin}, {Davies}, {Kuntschner}, {Miller}, {Monnet}, {Peletier}, \& {Verolme}}]{2002MNRAS.329..513D}
{de Zeeuw}, P.~T., {Bureau}, M., {Emsellem}, E., {et~al.} 2002, \href{http://dx.doi.org/10.1046/j.1365-8711.2002.05059.x}{\color{magenta}\mnras}, \href{https://ui.adsabs.harvard.edu/abs/2002MNRAS.329..513D}{329, 513}

\bibitem[{{Debattista} {et~al.}(2017){Debattista}, {Ness}, {Gonzalez}, {Freeman}, {Zoccali}, \& {Minniti}}]{2017MNRAS.469.1587D}
{Debattista}, V.~P., {Ness}, M., {Gonzalez}, O.~A., {et~al.} 2017, \href{http://dx.doi.org/10.1093/mnras/stx947}{\color{magenta}\mnras}, \href{https://ui.adsabs.harvard.edu/abs/2017MNRAS.469.1587D}{469, 1587}

\bibitem[{{Di Matteo} {et~al.}(2013){Di Matteo}, {Haywood}, {Combes}, {Semelin}, \& {Snaith}}]{2013A&A...553A.102D}
{Di Matteo}, P., {Haywood}, M., {Combes}, F., {Semelin}, B., \& {Snaith}, O.~N. 2013, \href{http://dx.doi.org/10.1051/0004-6361/201220539}{\color{magenta}\aap}, \href{https://ui.adsabs.harvard.edu/abs/2013A&A...553A.102D}{553, A102}

\bibitem[{{Emsellem} {et~al.}(2022){Emsellem}, {Schinnerer}, {Santoro}, {Belfiore}, {Pessa}, {McElroy}, {Blanc}, {Congiu}, {Groves}, {Ho}, {Kreckel}, {Razza}, {Sanchez-Blazquez}, {Egorov}, {Faesi}, {Klessen}, {Leroy}, {Meidt}, {Querejeta}, {Rosolowsky}, {Scheuermann}, {Anand}, {Barnes}, {Be{\v{s}}li{\'c}}, {Bigiel}, {Boquien}, {Cao}, {Chevance}, {Dale}, {Eibensteiner}, {Glover}, {Grasha}, {Henshaw}, {Hughes}, {Koch}, {Kruijssen}, {Lee}, {Liu}, {Pan}, {Pety}, {Saito}, {Sandstrom}, {Schruba}, {Sun}, {Thilker}, {Usero}, {Watkins}, \& {Williams}}]{2022A&A...659A.191E}
{Emsellem}, E., {Schinnerer}, E., {Santoro}, F., {et~al.} 2022, \href{http://dx.doi.org/10.1051/0004-6361/202141727}{\color{magenta}\aap}, \href{https://ui.adsabs.harvard.edu/abs/2022A&A...659A.191E}{659, A191}

\bibitem[{{Falc{\'o}n-Barroso} \& {Martig}(2021)}]{2021A&A...646A..31F}
{Falc{\'o}n-Barroso}, J. \& {Martig}, M. 2021, \href{http://dx.doi.org/10.1051/0004-6361/202039624}{\color{magenta}\aap}, \href{https://ui.adsabs.harvard.edu/abs/2021A&A...646A..31F}{646, A31}

\bibitem[{{Faucher} {et~al.}(2023){Faucher}, {Blanton}, \& {Macci{\`o}}}]{2023ApJ...957....7F}
{Faucher}, N., {Blanton}, M.~R., \& {Macci{\`o}}, A.~V. 2023, \href{http://dx.doi.org/10.3847/1538-4357/acf9f0}{\color{magenta}\apj}, \href{https://ui.adsabs.harvard.edu/abs/2023ApJ...957....7F}{957, 7}

\bibitem[{{Fielder} {et~al.}(2021){Fielder}, {Newman}, {Andrews}, {Zasowski}, {Boardman}, {Licquia}, {Masters}, \& {Salim}}]{2021MNRAS.508.4459F}
{Fielder}, C.~E., {Newman}, J.~A., {Andrews}, B.~H., {et~al.} 2021, \href{http://dx.doi.org/10.1093/mnras/stab2618}{\color{magenta}\mnras}, \href{https://ui.adsabs.harvard.edu/abs/2021MNRAS.508.4459F}{508, 4459}

\bibitem[{{Fraser-McKelvie} {et~al.}(2025{\natexlab{a}}){Fraser-McKelvie}, {Gadotti}, {Fragkoudi}, {de S{\'a}-Freitas}, {Martig}, {Bureau}, {Davis}, {Emsellem}, {Elliott}, {Fisher}, {Hayden}, {van de Sande}, \& {Watts}}]{2025A&A...705A...1F}
{Fraser-McKelvie}, A., {Gadotti}, D.~A., {Fragkoudi}, F., {et~al.} 2025{\natexlab{a}}, \href{http://dx.doi.org/10.1051/0004-6361/202557327}{\color{magenta}\aap}, \href{https://ui.adsabs.harvard.edu/abs/2025A&A...705A...1F}{705, A1}

\bibitem[{{Fraser-McKelvie} {et~al.}(2019){Fraser-McKelvie}, {Merrifield}, \& {Arag{\'o}n-Salamanca}}]{2019MNRAS.489.5030F}
{Fraser-McKelvie}, A., {Merrifield}, M., \& {Arag{\'o}n-Salamanca}, A. 2019, \href{http://dx.doi.org/10.1093/mnras/stz2493}{\color{magenta}\mnras}, \href{https://ui.adsabs.harvard.edu/abs/2019MNRAS.489.5030F}{489, 5030}

\bibitem[{{Fraser-McKelvie} {et~al.}(2025{\natexlab{b}}){Fraser-McKelvie}, {van de Sande}, {Gadotti}, {Emsellem}, {Brown}, {Fisher}, {Martig}, {Bureau}, {Gerhard}, {Battisti}, {Bland-Hawthorn}, {Boecker}, {Catinella}, {Combes}, {Cortese}, {Croom}, {Davis}, {Falc{\'o}n-Barroso}, {Fragkoudi}, {Freeman}, {Hayden}, {McDermid}, {Mazzilli Ciraulo}, {Mendel}, {Pinna}, {Poci}, {Rutherford}, {de S{\'a}-Freitas}, {Silva-Lima}, {Valenzuela}, {van de Ven}, {Wang}, \& {Watts}}]{2025A&A...700A.237F}
{Fraser-McKelvie}, A., {van de Sande}, J., {Gadotti}, D.~A., {et~al.} 2025{\natexlab{b}}, \href{http://dx.doi.org/10.1051/0004-6361/202452891}{\color{magenta}\aap}, \href{https://ui.adsabs.harvard.edu/abs/2025A&A...700A.237F}{700, A237}

\bibitem[{{Gaia Collaboration} {et~al.}(2016){Gaia Collaboration}, {Brown}, {Vallenari}, {Prusti}, {de Bruijne}, {Mignard}, {Drimmel}, {Babusiaux}, {Bailer-Jones}, {Bastian}, {Biermann}, {Evans}, {Eyer}, {Jansen}, {Jordi}, {Katz}, {Klioner}, {Lammers}, {Lindegren}, {Luri}, {O'Mullane}, {Panem}, {Pourbaix}, {Randich}, {Sartoretti}, {Siddiqui}, {Soubiran}, {Valette}, {van Leeuwen}, {Walton}, {Aerts}, {Arenou}, {Cropper}, {H{\o}g}, {Lattanzi}, {Grebel}, {Holland}, {Huc}, {Passot}, {Perryman}, {Bramante}, {Cacciari}, {Casta{\~n}eda}, {Chaoul}, {Cheek}, {De Angeli}, {Fabricius}, {Guerra}, {Hern{\'a}ndez}, {Jean-Antoine-Piccolo}, {Masana}, {Messineo}, {Mowlavi}, {Nienartowicz}, {Ord{\'o}{\~n}ez-Blanco}, {Panuzzo}, {Portell}, {Richards}, {Riello}, {Seabroke}, {Tanga}, {Th{\'e}venin}, {Torra}, {Els}, {Gracia-Abril}, {Comoretto}, {Garcia-Reinaldos}, {Lock}, {Mercier}, {Altmann}, {Andrae}, {Astraatmadja}, {Bellas-Velidis}, {Benson}, {Berthier}, {Blomme}, {Busso}, {Carry}, {Cellino}, {Clementini}, {Cowell}, {Creevey},
  {Cuypers}, {Davidson}, {De Ridder}, {de Torres}, {Delchambre}, {Dell'Oro}, {Ducourant}, {Fr{\'e}mat}, {Garc{\'\i}a-Torres}, {Gosset}, {Halbwachs}, {Hambly}, {Harrison}, {Hauser}, {Hestroffer}, {Hodgkin}, {Huckle}, {Hutton}, {Jasniewicz}, {Jordan}, {Kontizas}, {Korn}, {Lanzafame}, {Manteiga}, {Moitinho}, {Muinonen}, {Osinde}, {Pancino}, {Pauwels}, {Petit}, {Recio-Blanco}, {Robin}, {Sarro}, {Siopis}, {Smith}, {Smith}, {Sozzetti}, {Thuillot}, {van Reeven}, {Viala}, {Abbas}, {Abreu Aramburu}, {Accart}, {Aguado}, {Allan}, {Allasia}, {Altavilla}, {{\'A}lvarez}, {Alves}, {Anderson}, {Andrei}, {Anglada Varela}, {Antiche}, {Antoja}, {Ant{\'o}n}, {Arcay}, {Bach}, {Baker}, {Balaguer-N{\'u}{\~n}ez}, {Barache}, {Barata}, {Barbier}, {Barblan}, {Barrado y Navascu{\'e}s}, {Barros}, {Barstow}, {Becciani}, {Bellazzini}, {Bello Garc{\'\i}a}, {Belokurov}, {Bendjoya}, {Berihuete}, {Bianchi}, {Bienaym{\'e}}, {Billebaud}, {Blagorodnova}, {Blanco-Cuaresma}, {Boch}, {Bombrun}, {Borrachero}, {Bouquillon}, {Bourda}, {Bouy},
  {Bragaglia}, {Breddels}, {Brouillet}, {Br{\"u}semeister}, {Bucciarelli}, {Burgess}, {Burgon}, {Burlacu}, {Busonero}, {Buzzi}, {Caffau}, {Cambras}, {Campbell}, {Cancelliere}, {Cantat-Gaudin}, {Carlucci}, {Carrasco}, {Castellani}, {Charlot}, {Charnas}, {Chiavassa}, {Clotet}, {Cocozza}, {Collins}, {Costigan}, {Crifo}, {Cross}, {Crosta}, {Crowley}, {Dafonte}, {Damerdji}, {Dapergolas}, {David}, {David}, \& {De Cat}}]{2016A&A...595A...2G}
{Gaia Collaboration}, {Brown}, A.~G.~A., {Vallenari}, A., {et~al.} 2016, \href{http://dx.doi.org/10.1051/0004-6361/201629512}{\color{magenta}\aap}, \href{https://ui.adsabs.harvard.edu/abs/2016A&A...595A...2G}{595, A2}

\bibitem[{{Gaia Collaboration} {et~al.}(2023{\natexlab{a}}){Gaia Collaboration}, {Recio-Blanco}, {Kordopatis}, {de Laverny}, {Palicio}, {Spagna}, {Spina}, {Katz}, {Re Fiorentin}, {Poggio}, {McMillan}, {Vallenari}, {Lattanzi}, {Seabroke}, {Casamiquela}, {Bragaglia}, {Antoja}, {Bailer-Jones}, {Schultheis}, {Andrae}, {Fouesneau}, {Cropper}, {Cantat-Gaudin}, {Bijaoui}, {Heiter}, {Brown}, {Prusti}, {de Bruijne}, {Arenou}, {Babusiaux}, {Biermann}, {Creevey}, {Ducourant}, {Evans}, {Eyer}, {Guerra}, {Hutton}, {Jordi}, {Klioner}, {Lammers}, {Lindegren}, {Luri}, {Mignard}, {Panem}, {Pourbaix}, {Randich}, {Sartoretti}, {Soubiran}, {Tanga}, {Walton}, {Bastian}, {Drimmel}, {Jansen}, {van Leeuwen}, {Bakker}, {Cacciari}, {Casta{\~n}eda}, {De Angeli}, {Fabricius}, {Fr{\'e}mat}, {Galluccio}, {Guerrier}, {Masana}, {Messineo}, {Mowlavi}, {Nicolas}, {Nienartowicz}, {Pailler}, {Panuzzo}, {Riclet}, {Roux}, {Sordo}, {Th{\'e}venin}, {Gracia-Abril}, {Portell}, {Teyssier}, {Altmann}, {Audard}, {Bellas-Velidis}, {Benson}, {Berthier},
  {Blomme}, {Burgess}, {Busonero}, {Busso}, {C{\'a}novas}, {Carry}, {Cellino}, {Cheek}, {Clementini}, {Damerdji}, {Davidson}, {de Teodoro}, {Nu{\~n}ez Campos}, {Delchambre}, {Dell'Oro}, {Esquej}, {Fern{\'a}ndez-Hern{\'a}ndez}, {Fraile}, {Garabato}, {Garc{\'\i}a-Lario}, {Gosset}, {Haigron}, {Halbwachs}, {Hambly}, {Harrison}, {Hern{\'a}ndez}, {Hestroffer}, {Hodgkin}, {Holl}, {Jan{\ss}en}, {Jevardat de Fombelle}, {Jordan}, {Krone-Martins}, {Lanzafame}, {L{\"o}ffler}, {Marchal}, {Marrese}, {Moitinho}, {Muinonen}, {Osborne}, {Pancino}, {Pauwels}, {Reyl{\'e}}, {Riello}, {Rimoldini}, {Roegiers}, {Rybizki}, {Sarro}, {Siopis}, {Smith}, {Sozzetti}, {Utrilla}, {van Leeuwen}, {Abbas}, {{\'A}brah{\'a}m}, {Abreu Aramburu}, {Aerts}, {Aguado}, {Ajaj}, {Aldea-Montero}, {Altavilla}, {{\'A}lvarez}, {Alves}, {Anders}, {Anderson}, {Anglada Varela}, {Baines}, {Baker}, {Balaguer-N{\'u}{\~n}ez}, {Balbinot}, {Balog}, {Barache}, {Barbato}, {Barros}, {Barstow}, {Bartolom{\'e}}, {Bassilana}, {Bauchet}, {Becciani}, {Bellazzini},
  {Berihuete}, {Bernet}, {Bertone}, {Bianchi}, {Binnenfeld}, {Blanco-Cuaresma}, {Boch}, {Bombrun}, {Bossini}, {Bouquillon}, {Bramante}, {Breedt}, {Bressan}, {Brouillet}, {Brugaletta}, {Bucciarelli}, {Burlacu}, {Butkevich}, {Buzzi}, {Caffau}, {Cancelliere}, {Carballo}, {Carlucci}, {Carnerero}, {Carrasco}, {Castellani}, {Castro-Ginard}, {Chaoul}, {Charlot}, {Chemin}, {Chiaramida}, {Chiavassa}, {Chornay}, {Comoretto}, {Contursi}, {Cooper}, {Cornez}, {Cowell}, {Crifo}, {Crosta}, {Crowley}, {Dafonte}, {Dapergolas}, {David}, {De Luise}, {De March}, {De Ridder}, {de Souza}, {de Torres}, {del Peloso}, {del Pozo}, {Delbo}, {Delgado}, {Delisle}, {Demouchy}, {Dharmawardena}, {Di Matteo}, {Diakite}, {Diener}, {Distefano}, {Dolding}, {Edvardsson}, {Enke}, {Fabre}, {Fabrizio}, {Faigler}, {Fedorets}, {Fernique}, {Figueras}, {Fournier}, {Fouron}, {Fragkoudi}, {Gai}, {Garcia-Gutierrez}, {Garcia-Reinaldos}, {Garc{\'\i}a-Torres}, {Garofalo}, {Gavel}, {Gavras}, {Gerlach}, {Geyer}, {Giacobbe}, {Gilmore}, {Girona}, {Giuffrida},
  {Gomel}, {Gomez}, {Gonz{\'a}lez-N{\'u}{\~n}ez}, {Gonz{\'a}lez-Santamar{\'\i}a}, {Gonz{\'a}lez-Vidal}, {Granvik}, {Guillout}, {Guiraud}, {Guti{\'e}rrez-S{\'a}nchez}, {Guy}, {Hatzidimitriou}, {Hauser}, {Haywood}, {Helmer}, {Helmi}, {Sarmiento}, {Hidalgo}, {H{\l}adczuk}, {Hobbs}, {Holland}, {Huckle}, {Jardine}, {Jasniewicz}, {Jean-Antoine Piccolo}, {Jim{\'e}nez-Arranz}, {Juaristi Campillo}, {Julbe}, {Karbevska}, {Kervella}, {Khanna}, {Korn}, {K{\'o}sp{\'a}l}, {Kostrzewa-Rutkowska}, {Kruszy{\'n}ska}, {Kun}, {Laizeau}, {Lambert}, {Lanza}, {Lasne}, {Le Campion}, {Lebreton}, {Lebzelter}, {Leccia}, {Leclerc}, {Lecoeur-Taibi}, {Liao}, {Licata}, {Lindstr{\o}m}, {Lister}, {Livanou}, {Lobel}, {Lorca}, {Loup}, {Madrero Pardo}, {Magdaleno Romeo}, {Managau}, {Mann}, {Manteiga}, {Marchant}, {Marconi}, {Marcos}, {Marcos Santos}, {Mar{\'\i}n Pina}, {Marinoni}, {Marocco}, {Marshall}, {Martin Polo}, {Mart{\'\i}n-Fleitas}, {Marton}, {Mary}, {Masip}, {Massari}, {Mastrobuono-Battisti}, {Mazeh}, {Messina}, {Michalik}, {Millar},
  {Mints}, {Molina}, {Molinaro}, {Moln{\'a}r}, {Monari}, {Mongui{\'o}}, {Montegriffo}, {Montero}, {Mor}, {Mora}, {Morbidelli}, {Morel}, {Morris}, {Muraveva}, {Murphy}, {Musella}, {Nagy}, {Noval}, {Oca{\~n}a}, {Ogden}, {Ordenovic}, {Osinde}, {Pagani}, {Pagano}, {Palaversa}, {Pallas-Quintela}, {Panahi}, {Payne-Wardenaar}, {Pe{\~n}alosa Esteller}, {Penttil{\"a}}, {Pichon}, {Piersimoni}, {Pineau}, {Plachy}, {Plum}, {Pr{\v{s}}a}, {Pulone}, {Racero}, {Ragaini}, {Rainer}, {Raiteri}, {Ramos}, {Ramos-Lerate}, {Regibo}, {Richards}, {Rios Diaz}, {Ripepi}, {Riva}, {Rix}, {Rixon}, {Robichon}, {Robin}, {Robin}, {Roelens}, {Rogues}, {Rohrbasser}, {Romero-G{\'o}mez}, {Rowell}, {Royer}, {Ruz Mieres}, {Rybicki}, {Sadowski}, {S{\'a}ez N{\'u}{\~n}ez}, {Sagrist{\`a} Sell{\'e}s}, {Sahlmann}, {Salguero}, {Samaras}, {Sanchez Gimenez}, {Sanna}, {Santove{\~n}a}, {Sarasso}, {Sciacca}, {Segol}, {Segovia}, {S{\'e}gransan}, {Semeux}, {Shahaf}, {Siddiqui}, {Siebert}, {Siltala}, {Silvelo}, {Slezak}, {Slezak}, {Smart}, {Snaith}, {Solano},
  {Solitro}, {Souami}, {Souchay}, {Spoto}, {Steele}, {Steidelm{\"u}ller}, {Stephenson}, {S{\"u}veges}, {Surdej}, {Szabados}, {Szegedi-Elek}, {Taris}, {Taylor}, {Teixeira}, {Tolomei}, {Tonello}, {Torra}, {Torra}, {Torralba Elipe}, {Trabucchi}, {Tsounis}, {Turon}, {Ulla}, {Unger}, {Vaillant}, {van Dillen}, {van Reeven}, {Vanel}, {Vecchiato}, {Viala}, {Vicente}, {Voutsinas}, {Weiler}, {Wevers}, {Wyrzykowski}, {Yoldas}, {Yvard}, {Zhao}, {Zorec}, {Zucker}, \& {Zwitter}}]{2023A&A...674A..38G}
{Gaia Collaboration}, {Recio-Blanco}, A., {Kordopatis}, G., {et~al.} 2023{\natexlab{a}}, \href{http://dx.doi.org/10.1051/0004-6361/202243511}{\color{magenta}\aap}, \href{https://ui.adsabs.harvard.edu/abs/2023A&A...674A..38G}{674, A38}

\bibitem[{{Gaia Collaboration} {et~al.}(2023{\natexlab{b}}){Gaia Collaboration}, {Vallenari}, {Brown}, {Prusti}, {de Bruijne}, {Arenou}, {Babusiaux}, {Biermann}, {Creevey}, {Ducourant}, {Evans}, {Eyer}, {Guerra}, {Hutton}, {Jordi}, {Klioner}, {Lammers}, {Lindegren}, {Luri}, {Mignard}, {Panem}, {Pourbaix}, {Randich}, {Sartoretti}, {Soubiran}, {Tanga}, {Walton}, {Bailer-Jones}, {Bastian}, {Drimmel}, {Jansen}, {Katz}, {Lattanzi}, {van Leeuwen}, {Bakker}, {Cacciari}, {Casta{\~n}eda}, {De Angeli}, {Fabricius}, {Fouesneau}, {Fr{\'e}mat}, {Galluccio}, {Guerrier}, {Heiter}, {Masana}, {Messineo}, {Mowlavi}, {Nicolas}, {Nienartowicz}, {Pailler}, {Panuzzo}, {Riclet}, {Roux}, {Seabroke}, {Sordo}, {Th{\'e}venin}, {Gracia-Abril}, {Portell}, {Teyssier}, {Altmann}, {Andrae}, {Audard}, {Bellas-Velidis}, {Benson}, {Berthier}, {Blomme}, {Burgess}, {Busonero}, {Busso}, {C{\'a}novas}, {Carry}, {Cellino}, {Cheek}, {Clementini}, {Damerdji}, {Davidson}, {de Teodoro}, {Nu{\~n}ez Campos}, {Delchambre}, {Dell'Oro}, {Esquej},
  {Fern{\'a}ndez-Hern{\'a}ndez}, {Fraile}, {Garabato}, {Garc{\'\i}a-Lario}, {Gosset}, {Haigron}, {Halbwachs}, {Hambly}, {Harrison}, {Hern{\'a}ndez}, {Hestroffer}, {Hodgkin}, {Holl}, {Jan{\ss}en}, {Jevardat de Fombelle}, {Jordan}, {Krone-Martins}, {Lanzafame}, {L{\"o}ffler}, {Marchal}, {Marrese}, {Moitinho}, {Muinonen}, {Osborne}, {Pancino}, {Pauwels}, {Recio-Blanco}, {Reyl{\'e}}, {Riello}, {Rimoldini}, {Roegiers}, {Rybizki}, {Sarro}, {Siopis}, {Smith}, {Sozzetti}, {Utrilla}, {van Leeuwen}, {Abbas}, {{\'A}brah{\'a}m}, {Abreu Aramburu}, {Aerts}, {Aguado}, {Ajaj}, {Aldea-Montero}, {Altavilla}, {{\'A}lvarez}, {Alves}, {Anders}, {Anderson}, {Anglada Varela}, {Antoja}, {Baines}, {Baker}, {Balaguer-N{\'u}{\~n}ez}, {Balbinot}, {Balog}, {Barache}, {Barbato}, {Barros}, {Barstow}, {Bartolom{\'e}}, {Bassilana}, {Bauchet}, {Becciani}, {Bellazzini}, {Berihuete}, {Bernet}, {Bertone}, {Bianchi}, {Binnenfeld}, {Blanco-Cuaresma}, {Blazere}, {Boch}, {Bombrun}, {Bossini}, {Bouquillon}, {Bragaglia}, {Bramante}, {Breedt},
  {Bressan}, {Brouillet}, {Brugaletta}, {Bucciarelli}, {Burlacu}, {Butkevich}, {Buzzi}, {Caffau}, {Cancelliere}, {Cantat-Gaudin}, {Carballo}, {Carlucci}, {Carnerero}, {Carrasco}, {Casamiquela}, {Castellani}, {Castro-Ginard}, {Chaoul}, {Charlot}, {Chemin}, {Chiaramida}, {Chiavassa}, {Chornay}, {Comoretto}, {Contursi}, {Cooper}, {Cornez}, {Cowell}, {Crifo}, {Cropper}, {Crosta}, {Crowley}, {Dafonte}, {Dapergolas}, {David}, {David}, {de Laverny}, {De Luise}, {De March}, {De Ridder}, {de Souza}, {de Torres}, {del Peloso}, {del Pozo}, {Delbo}, {Delgado}, {Delisle}, {Demouchy}, {Dharmawardena}, {Di Matteo}, {Diakite}, {Diener}, {Distefano}, {Dolding}, {Edvardsson}, {Enke}, {Fabre}, {Fabrizio}, {Faigler}, {Fedorets}, {Fernique}, {Fienga}, {Figueras}, {Fournier}, {Fouron}, {Fragkoudi}, {Gai}, {Garcia-Gutierrez}, {Garcia-Reinaldos}, {Garc{\'\i}a-Torres}, {Garofalo}, {Gavel}, {Gavras}, {Gerlach}, {Geyer}, {Giacobbe}, {Gilmore}, {Girona}, {Giuffrida}, {Gomel}, {Gomez}, {Gonz{\'a}lez-N{\'u}{\~n}ez},
  {Gonz{\'a}lez-Santamar{\'\i}a}, {Gonz{\'a}lez-Vidal}, {Granvik}, {Guillout}, {Guiraud}, {Guti{\'e}rrez-S{\'a}nchez}, {Guy}, {Hatzidimitriou}, {Hauser}, {Haywood}, {Helmer}, {Helmi}, {Sarmiento}, {Hidalgo}, {Hilger}, {H{\l}adczuk}, {Hobbs}, {Holland}, {Huckle}, {Jardine}, {Jasniewicz}, {Jean-Antoine Piccolo}, {Jim{\'e}nez-Arranz}, {Jorissen}, {Juaristi Campillo}, {Julbe}, {Karbevska}, {Kervella}, {Khanna}, {Kontizas}, {Kordopatis}, {Korn}, {K{\'o}sp{\'a}l}, {Kostrzewa-Rutkowska}, {Kruszy{\'n}ska}, {Kun}, {Laizeau}, {Lambert}, {Lanza}, {Lasne}, {Le Campion}, {Lebreton}, {Lebzelter}, {Leccia}, {Leclerc}, {Lecoeur-Taibi}, {Liao}, {Licata}, {Lindstr{\o}m}, {Lister}, {Livanou}, {Lobel}, {Lorca}, {Loup}, {Madrero Pardo}, {Magdaleno Romeo}, {Managau}, {Mann}, {Manteiga}, {Marchant}, {Marconi}, {Marcos}, {Marcos Santos}, {Mar{\'\i}n Pina}, {Marinoni}, {Marocco}, {Marshall}, {Martin Polo}, {Mart{\'\i}n-Fleitas}, {Marton}, {Mary}, {Masip}, {Massari}, {Mastrobuono-Battisti}, {Mazeh}, {McMillan}, {Messina}, {Michalik},
  {Millar}, {Mints}, {Molina}, {Molinaro}, {Moln{\'a}r}, {Monari}, {Mongui{\'o}}, {Montegriffo}, {Montero}, {Mor}, {Mora}, {Morbidelli}, {Morel}, {Morris}, {Muraveva}, {Murphy}, {Musella}, {Nagy}, {Noval}, {Oca{\~n}a}, {Ogden}, {Ordenovic}, {Osinde}, {Pagani}, {Pagano}, {Palaversa}, {Palicio}, {Pallas-Quintela}, {Panahi}, {Payne-Wardenaar}, {Pe{\~n}alosa Esteller}, {Penttil{\"a}}, {Pichon}, {Piersimoni}, {Pineau}, {Plachy}, {Plum}, {Poggio}, {Pr{\v{s}}a}, {Pulone}, {Racero}, {Ragaini}, {Rainer}, {Raiteri}, {Rambaux}, {Ramos}, {Ramos-Lerate}, {Re Fiorentin}, {Regibo}, {Richards}, {Rios Diaz}, {Ripepi}, {Riva}, {Rix}, {Rixon}, {Robichon}, {Robin}, {Robin}, {Roelens}, {Rogues}, {Rohrbasser}, {Romero-G{\'o}mez}, {Rowell}, {Royer}, {Ruz Mieres}, {Rybicki}, {Sadowski}, {S{\'a}ez N{\'u}{\~n}ez}, {Sagrist{\`a} Sell{\'e}s}, {Sahlmann}, {Salguero}, {Samaras}, {Sanchez Gimenez}, {Sanna}, {Santove{\~n}a}, {Sarasso}, {Schultheis}, {Sciacca}, {Segol}, {Segovia}, {S{\'e}gransan}, {Semeux}, {Shahaf}, {Siddiqui}, {Siebert},
  {Siltala}, {Silvelo}, {Slezak}, {Slezak}, {Smart}, {Snaith}, {Solano}, {Solitro}, {Souami}, {Souchay}, {Spagna}, {Spina}, {Spoto}, {Steele}, {Steidelm{\"u}ller}, {Stephenson}, {S{\"u}veges}, {Surdej}, {Szabados}, {Szegedi-Elek}, {Taris}, {Taylor}, {Teixeira}, {Tolomei}, {Tonello}, {Torra}, {Torra}, {Torralba Elipe}, {Trabucchi}, {Tsounis}, {Turon}, {Ulla}, {Unger}, {Vaillant}, {van Dillen}, {van Reeven}, {Vanel}, {Vecchiato}, {Viala}, {Vicente}, {Voutsinas}, {Weiler}, {Wevers}, {Wyrzykowski}, {Yoldas}, {Yvard}, {Zhao}, {Zorec}, {Zucker}, \& {Zwitter}}]{2023A&A...674A...1G}
{Gaia Collaboration}, {Vallenari}, A., {Brown}, A.~G.~A., {et~al.} 2023{\natexlab{b}}, \href{http://dx.doi.org/10.1051/0004-6361/202243940}{\color{magenta}\aap}, \href{https://ui.adsabs.harvard.edu/abs/2023A&A...674A...1G}{674, A1}

\bibitem[{{Ge} {et~al.}(2019){Ge}, {Mao}, {Lu}, {Cappellari}, \& {Yan}}]{2019MNRAS.485.1675G}
{Ge}, J., {Mao}, S., {Lu}, Y., {Cappellari}, M., \& {Yan}, R. 2019, \href{http://dx.doi.org/10.1093/mnras/stz418}{\color{magenta}\mnras}, \href{https://ui.adsabs.harvard.edu/abs/2019MNRAS.485.1675G}{485, 1675}

\bibitem[{{Gerhard}(1993)}]{1993MNRAS.265..213G}
{Gerhard}, O.~E. 1993, \href{http://dx.doi.org/10.1093/mnras/265.1.213}{\color{magenta}\mnras}, \href{https://ui.adsabs.harvard.edu/abs/1993MNRAS.265..213G}{265, 213}

\bibitem[{{Gonzalez} {et~al.}(2017){Gonzalez}, {Debattista}, {Ness}, {Erwin}, \& {Gadotti}}]{2017MNRAS.466L..93G}
{Gonzalez}, O.~A., {Debattista}, V.~P., {Ness}, M., {Erwin}, P., \& {Gadotti}, D.~A. 2017, \href{http://dx.doi.org/10.1093/mnrasl/slw232}{\color{magenta}\mnras}, \href{https://ui.adsabs.harvard.edu/abs/2017MNRAS.466L..93G}{466, L93}

\bibitem[{{Gonzalez} {et~al.}(2016){Gonzalez}, {Gadotti}, {Debattista}, {Rejkuba}, {Valenti}, {Zoccali}, {Coccato}, {Minniti}, \& {Ness}}]{2016A&A...591A...7G}
{Gonzalez}, O.~A., {Gadotti}, D.~A., {Debattista}, V.~P., {et~al.} 2016, \href{http://dx.doi.org/10.1051/0004-6361/201527806}{\color{magenta}\aap}, \href{https://ui.adsabs.harvard.edu/abs/2016A&A...591A...7G}{591, A7}

\bibitem[{{Gonz{\'a}lez Delgado} {et~al.}(2017){Gonz{\'a}lez Delgado}, {P{\'e}rez}, {Cid Fernandes}, {Garc{\'\i}a-Benito}, {L{\'o}pez Fern{\'a}ndez}, {Vale Asari}, {Cortijo-Ferrero}, {de Amorim}, {Lacerda}, {S{\'a}nchez}, {Lehnert}, \& {Walcher}}]{2017A&A...607A.128G}
{Gonz{\'a}lez Delgado}, R.~M., {P{\'e}rez}, E., {Cid Fernandes}, R., {et~al.} 2017, \href{http://dx.doi.org/10.1051/0004-6361/201730883}{\color{magenta}\aap}, \href{https://ui.adsabs.harvard.edu/abs/2017A&A...607A.128G}{607, A128}

\bibitem[{{Groves} {et~al.}(2008){Groves}, {Dopita}, {Sutherland}, {Kewley}, {Fischera}, {Leitherer}, {Brandl}, \& {van Breugel}}]{2008ApJS..176..438G}
{Groves}, B., {Dopita}, M.~A., {Sutherland}, R.~S., {et~al.} 2008, \href{http://dx.doi.org/10.1086/528711}{\color{magenta}\apjs}, \href{https://ui.adsabs.harvard.edu/abs/2008ApJS..176..438G}{176, 438}

\bibitem[{{Halle} {et~al.}(2018){Halle}, {Di Matteo}, {Haywood}, \& {Combes}}]{2018A&A...616A..86H}
{Halle}, A., {Di Matteo}, P., {Haywood}, M., \& {Combes}, F. 2018, \href{http://dx.doi.org/10.1051/0004-6361/201832603}{\color{magenta}\aap}, \href{https://ui.adsabs.harvard.edu/abs/2018A&A...616A..86H}{616, A86}

\bibitem[{{Hayden} {et~al.}(2015){Hayden}, {Bovy}, {Holtzman}, {Nidever}, {Bird}, {Weinberg}, {Andrews}, {Majewski}, {Allende Prieto}, {Anders}, {Beers}, {Bizyaev}, {Chiappini}, {Cunha}, {Frinchaboy}, {Garc{\'\i}a-Her{\'n}andez}, {Garc{\'\i}a P{\'e}rez}, {Girardi}, {Harding}, {Hearty}, {Johnson}, {M{\'e}sz{\'a}ros}, {Minchev}, {O'Connell}, {Pan}, {Robin}, {Schiavon}, {Schneider}, {Schultheis}, {Shetrone}, {Skrutskie}, {Steinmetz}, {Smith}, {Wilson}, {Zamora}, \& {Zasowski}}]{2015ApJ...808..132H}
{Hayden}, M.~R., {Bovy}, J., {Holtzman}, J.~A., {et~al.} 2015, \href{http://dx.doi.org/10.1088/0004-637X/808/2/132}{\color{magenta}\apj}, \href{https://ui.adsabs.harvard.edu/abs/2015ApJ...808..132H}{808, 132}

\bibitem[{{Haywood} {et~al.}(2018){Haywood}, {Di Matteo}, {Lehnert}, {Snaith}, {Fragkoudi}, \& {Khoperskov}}]{2018A&A...618A..78H}
{Haywood}, M., {Di Matteo}, P., {Lehnert}, M., {et~al.} 2018, \href{http://dx.doi.org/10.1051/0004-6361/201731363}{\color{magenta}\aap}, \href{https://ui.adsabs.harvard.edu/abs/2018A&A...618A..78H}{618, A78}

\bibitem[{{Haywood} {et~al.}(2016){Haywood}, {Lehnert}, {Di Matteo}, {Snaith}, {Schultheis}, {Katz}, \& {G{\'o}mez}}]{2016A&A...589A..66H}
{Haywood}, M., {Lehnert}, M.~D., {Di Matteo}, P., {et~al.} 2016, \href{http://dx.doi.org/10.1051/0004-6361/201527567}{\color{magenta}\aap}, \href{https://ui.adsabs.harvard.edu/abs/2016A&A...589A..66H}{589, A66}

\bibitem[{{Helmi}(2020)}]{2020ARA&A..58..205H}
{Helmi}, A. 2020, \href{http://dx.doi.org/10.1146/annurev-astro-032620-021917}{\color{magenta}\araa}, \href{https://ui.adsabs.harvard.edu/abs/2020ARA&A..58..205H}{58, 205}

\bibitem[{{Helmi} {et~al.}(2018){Helmi}, {Babusiaux}, {Koppelman}, {Massari}, {Veljanoski}, \& {Brown}}]{2018Natur.563...85H}
{Helmi}, A., {Babusiaux}, C., {Koppelman}, H.~H., {et~al.} 2018, \href{http://dx.doi.org/10.1038/s41586-018-0625-x}{\color{magenta}\nat}, \href{https://ui.adsabs.harvard.edu/abs/2018Natur.563...85H}{563, 85}

\bibitem[{{Hinterer} {et~al.}(2022){Hinterer}, {Hubmer}, {Jethwa}, {Soodhalter}, {van de Ven}, \& {Ramlau}}]{2022arXiv220603925H}
{Hinterer}, F., {Hubmer}, S., {Jethwa}, P., {et~al.} 2022, \href{https://ui.adsabs.harvard.edu/abs/2022arXiv220603925H}{\href{http://dx.doi.org/10.48550/arXiv.2206.03925}{\color{magenta}arXiv e-prints}, arXiv:2206.03925}

\bibitem[{{Hunter} {et~al.}(2024){Hunter}, {Sormani}, {Beckmann}, {Vasiliev}, {Glover}, {Klessen}, {Soler}, {Brucy}, {Girichidis}, {G{\"o}ller}, {Ohlin}, {Tress}, {Molinari}, {Gerhard}, {Benedettini}, {Smith}, {Hennebelle}, \& {Testi}}]{2024A&A...692A.216H}
{Hunter}, G.~H., {Sormani}, M.~C., {Beckmann}, J.~P., {et~al.} 2024, \href{http://dx.doi.org/10.1051/0004-6361/202450000}{\color{magenta}\aap}, \href{https://ui.adsabs.harvard.edu/abs/2024A&A...692A.216H}{692, A216}

\bibitem[{{Iannuzzi} \& {Athanassoula}(2015)}]{2015MNRAS.450.2514I}
{Iannuzzi}, F. \& {Athanassoula}, E. 2015, \href{http://dx.doi.org/10.1093/mnras/stv764}{\color{magenta}\mnras}, \href{https://ui.adsabs.harvard.edu/abs/2015MNRAS.450.2514I}{450, 2514}

\bibitem[{{Imig} {et~al.}(2025{\natexlab{a}}){Imig}, {Holtzman}, {Zasowski}, {Lian}, {Boardman}, {Stone-Martinez}, {Mackereth}, {Prescott}, {Beaton}, {Beers}, {Bizyaev}, {Blanton}, {Cunha}, {Fern{\'a}ndez-Trincado}, {Fielder}, {Hasselquist}, {Hayes}, {Haywood}, {J{\"o}nsson}, {Lane}, {Majewski}, {M{\'e}sz{\'a}ros}, {Minchev}, {Nidever}, {Nitschelm}, \& {Sobeck}}]{2025ApJ...990..203I}
{Imig}, J., {Holtzman}, J.~A., {Zasowski}, G., {et~al.} 2025{\natexlab{a}}, \href{http://dx.doi.org/10.3847/1538-4357/adf723}{\color{magenta}\apj}, \href{https://ui.adsabs.harvard.edu/abs/2025ApJ...990..203I}{990, 203}

\bibitem[{{Imig} {et~al.}(2025{\natexlab{b}}){Imig}, {Holtzman}, {Zasowski}, {Lian}, {Boardman}, {Stone-Martinez}, {Mackereth}, {Prescott}, {Beaton}, {Beers}, {Bizyaev}, {Blanton}, {Cunha}, {Fern{\'a}ndez-Trincado}, {Fielder}, {Hasselquist}, {Hayes}, {Haywood}, {J{\"o}nsson}, {Lane}, {Majewski}, {M{\'e}sz{\'a}ros}, {Minchev}, {Nidever}, {Nitschelm}, \& {Sobeck}}]{Imig2025}
{Imig}, J., {Holtzman}, J.~A., {Zasowski}, G., {et~al.} 2025{\natexlab{b}}, \href{http://dx.doi.org/10.3847/1538-4357/adf723}{\color{magenta}\apj}, \href{https://ui.adsabs.harvard.edu/abs/2025ApJ...990..203I}{990, 203}

\bibitem[{{Imig} {et~al.}(2023){Imig}, {Price}, {Holtzman}, {Stone-Martinez}, {Majewski}, {Weinberg}, {Johnson}, {Allende Prieto}, {Beaton}, {Beers}, {Bizyaev}, {Blanton}, {Brownstein}, {Cunha}, {Fern{\'a}ndez-Trincado}, {Feuillet}, {Hasselquist}, {Hayes}, {J{\"o}nsson}, {Lane}, {Lian}, {M{\'e}sz{\'a}ros}, {Nidever}, {Robin}, {Shetrone}, {Smith}, \& {Wilson}}]{2023ApJ...954..124I}
{Imig}, J., {Price}, C., {Holtzman}, J.~A., {et~al.} 2023, \href{http://dx.doi.org/10.3847/1538-4357/ace9b8}{\color{magenta}\apj}, \href{https://ui.adsabs.harvard.edu/abs/2023ApJ...954..124I}{954, 124}

\bibitem[{{Jethwa} {et~al.}(2025){Jethwa}, {Hubmer}, {Ramlau}, \& {Van de Ven}}]{2025arXiv251103663J}
{Jethwa}, P., {Hubmer}, S., {Ramlau}, R., \& {Van de Ven}, G. 2025, \href{https://ui.adsabs.harvard.edu/abs/2025arXiv251103663J}{\href{http://dx.doi.org/10.48550/arXiv.2511.03663}{\color{magenta}arXiv e-prints}, arXiv:2511.03663}

\bibitem[{{Jin} {et~al.}(2024){Jin}, {Trager}, {Dalton}, {Aguerri}, {Drew}, {Falc{\'o}n-Barroso}, {G{\"a}nsicke}, {Hill}, {Iovino}, {Pieri}, {Poggianti}, {Smith}, {Vallenari}, {Abrams}, {Aguado}, {Antoja}, {Arag{\'o}n-Salamanca}, {Ascasibar}, {Babusiaux}, {Balcells}, {Barrena}, {Battaglia}, {Belokurov}, {Bensby}, {Bonifacio}, {Bragaglia}, {Carrasco}, {Carrera}, {Cornwell}, {Dom{\'\i}nguez-Palmero}, {Duncan}, {Famaey}, {Fari{\~n}a}, {Gonzalez}, {Guest}, {Hatch}, {Hess}, {Hoskin}, {Irwin}, {Knapen}, {Koposov}, {Kuchner}, {Laigle}, {Lewis}, {Longhetti}, {Lucatello}, {M{\'e}ndez-Abreu}, {Mercurio}, {Molaeinezhad}, {Mongui{\'o}}, {Morrison}, {Murphy}, {Peralta de Arriba}, {P{\'e}rez}, {P{\'e}rez-R{\`a}fols}, {Pic{\'o}}, {Raddi}, {Romero-G{\'o}mez}, {Royer}, {Siebert}, {Seabroke}, {Som}, {Terrett}, {Thomas}, {Wesson}, {Worley}, {Alfaro}, {Allende Prieto}, {Alonso-Santiago}, {Amos}, {Ashley}, {Balaguer-N{\'u}{\~n}ez}, {Balbinot}, {Bellazzini}, {Benn}, {Berlanas}, {Bernard}, {Best}, {Bettoni}, {Bianco}, {Bishop},
  {Blomqvist}, {Boeche}, {Bolzonella}, {Bonoli}, {Bosma}, {Britavskiy}, {Busarello}, {Caffau}, {Cantat-Gaudin}, {Castro-Ginard}, {Couto}, {Carbajo-Hijarrubia}, {Carter}, {Casamiquela}, {Conrado}, {Corcho-Caballero}, {Costantin}, {Deason}, {de Burgos}, {De Grandi}, {Di Matteo}, {Dom{\'\i}nguez-G{\'o}mez}, {Dorda}, {Drake}, {Dutta}, {Erkal}, {Feltzing}, {Ferr{\'e}-Mateu}, {Feuillet}, {Figueras}, {Fossati}, {Franciosini}, {Frasca}, {Fumagalli}, {Gallazzi}, {Garc{\'\i}a-Benito}, {Gentile Fusillo}, {Gebran}, {Gilbert}, {Gledhill}, {Gonz{\'a}lez Delgado}, {Greimel}, {Guarcello}, {Guerra}, {Gullieuszik}, {Haines}, {Hardcastle}, {Harris}, {Haywood}, {Helmi}, {Hernandez}, {Herrero}, {Hughes}, {Ir{\v{s}}i{\v{c}}}, {Jablonka}, {Jarvis}, {Jordi}, {Kondapally}, {Kordopatis}, {Krogager}, {La Barbera}, {Lam}, {Larsen}, {Lemasle}, {Lewis}, {Lhom{\'e}}, {Lind}, {Lodi}, {Longobardi}, {Lonoce}, {Magrini}, {Ma{\'\i}z Apell{\'a}niz}, {Marchal}, {Marco}, {Martin}, {Matsuno}, {Maurogordato}, {Merluzzi}, {Miralda-Escud{\'e}},
  {Molinari}, {Monari}, {Morelli}, {Mottram}, {Naylor}, {Negueruela}, {O{\~n}orbe}, {Pancino}, {Peirani}, {Peletier}, {Pozzetti}, {Rainer}, {Ramos}, {Read}, {Rossi}, {R{\"o}ttgering}, {Rubi{\~n}o-Mart{\'\i}n}, {Sabater}, {San Juan}, {Sanna}, {Schallig}, {Schiavon}, {Schultheis}, {Serra}, {Shimwell}, {Sim{\'o}n-D{\'\i}az}, {Smith}, {Sordo}, {Sorini}, {Soubiran}, {Starkenburg}, {Steele}, {Stott}, {Stuik}, {Tolstoy}, {Tortora}, {Tsantaki}, {Van der Swaelmen}, {van Weeren}, \& {Vergani}}]{2024MNRAS.530.2688J}
{Jin}, S., {Trager}, S.~C., {Dalton}, G.~B., {et~al.} 2024, \href{http://dx.doi.org/10.1093/mnras/stad557}{\color{magenta}\mnras}, \href{https://ui.adsabs.harvard.edu/abs/2024MNRAS.530.2688J}{530, 2688}

\bibitem[{{Kacharov} {et~al.}(2018){Kacharov}, {Neumayer}, {Seth}, {Cappellari}, {McDermid}, {Walcher}, \& {B{\"o}ker}}]{2018MNRAS.480.1973K}
{Kacharov}, N., {Neumayer}, N., {Seth}, A.~C., {et~al.} 2018, \href{http://dx.doi.org/10.1093/mnras/sty1985}{\color{magenta}\mnras}, \href{https://ui.adsabs.harvard.edu/abs/2018MNRAS.480.1973K}{480, 1973}

\bibitem[{{Katkov} {et~al.}(2013){Katkov}, {Sil'chenko}, \& {Afanasiev}}]{2013ApJ...769..105K}
{Katkov}, I.~Y., {Sil'chenko}, O.~K., \& {Afanasiev}, V.~L. 2013, \href{http://dx.doi.org/10.1088/0004-637X/769/2/105}{\color{magenta}\apj}, \href{https://ui.adsabs.harvard.edu/abs/2013ApJ...769..105K}{769, 105}

\bibitem[{{Kawata} {et~al.}(2018){Kawata}, {Baba}, {Ciuc{\v{a}}}, {Cropper}, {Grand}, {Hunt}, \& {Seabroke}}]{2018MNRAS.479L.108K}
{Kawata}, D., {Baba}, J., {Ciuc{\v{a}}}, I., {et~al.} 2018, \href{http://dx.doi.org/10.1093/mnrasl/sly107}{\color{magenta}\mnras}, \href{https://ui.adsabs.harvard.edu/abs/2018MNRAS.479L.108K}{479, L108}

\bibitem[{{Khoperskov} {et~al.}(2025{\natexlab{a}}){Khoperskov}, {Di Matteo}, {Steinmetz}, {Ratcliffe}, {van de Ven}, {Boin}, {Haywood}, {Kacharov}, {Minchev}, {Krajnovi{\'c}}, {Valentini}, \& {de Jong}}]{2025A&A...700A..90K}
{Khoperskov}, S., {Di Matteo}, P., {Steinmetz}, M., {et~al.} 2025{\natexlab{a}}, \href{http://dx.doi.org/10.1051/0004-6361/202453306}{\color{magenta}\aap}, \href{https://ui.adsabs.harvard.edu/abs/2025A&A...700A..90K}{700, A90}

\bibitem[{{Khoperskov} {et~al.}(2024){Khoperskov}, {Minchev}, {Steinmetz}, {Ratcliffe}, {Walcher}, \& {Libeskind}}]{2024MNRAS.533.3975K}
{Khoperskov}, S., {Minchev}, I., {Steinmetz}, M., {et~al.} 2024, \href{http://dx.doi.org/10.1093/mnras/stae1902}{\color{magenta}\mnras}, \href{https://ui.adsabs.harvard.edu/abs/2024MNRAS.533.3975K}{533, 3975}

\bibitem[{{Khoperskov} {et~al.}(2025{\natexlab{b}}){Khoperskov}, {Steinmetz}, {Haywood}, {van de Ven}, {Krajnovi{\'c}}, {Ratcliffe}, {Minchev}, {Di Matteo}, {Kacharov}, {Marques}, {Valentini}, \& {de Jong}}]{2025A&A...700A..89K}
{Khoperskov}, S., {Steinmetz}, M., {Haywood}, M., {et~al.} 2025{\natexlab{b}}, \href{http://dx.doi.org/10.1051/0004-6361/202453305}{\color{magenta}\aap}, \href{https://ui.adsabs.harvard.edu/abs/2025A&A...700A..89K}{700, A89}

\bibitem[{{Khoperskov} {et~al.}(2025{\natexlab{c}}){Khoperskov}, {van de Ven}, {Steinmetz}, {Ratcliffe}, {Minchev}, {Krajnovi{\'c}}, {Haywood}, {Di Matteo}, {Kacharov}, {Marques}, {Valentini}, \& {de Jong}}]{2025A&A...695A.220K}
{Khoperskov}, S., {van de Ven}, G., {Steinmetz}, M., {et~al.} 2025{\natexlab{c}}, \href{http://dx.doi.org/10.1051/0004-6361/202453304}{\color{magenta}\aap}, \href{https://ui.adsabs.harvard.edu/abs/2025A&A...695A.220K}{695, A220}

\bibitem[{{Knowles} {et~al.}(2023){Knowles}, {Sansom}, {Vazdekis}, \& {Allende Prieto}}]{2023MNRAS.523.3450K}
{Knowles}, A.~T., {Sansom}, A.~E., {Vazdekis}, A., \& {Allende Prieto}, C. 2023, \href{http://dx.doi.org/10.1093/mnras/stad1647}{\color{magenta}\mnras}, \href{https://ui.adsabs.harvard.edu/abs/2023MNRAS.523.3450K}{523, 3450}

\bibitem[{{Koleva} {et~al.}(2008){Koleva}, {Prugniel}, {Ocvirk}, {Le Borgne}, \& {Soubiran}}]{2008MNRAS.385.1998K}
{Koleva}, M., {Prugniel}, P., {Ocvirk}, P., {Le Borgne}, D., \& {Soubiran}, C. 2008, \href{http://dx.doi.org/10.1111/j.1365-2966.2008.12908.x}{\color{magenta}\mnras}, \href{https://ui.adsabs.harvard.edu/abs/2008MNRAS.385.1998K}{385, 1998}

\bibitem[{{Kompaniiets} {et~al.}(2025){Kompaniiets}, {Vavilova}, {Vasylkivskyi}, {Konovalenko}, {Pastoven}, {Izviekova}, {Dmytrenko}, {Dobrycheva}, {Fedorov}, {Khramtsov}, {Sergijenko}, \& {Vasylenko}}]{2025arXiv251214664K}
{Kompaniiets}, O.~V., {Vavilova}, I.~B., {Vasylkivskyi}, Y.~V., {et~al.} 2025, \href{https://ui.adsabs.harvard.edu/abs/2025arXiv251214664K}{\href{http://dx.doi.org/10.48550/arXiv.2512.14664}{\color{magenta}arXiv e-prints}, arXiv:2512.14664}

\bibitem[{{Kormendy} \& {Bender}(2019)}]{2019ApJ...872..106K}
{Kormendy}, J. \& {Bender}, R. 2019, \href{http://dx.doi.org/10.3847/1538-4357/aafdff}{\color{magenta}\apj}, \href{https://ui.adsabs.harvard.edu/abs/2019ApJ...872..106K}{872, 106}

\bibitem[{{Krishnarao} {et~al.}(2020){Krishnarao}, {Tremonti}, {Fraser-McKelvie}, {Kraljic}, {Boardman}, {Masters}, {Benjamin}, {Haffner}, {Jones}, {Pace}, {Zasowski}, {Bershady}, {Bizyaev}, {Brinkmann}, {Brownstein}, {Drory}, {Pan}, \& {Zhang}}]{2020ApJ...898..116K}
{Krishnarao}, D., {Tremonti}, C., {Fraser-McKelvie}, A., {et~al.} 2020, \href{http://dx.doi.org/10.3847/1538-4357/ab9fa3}{\color{magenta}\apj}, \href{https://ui.adsabs.harvard.edu/abs/2020ApJ...898..116K}{898, 116}

\bibitem[{{Kroupa}(2001)}]{2001MNRAS.322..231K}
{Kroupa}, P. 2001, \href{http://dx.doi.org/10.1046/j.1365-8711.2001.04022.x}{\color{magenta}\mnras}, \href{https://ui.adsabs.harvard.edu/abs/2001MNRAS.322..231K}{322, 231}

\bibitem[{{Kuijken} \& {Merrifield}(1993)}]{1993MNRAS.264..712K}
{Kuijken}, K. \& {Merrifield}, M.~R. 1993, \href{http://dx.doi.org/10.1093/mnras/264.3.712}{\color{magenta}\mnras}, \href{https://ui.adsabs.harvard.edu/abs/1993MNRAS.264..712K}{264, 712}

\bibitem[{{Lee} {et~al.}(2023){Lee}, {Sandstrom}, {Leroy}, {Thilker}, {Schinnerer}, {Rosolowsky}, {Larson}, {Egorov}, {Williams}, {Schmidt}, {Emsellem}, {Anand}, {Barnes}, {Belfiore}, {Be{\v{s}}li{\'c}}, {Bigiel}, {Blanc}, {Bolatto}, {Boquien}, {den Brok}, {Cao}, {Chandar}, {Chastenet}, {Chevance}, {Chiang}, {Congiu}, {Dale}, {Deger}, {Eibensteiner}, {Faesi}, {Glover}, {Grasha}, {Groves}, {Hassani}, {Henny}, {Henshaw}, {Hoyer}, {Hughes}, {Jeffreson}, {Jim{\'e}nez-Donaire}, {Kim}, {Kim}, {Klessen}, {Koch}, {Kreckel}, {Kruijssen}, {Li}, {Liu}, {Lopez}, {Maschmann}, {Chen}, {Meidt}, {Murphy}, {Neumann}, {Neumayer}, {Pan}, {Pessa}, {Pety}, {Querejeta}, {Pinna}, {Rodr{\'\i}guez}, {Saito}, {S{\'a}nchez-Bl{\'a}zquez}, {Santoro}, {Sardone}, {Smith}, {Sormani}, {Scheuermann}, {Stuber}, {Sutter}, {Sun}, {Teng}, {Tre{\ss}}, {Usero}, {Watkins}, {Whitmore}, \& {Razza}}]{2023ApJ...944L..17L}
{Lee}, J.~C., {Sandstrom}, K.~M., {Leroy}, A.~K., {et~al.} 2023, \href{http://dx.doi.org/10.3847/2041-8213/acaaae}{\color{magenta}\apjl}, \href{https://ui.adsabs.harvard.edu/abs/2023ApJ...944L..17L}{944, L17}

\bibitem[{{Licquia} {et~al.}(2015){Licquia}, {Newman}, \& {Brinchmann}}]{2015ApJ...809...96L}
{Licquia}, T.~C., {Newman}, J.~A., \& {Brinchmann}, J. 2015, \href{http://dx.doi.org/10.1088/0004-637X/809/1/96}{\color{magenta}\apj}, \href{https://ui.adsabs.harvard.edu/abs/2015ApJ...809...96L}{809, 96}

\bibitem[{{Loebman} {et~al.}(2016){Loebman}, {Debattista}, {Nidever}, {Hayden}, {Holtzman}, {Clarke}, {Ro{\v{s}}kar}, \& {Valluri}}]{2016ApJ...818L...6L}
{Loebman}, S.~R., {Debattista}, V.~P., {Nidever}, D.~L., {et~al.} 2016, \href{http://dx.doi.org/10.3847/2041-8205/818/1/L6}{\color{magenta}\apjl}, \href{https://ui.adsabs.harvard.edu/abs/2016ApJ...818L...6L}{818, L6}

\bibitem[{{Mainieri} {et~al.}(2024){Mainieri}, {Anderson}, {Brinchmann}, {Cimatti}, {Ellis}, {Hill}, {Kneib}, {McLeod}, {Opitom}, {Roth}, {Sanchez-Saez}, {Smiljanic}, {Tolstoy}, {Bacon}, {Randich}, {Adamo}, {Annibali}, {Arevalo}, {Audard}, {Barsanti}, {Battaglia}, {Bayo Aran}, {Belfiore}, {Bellazzini}, {Bellini}, {Beltran}, {Berni}, {Bianchi}, {Biazzo}, {Bisero}, {Bisogni}, {Bland-Hawthorn}, {Blondin}, {Bodensteiner}, {Boffin}, {Bonito}, {Bono}, {Bouche}, {Bowman}, {Braga}, {Bragaglia}, {Branchesi}, {Brucalassi}, {Bryant}, {Bryson}, {Busa}, {Camera}, {Carbone}, {Casali}, {Casali}, {Casasola}, {Castro}, {Catelan}, {Cavallo}, {Chiappini}, {Cioni}, {Colless}, {Colzi}, {Contarini}, {Couch}, {D'Ammando}, {d'Assignies D.}, {D'Orazi}, {da Silva}, {Dainotti}, {Damiani}, {Danielski}, {De Cia}, {de Jong}, {Dhawan}, {Dierickx}, {Driver}, {Dupletsa}, {Escoffier}, {Escorza}, {Fabrizio}, {Fiorentino}, {Fontana}, {Fontani}, {Forero Sanchez}, {Franois}, {Galindo-Guil}, {Gallazzi}, {Galli}, {Garcia}, {Garcia-Rojas},
  {Garilli}, {Grand}, {Guarcello}, {Hazra}, {Helmi}, {Herrero}, {Iglesias}, {Ilic}, {Irsic}, {Ivanov}, {Izzo}, {Jablonka}, {Joachimi}, {Kakkad}, {Kamann}, {Koposov}, {Kordopatis}, {Kovacevic}, {Kraljic}, {Kuncarayakti}, {Kwon}, {La Forgia}, {Lahav}, {Laigle}, {Lazzarin}, {Leaman}, {Leclercq}, {Lee}, {Lee}, {Lehnert}, {Lira}, {Loffredo}, {Lucatello}, {Magrini}, {Maguire}, {Mahler}, {Zahra Majidi}, {Malavasi}, {Mannucci}, {Marconi}, {Martin}, {Marulli}, {Massari}, {Matsuno}, {Mattheee}, {McGee}, {Merc}, {Merle}, {Miglio}, {Migliorini}, {Minchev}, {Minniti}, {Miret-Roig}, {Monreal Ibero}, {Montano}, {Montet}, {Moresco}, {Moretti}, {Moscardini}, {Moya}, {Mueller}, {Nanayakkara}, {Nicholl}, {Nordlander}, {Onori}, {Padovani}, {Pala}, {Panda}, {Pandey-Pommier}, {Pasquini}, {Pawlak}, {Pessi}, {Pisani}, {Popovic}, {Prisinzano}, {Raddi}, {Rainer}, {Rebassa-Mansergas}, {Richard}, {Rigault}, {Rocher}, {Romano}, {Rosati}, {Sacco}, {Sanchez-Janssen}, {Sander}, {Sanders}, {Sargent}, {Sarpa}, {Schimd}, {Schipani},
  {Sefusatti}, {Smith}, {Spina}, {Steinmetz}, {Tacchella}, {Tautvaisiene}, {Theissen}, {Thomas}, {Ting}, {Travouillon}, {Tresse}, {Trivedi}, {Tsantaki}, {Tsedrik}, {Urrutia}, {Valenti}, {Van der Swaelmen}, {Van Eck}, {Verdiani}, {Verdier}, {Vergani}, {Verhamme}, \& {Vernet}}]{2024arXiv240305398M}
{Mainieri}, V., {Anderson}, R.~I., {Brinchmann}, J., {et~al.} 2024, \href{https://ui.adsabs.harvard.edu/abs/2024arXiv240305398M}{\href{http://dx.doi.org/10.48550/arXiv.2403.05398}{\color{magenta}arXiv e-prints}, arXiv:2403.05398}

\bibitem[{{Mart{\'\i}n-Navarro} {et~al.}(2026){Mart{\'\i}n-Navarro}, {Vazdekis}, {Peralta de Arriba}, {Alonso Asensio}, {Angeloudi}, {Iglesias Navarro}, {La Barbera}, {Fahrion}, {Jerabkova}, {Beasley}, {Falc{\'o}n-Barroso}, {Huertas-Company}, {S{\'a}nchez}, \& {Jethwa}}]{2026arXiv260524093M}
{Mart{\'\i}n-Navarro}, I., {Vazdekis}, A., {Peralta de Arriba}, L., {et~al.} 2026, \href{https://ui.adsabs.harvard.edu/abs/2026arXiv260524093M}{\href{http://dx.doi.org/10.48550/arXiv.2605.24093}{\color{magenta}arXiv e-prints}, arXiv:2605.24093}

\bibitem[{{Mazzi} {et~al.}(2024){Mazzi}, {Girardi}, {Trabucchi}, {Dalcanton}, {Luger}, {Marigo}, {Miglio}, {Costa}, {Chen}, {Pastorelli}, {Fouesneau}, {Zaggia}, {Bressan}, \& {Dal Tio}}]{2024MNRAS.527..583M}
{Mazzi}, A., {Girardi}, L., {Trabucchi}, M., {et~al.} 2024, \href{http://dx.doi.org/10.1093/mnras/stad2952}{\color{magenta}\mnras}, \href{https://ui.adsabs.harvard.edu/abs/2024MNRAS.527..583M}{527, 583}

\bibitem[{{Minchev} {et~al.}(2012){Minchev}, {Famaey}, {Quillen}, {Di Matteo}, {Combes}, {Vlaji{\'c}}, {Erwin}, \& {Bland-Hawthorn}}]{2012A&A...548A.126M}
{Minchev}, I., {Famaey}, B., {Quillen}, A.~C., {et~al.} 2012, \href{http://dx.doi.org/10.1051/0004-6361/201219198}{\color{magenta}\aap}, \href{https://ui.adsabs.harvard.edu/abs/2012A&A...548A.126M}{548, A126}

\bibitem[{{Molaeinezhad} {et~al.}(2016){Molaeinezhad}, {Falc{\'o}n-Barroso}, {Mart{\'\i}nez-Valpuesta}, {Khosroshahi}, {Balcells}, \& {Peletier}}]{2016MNRAS.456..692M}
{Molaeinezhad}, A., {Falc{\'o}n-Barroso}, J., {Mart{\'\i}nez-Valpuesta}, I., {et~al.} 2016, \href{http://dx.doi.org/10.1093/mnras/stv2697}{\color{magenta}\mnras}, \href{https://ui.adsabs.harvard.edu/abs/2016MNRAS.456..692M}{456, 692}

\bibitem[{{Mor} {et~al.}(2019){Mor}, {Robin}, {Figueras}, {Roca-F{\`a}brega}, \& {Luri}}]{2019A&A...624L...1M}
{Mor}, R., {Robin}, A.~C., {Figueras}, F., {Roca-F{\`a}brega}, S., \& {Luri}, X. 2019, \href{http://dx.doi.org/10.1051/0004-6361/201935105}{\color{magenta}\aap}, \href{https://ui.adsabs.harvard.edu/abs/2019A&A...624L...1M}{624, L1}

\bibitem[{{Moreno} {et~al.}(2015){Moreno}, {Torrey}, {Ellison}, {Patton}, {Bluck}, {Bansal}, \& {Hernquist}}]{2015MNRAS.448.1107M}
{Moreno}, J., {Torrey}, P., {Ellison}, S.~L., {et~al.} 2015, \href{http://dx.doi.org/10.1093/mnras/stv094}{\color{magenta}\mnras}, \href{https://ui.adsabs.harvard.edu/abs/2015MNRAS.448.1107M}{448, 1107}

\bibitem[{{Neumann} {et~al.}(2020){Neumann}, {Fragkoudi}, {P{\'e}rez}, {Gadotti}, {Falc{\'o}n-Barroso}, {S{\'a}nchez-Bl{\'a}zquez}, {Bittner}, {Husemann}, {G{\'o}mez}, {Grand}, {Donohoe-Keyes}, {Kim}, {de Lorenzo-C{\'a}ceres}, {Martig}, {M{\'e}ndez-Abreu}, {Pakmor}, {Seidel}, \& {van de Ven}}]{2020A&A...637A..56N}
{Neumann}, J., {Fragkoudi}, F., {P{\'e}rez}, I., {et~al.} 2020, \href{http://dx.doi.org/10.1051/0004-6361/202037604}{\color{magenta}\aap}, \href{https://ui.adsabs.harvard.edu/abs/2020A&A...637A..56N}{637, A56}

\bibitem[{{Neumann} {et~al.}(2024){Neumann}, {Thomas}, {Maraston}, {Gleis}, {Mao}, {Schinnerer}, \& {Stuber}}]{2024MNRAS.534.2438N}
{Neumann}, J., {Thomas}, D., {Maraston}, C., {et~al.} 2024, \href{http://dx.doi.org/10.1093/mnras/stae2252}{\color{magenta}\mnras}, \href{https://ui.adsabs.harvard.edu/abs/2024MNRAS.534.2438N}{534, 2438}

\bibitem[{{Nissen} {et~al.}(2020){Nissen}, {Christensen-Dalsgaard}, {Mosumgaard}, {Silva Aguirre}, {Spitoni}, \& {Verma}}]{2020A&A...640A..81N}
{Nissen}, P.~E., {Christensen-Dalsgaard}, J., {Mosumgaard}, J.~R., {et~al.} 2020, \href{http://dx.doi.org/10.1051/0004-6361/202038300}{\color{magenta}\aap}, \href{https://ui.adsabs.harvard.edu/abs/2020A&A...640A..81N}{640, A81}

\bibitem[{{Ocvirk} {et~al.}(2006){Ocvirk}, {Pichon}, {Lan{\c{c}}on}, \& {Thi{\'e}baut}}]{2006MNRAS.365...46O}
{Ocvirk}, P., {Pichon}, C., {Lan{\c{c}}on}, A., \& {Thi{\'e}baut}, E. 2006, \href{http://dx.doi.org/10.1111/j.1365-2966.2005.09182.x}{\color{magenta}\mnras}, \href{https://ui.adsabs.harvard.edu/abs/2006MNRAS.365...46O}{365, 46}

\bibitem[{{Orkney} {et~al.}(2026){Orkney}, {Laporte}, {Grand}, \& {Springel}}]{2026MNRAS.545f1551O}
{Orkney}, M. D.~A., {Laporte}, C. F.~P., {Grand}, R. J.~J., \& {Springel}, V. 2026, \href{http://dx.doi.org/10.1093/mnras/staf1551}{\color{magenta}\mnras}, \href{https://ui.adsabs.harvard.edu/abs/2026MNRAS.545f1551O}{545, staf1551}

\bibitem[{{Parul} {et~al.}(2025){Parul}, {Bailin}, {Loebman}, {Wetzel}, {Barry}, \& {Bhattarai}}]{2025MNRAS.537.1571P}
{Parul}, H., {Bailin}, J., {Loebman}, S.~R., {et~al.} 2025, \href{http://dx.doi.org/10.1093/mnras/staf137}{\color{magenta}\mnras}, \href{https://ui.adsabs.harvard.edu/abs/2025MNRAS.537.1571P}{537, 1571}

\bibitem[{{Pernet} {et~al.}(2024){Pernet}, {Boecker}, \& {Mart{\'\i}n-Navarro}}]{2024A&A...687L..14P}
{Pernet}, E., {Boecker}, A., \& {Mart{\'\i}n-Navarro}, I. 2024, \href{http://dx.doi.org/10.1051/0004-6361/202449308}{\color{magenta}\aap}, \href{https://ui.adsabs.harvard.edu/abs/2024A&A...687L..14P}{687, L14}

\bibitem[{{Pettitt} {et~al.}(2017){Pettitt}, {Tasker}, {Wadsley}, {Keller}, \& {Benincasa}}]{2017MNRAS.468.4189P}
{Pettitt}, A.~R., {Tasker}, E.~J., {Wadsley}, J.~W., {Keller}, B.~W., \& {Benincasa}, S.~M. 2017, \href{http://dx.doi.org/10.1093/mnras/stx736}{\color{magenta}\mnras}, \href{https://ui.adsabs.harvard.edu/abs/2017MNRAS.468.4189P}{468, 4189}

\bibitem[{{Pilyugin} {et~al.}(2025){Pilyugin}, {Lara-L{\'o}pez}, {Tautvai{\v{s}}ien{\.{e}}}, {Zinchenko}, {Gardu{\~n}o}, {De Rossi}, {Zaragoza-Cardiel}, {Dib}, \& {Val{\'e}}}]{2025A&A...694A.113P}
{Pilyugin}, L.~S., {Lara-L{\'o}pez}, M.~A., {Tautvai{\v{s}}ien{\.{e}}}, G., {et~al.} 2025, \href{http://dx.doi.org/10.1051/0004-6361/202452605}{\color{magenta}\aap}, \href{https://ui.adsabs.harvard.edu/abs/2025A&A...694A.113P}{694, A113}

\bibitem[{{Pilyugin} {et~al.}(2023){Pilyugin}, {Tautvai{\v{s}}ien{\.{e}}}, \& {Lara-L{\'o}pez}}]{2023A&A...676A..57P}
{Pilyugin}, L.~S., {Tautvai{\v{s}}ien{\.{e}}}, G., \& {Lara-L{\'o}pez}, M.~A. 2023, \href{http://dx.doi.org/10.1051/0004-6361/202346503}{\color{magenta}\aap}, \href{https://ui.adsabs.harvard.edu/abs/2023A&A...676A..57P}{676, A57}

\bibitem[{{Pinna} {et~al.}(2019{\natexlab{a}}){Pinna}, {Falc{\'o}n-Barroso}, {Martig}, {Coccato}, {Corsini}, {de Zeeuw}, {Gadotti}, {Iodice}, {Leaman}, {Lyubenova}, {Mart{\'\i}n-Navarro}, {Morelli}, {Sarzi}, {van de Ven}, {Viaene}, \& {McDermid}}]{2019A&A...625A..95P}
{Pinna}, F., {Falc{\'o}n-Barroso}, J., {Martig}, M., {et~al.} 2019{\natexlab{a}}, \href{http://dx.doi.org/10.1051/0004-6361/201935154}{\color{magenta}\aap}, \href{https://ui.adsabs.harvard.edu/abs/2019A&A...625A..95P}{625, A95}

\bibitem[{{Pinna} {et~al.}(2019{\natexlab{b}}){Pinna}, {Falc{\'o}n-Barroso}, {Martig}, {Sarzi}, {Coccato}, {Iodice}, {Corsini}, {de Zeeuw}, {Gadotti}, {Leaman}, {Lyubenova}, {McDermid}, {Minchev}, {Morelli}, {van de Ven}, \& {Viaene}}]{2019A&A...623A..19P}
{Pinna}, F., {Falc{\'o}n-Barroso}, J., {Martig}, M., {et~al.} 2019{\natexlab{b}}, \href{http://dx.doi.org/10.1051/0004-6361/201833193}{\color{magenta}\aap}, \href{https://ui.adsabs.harvard.edu/abs/2019A&A...623A..19P}{623, A19}

\bibitem[{{Ploeckinger} \& {Schaye}(2020)}]{2020MNRAS.497.4857P}
{Ploeckinger}, S. \& {Schaye}, J. 2020, \href{http://dx.doi.org/10.1093/mnras/staa2172}{\color{magenta}\mnras}, \href{https://ui.adsabs.harvard.edu/abs/2020MNRAS.497.4857P}{497, 4857}

\bibitem[{{Portail} {et~al.}(2017){Portail}, {Gerhard}, {Wegg}, \& {Ness}}]{2017MNRAS.465.1621P}
{Portail}, M., {Gerhard}, O., {Wegg}, C., \& {Ness}, M. 2017, \href{http://dx.doi.org/10.1093/mnras/stw2819}{\color{magenta}\mnras}, \href{https://ui.adsabs.harvard.edu/abs/2017MNRAS.465.1621P}{465, 1621}

\bibitem[{{Ratcliffe} {et~al.}(2026){Ratcliffe}, {Khoperskov}, {Lee}, {Minchev}, {Di Matteo}, {van de Ven}, {Haywood}, {Marques}, {Bernaldez}, {Krajnovi{\'c}}, \& {Steinmetz}}]{2026A&A...706A.103R}
{Ratcliffe}, B., {Khoperskov}, S., {Lee}, N., {et~al.} 2026, \href{http://dx.doi.org/10.1051/0004-6361/202557057}{\color{magenta}\aap}, \href{https://ui.adsabs.harvard.edu/abs/2026A&A...706A.103R}{706, A103}

\bibitem[{{Reiter} {et~al.}(2025){Reiter}, {Jethwa}, {van de Ven}, {Thater}, \& {Leaman}}]{2025A&A...701A..12R}
{Reiter}, S., {Jethwa}, P., {van de Ven}, G., {Thater}, S., \& {Leaman}, R. 2025, \href{http://dx.doi.org/10.1051/0004-6361/202554394}{\color{magenta}\aap}, \href{https://ui.adsabs.harvard.edu/abs/2025A&A...701A..12R}{701, A12}

\bibitem[{{Renaud} {et~al.}(2019){Renaud}, {Bournaud}, {Agertz}, {Kraljic}, {Schinnerer}, {Bolatto}, {Daddi}, \& {Hughes}}]{2019A&A...625A..65R}
{Renaud}, F., {Bournaud}, F., {Agertz}, O., {et~al.} 2019, \href{http://dx.doi.org/10.1051/0004-6361/201935222}{\color{magenta}\aap}, \href{https://ui.adsabs.harvard.edu/abs/2019A&A...625A..65R}{625, A65}

\bibitem[{{Richard} {et~al.}(2019){Richard}, {Bacon}, {Blaizot}, {Boissier}, {Boselli}, {NicolasBouch{\'e}}, {Brinchmann}, {Castro}, {Ciesla}, {Crowther}, {Daddi}, {Dreizler}, {Duc}, {Elbaz}, {Epinat}, {Evans}, {Fossati}, {Fumagalli}, {Garcia}, {Garel}, {Hayes}, {Adamo}, {Herrero}, {Hugot}, {Humphrey}, {Jablonka}, {Kamann}, {Kaper}, {Kelz}, {Kneib}, {de Koter}, {Krajnovi{\'c}}, {Kudritzki}, {Langer}, {Lardo}, {Leclercq}, {Lennon}, {Mahler}, {Martins}, {Massey}, {Mitchell}, {Monreal-Ibero}, {Najarro}, {Opitom}, {Papaderos}, {P{\'e}roux}, {Revaz}, {Roth}, {Rousselot}, {Sander}, {Simmonds Wagemann}, {Smail}, {Swinbank}, {Tramper}, {Urrutia}, {Verhamme}, {Vink}, {Walsh}, {Weilbacher}, {Wendt}, {Wisotzki}, \& {Yang}}]{2019arXiv190601657R}
{Richard}, J., {Bacon}, R., {Blaizot}, J., {et~al.} 2019, \href{https://ui.adsabs.harvard.edu/abs/2019arXiv190601657R}{\href{http://dx.doi.org/10.48550/arXiv.1906.01657}{\color{magenta}arXiv e-prints}, arXiv:1906.01657}

\bibitem[{{Rix} {et~al.}(2024){Rix}, {Chandra}, {Zasowski}, {Pillepich}, {Khoperskov}, {Feltzing}, {Wyse}, {Frankel}, {Horta}, {Kollmeier}, {Stassun}, {Ness}, {Bird}, {Nidever}, {Fern{\'a}ndez-Trincado}, {Amarante}, {Laporte}, \& {Lian}}]{2024ApJ...975..293R}
{Rix}, H.-W., {Chandra}, V., {Zasowski}, G., {et~al.} 2024, \href{http://dx.doi.org/10.3847/1538-4357/ad7aee}{\color{magenta}\apj}, \href{https://ui.adsabs.harvard.edu/abs/2024ApJ...975..293R}{975, 293}

\bibitem[{{Ro{\v{s}}kar} {et~al.}(2013){Ro{\v{s}}kar}, {Debattista}, \& {Loebman}}]{2013MNRAS.433..976R}
{Ro{\v{s}}kar}, R., {Debattista}, V.~P., \& {Loebman}, S.~R. 2013, \href{http://dx.doi.org/10.1093/mnras/stt788}{\color{magenta}\mnras}, \href{https://ui.adsabs.harvard.edu/abs/2013MNRAS.433..976R}{433, 976}

\bibitem[{{Ro{\v{s}}kar} {et~al.}(2012){Ro{\v{s}}kar}, {Debattista}, {Quinn}, \& {Wadsley}}]{2012MNRAS.426.2089R}
{Ro{\v{s}}kar}, R., {Debattista}, V.~P., {Quinn}, T.~R., \& {Wadsley}, J. 2012, \href{http://dx.doi.org/10.1111/j.1365-2966.2012.21860.x}{\color{magenta}\mnras}, \href{https://ui.adsabs.harvard.edu/abs/2012MNRAS.426.2089R}{426, 2089}

\bibitem[{{Ruiz-Lara} {et~al.}(2020){Ruiz-Lara}, {Gallart}, {Bernard}, \& {Cassisi}}]{2020NatAs...4..965R}
{Ruiz-Lara}, T., {Gallart}, C., {Bernard}, E.~J., \& {Cassisi}, S. 2020, \href{http://dx.doi.org/10.1038/s41550-020-1097-0}{\color{magenta}Nature Astronomy}, \href{https://ui.adsabs.harvard.edu/abs/2020NatAs...4..965R}{4, 965}

\bibitem[{{Ruiz-Lara} {et~al.}(2017){Ruiz-Lara}, {P{\'e}rez}, {Florido}, {S{\'a}nchez-Bl{\'a}zquez}, {M{\'e}ndez-Abreu}, {S{\'a}nchez-Menguiano}, {S{\'a}nchez}, {Lyubenova}, {Falc{\'o}n-Barroso}, {van de Ven}, {Marino}, {de Lorenzo-C{\'a}ceres}, {Catal{\'a}n-Torrecilla}, {Costantin}, {Bland-Hawthorn}, {Galbany}, {Garc{\'\i}a-Benito}, {Husemann}, {Kehrig}, {M{\'a}rquez}, {Mast}, {Walcher}, {Zibetti}, {Ziegler}, \& {CALIFA Team}}]{2017A&A...604A...4R}
{Ruiz-Lara}, T., {P{\'e}rez}, I., {Florido}, E., {et~al.} 2017, \href{http://dx.doi.org/10.1051/0004-6361/201730705}{\color{magenta}\aap}, \href{https://ui.adsabs.harvard.edu/abs/2017A&A...604A...4R}{604, A4}

\bibitem[{{S{\'a}nchez} {et~al.}(2019){S{\'a}nchez}, {Avila-Reese}, {Rodr{\'\i}guez-Puebla}, {Ibarra-Medel}, {Calette}, {Bershady}, {Hern{\'a}ndez-Toledo}, {Pan}, \& {Bizyaev}}]{2019MNRAS.482.1557S}
{S{\'a}nchez}, S.~F., {Avila-Reese}, V., {Rodr{\'\i}guez-Puebla}, A., {et~al.} 2019, \href{http://dx.doi.org/10.1093/mnras/sty2730}{\color{magenta}\mnras}, \href{https://ui.adsabs.harvard.edu/abs/2019MNRAS.482.1557S}{482, 1557}

\bibitem[{{S{\'a}nchez} {et~al.}(2012){S{\'a}nchez}, {Kennicutt}, {Gil de Paz}, {van de Ven}, {V{\'\i}lchez}, {Wisotzki}, {Walcher}, {Mast}, {Aguerri}, {Albiol-P{\'e}rez}, {Alonso-Herrero}, {Alves}, {Bakos}, {Bart{\'a}kov{\'a}}, {Bland-Hawthorn}, {Boselli}, {Bomans}, {Castillo-Morales}, {Cortijo-Ferrero}, {de Lorenzo-C{\'a}ceres}, {Del Olmo}, {Dettmar}, {D{\'\i}az}, {Ellis}, {Falc{\'o}n-Barroso}, {Flores}, {Gallazzi}, {Garc{\'\i}a-Lorenzo}, {Gonz{\'a}lez Delgado}, {Gruel}, {Haines}, {Hao}, {Husemann}, {Igl{\'e}sias-P{\'a}ramo}, {Jahnke}, {Johnson}, {Jungwiert}, {Kalinova}, {Kehrig}, {Kupko}, {L{\'o}pez-S{\'a}nchez}, {Lyubenova}, {Marino}, {M{\'a}rmol-Queralt{\'o}}, {M{\'a}rquez}, {Masegosa}, {Meidt}, {Mendez-Abreu}, {Monreal-Ibero}, {Montijo}, {Mour{\~a}o}, {Palacios-Navarro}, {Papaderos}, {Pasquali}, {Peletier}, {P{\'e}rez}, {P{\'e}rez}, {Quirrenbach}, {Rela{\~n}o}, {Rosales-Ortega}, {Roth}, {Ruiz-Lara}, {S{\'a}nchez-Bl{\'a}zquez}, {Sengupta}, {Singh}, {Stanishev}, {Trager}, {Vazdekis}, {Viironen}, {Wild},
  {Zibetti}, \& {Ziegler}}]{2012A&A...538A...8S}
{S{\'a}nchez}, S.~F., {Kennicutt}, R.~C., {Gil de Paz}, A., {et~al.} 2012, \href{http://dx.doi.org/10.1051/0004-6361/201117353}{\color{magenta}\aap}, \href{https://ui.adsabs.harvard.edu/abs/2012A&A...538A...8S}{538, A8}

\bibitem[{{Schaller} {et~al.}(2024){Schaller}, {Borrow}, {Draper}, {Ivkovic}, {McAlpine}, {Vandenbroucke}, {Bah{\'e}}, {Chaikin}, {Chalk}, {Chan}, {Correa}, {van Daalen}, {Elbers}, {Gonnet}, {Hausammann}, {Helly}, {Hu{\v{s}}ko}, {Kegerreis}, {Nobels}, {Ploeckinger}, {Revaz}, {Roper}, {Ruiz-Bonilla}, {Sandnes}, {Uyttenhove}, {Willis}, \& {Xiang}}]{2024MNRAS.530.2378S}
{Schaller}, M., {Borrow}, J., {Draper}, P.~W., {et~al.} 2024, \href{http://dx.doi.org/10.1093/mnras/stae922}{\color{magenta}\mnras}, \href{https://ui.adsabs.harvard.edu/abs/2024MNRAS.530.2378S}{530, 2378}

\bibitem[{{Sch{\"o}nrich} \& {Binney}(2009)}]{2009MNRAS.396..203S}
{Sch{\"o}nrich}, R. \& {Binney}, J. 2009, \href{http://dx.doi.org/10.1111/j.1365-2966.2009.14750.x}{\color{magenta}\mnras}, \href{https://ui.adsabs.harvard.edu/abs/2009MNRAS.396..203S}{396, 203}

\bibitem[{{Scott} {et~al.}(2021){Scott}, {van de Sande}, {Sharma}, {Bland-Hawthorn}, {Freeman}, {Gerhard}, {Hayden}, \& {McDermid}}]{2021ApJ...913L..11S}
{Scott}, N., {van de Sande}, J., {Sharma}, S., {et~al.} 2021, \href{http://dx.doi.org/10.3847/2041-8213/abfc57}{\color{magenta}\apjl}, \href{https://ui.adsabs.harvard.edu/abs/2021ApJ...913L..11S}{913, L11}

\bibitem[{{SDSS Collaboration} {et~al.}(2025){SDSS Collaboration}, {Adamane Pallathadka}, {Aghakhanloo}, {Aird}, {Almeida}, {Amrita}, {Anders}, {Anderson}, {Arseneau}, {Gonz{\'a}lez Avila}, {Aviram}, {Aydar}, {Badenes}, {Barrera-Ballesteros}, {Bauer}, {Behmard}, {Berg}, {Besser}, {Moni Bidin}, {Bizyaev}, {Blanc}, {Blanton}, {Bovy}, {Brandt}, {Brownstein}, {Buchner}, {Bulbul}, {Burchett}, {Carigi}, {Carlberg}, {Casey}, {Chakraborty}, {Chanam{\'e}}, {Chandra}, {Chiappini}, {Chilingarian}, {Comparat}, {Covey}, {Crumpler}, {Cunha}, {D'Onghia}, {Dai}, {Darling}, {Davis}, {De Lee}, {Deacon}, {M{\'e}ndez Delgado}, {Demasi}, {Demianenko}, {Demke}, {Donor}, {Drory}, {Villa Durango}, {Dwelly}, {Egorov}, {Egorova}, {El-Badry}, {Eracleous}, {Fan}, {Farr}, {Finkbeiner}, {Fries}, {Frinchaboy}, {Gentile Fusillo}, {Serrano F{\'e}lix}, {Gaensicke}, {Galligan}, {Garc{\'\i}a}, {Gelfand}, {Grabowski}, {Grebel}, {Green}, {Greve}, {Grier}, {Griffith}, {Guetzoyan}, {Gupta}, {Hackshaw}, {Hall}, {Hawkins}, {Heged{\H{u}}s}, {Hekker},
  {Herbst}, {Hermes}, {Hern{\'a}ndez-Garc{\'\i}a}, {Hiremath}, {Hogg}, {Holtzman}, {Horne}, {Horta}, {Huang}, {Hutchinson}, {H{\"a}berle}, {Ibarra-Medel}, {Ji}, {Jofre}, {Johnson}, {Johnson}, {Johnston}, {Kaldor}, {Katkov}, {Khalatyan}, {Khoperskov}, {Klessen}, {Kluge}, {Koekemoer}, {Kollmeier}, {Kounkel}, {Kreckel}, {Krishnarao}, {Krumpe}, {Lacerna}, {Laporte}, {Lepine}, {Li}, {Liang}, {Limberg}, {Liu}, {Loebman}, {Long}, {Lu}, {Lucey}, {Lugo-Aranda}, {Mart{\'\i}nez Martinez-Aldama}, {McKinnon}, {Medan}, {Merloni}, {Morrison}, {Myers}, {M{\'e}sz{\'a}ros}, {M{\"u}ller-Horn}, {Nepal}, {Ness}, {Nidever}, {Nitschelm}, {Oravetz}, {Otto}, {Pan}, {P{\'e}rez Paolino}, {Negrete Pe{\~n}aloza}, {Pinsonneault}, {Taghizadeh Popp}, {Price-Whelan}, {Pulatova}, {Queiroz}, {Raddick}, {Rankine}, {Rix}, {Rom{\'a}n-Z{\'u}{\~n}iga}, {Fern{\'a}ndez Rosso}, {Runnoe}, {Mahmud Saad}, {Salvato}, {Sanchez}, {Sattler}, {Saydjari}, {Sayres}, {Schlaufman}, {Schneider}, {Schwope}, {Seaton}, {Seeburger}, {Serna}, {Sharma}, {Shen}, {Sinha},
  {Sizemore}, {Sniegowska}, {Song}, {Souto}, {Stassun}, {Steinmetz}, {Stone}, {Stone-Martinez}, {Stringfellow}, {Mata S{\'a}nchez}, {S{\'a}nchez-Gallego}, {Tan}, {Tayar}, {Thai}, {Thakar}, {Thibodeaux}, {Ting}, {Tkachenko}, {Trakhtenbrot}, {Fernandez Trincado}, {Troup}, {Trump}, {Ulloa}, {Van der Marel}, {Vera}, {Villanova}, {Villase{\~n}or}, {Wang}, {Way}, {Weijmans}, {Wheeler}, {Wilson}, {Wofford}, \& {Wong}}]{2025arXiv250707093S}
{SDSS Collaboration}, {Adamane Pallathadka}, G., {Aghakhanloo}, M., {et~al.} 2025, \href{https://ui.adsabs.harvard.edu/abs/2025arXiv250707093S}{\href{http://dx.doi.org/10.48550/arXiv.2507.07093}{\color{magenta}arXiv e-prints}, arXiv:2507.07093}

\bibitem[{{Semczuk} {et~al.}(2025){Semczuk}, {Antoja}, {Gir{\'o}n-Soto}, \& {Laporte}}]{2025arXiv250800690S}
{Semczuk}, M., {Antoja}, T., {Gir{\'o}n-Soto}, A., \& {Laporte}, C. F.~P. 2025, \href{https://ui.adsabs.harvard.edu/abs/2025arXiv250800690S}{\href{http://dx.doi.org/10.48550/arXiv.2508.00690}{\color{magenta}arXiv e-prints}, arXiv:2508.00690}

\bibitem[{{Sharma} {et~al.}(2021){Sharma}, {Hayden}, \& {Bland-Hawthorn}}]{2021MNRAS.507.5882S}
{Sharma}, S., {Hayden}, M.~R., \& {Bland-Hawthorn}, J. 2021, \href{http://dx.doi.org/10.1093/mnras/stab2015}{\color{magenta}\mnras}, \href{https://ui.adsabs.harvard.edu/abs/2021MNRAS.507.5882S}{507, 5882}

\bibitem[{{Sormani} {et~al.}(2022){Sormani}, {Gerhard}, {Portail}, {Vasiliev}, \& {Clarke}}]{2022MNRAS.514L...1S}
{Sormani}, M.~C., {Gerhard}, O., {Portail}, M., {Vasiliev}, E., \& {Clarke}, J. 2022, \href{http://dx.doi.org/10.1093/mnrasl/slac046}{\color{magenta}\mnras}, \href{https://ui.adsabs.harvard.edu/abs/2022MNRAS.514L...1S}{514, L1}

\bibitem[{{Steinmetz} {et~al.}(2020){Steinmetz}, {Matijevi{\v{c}}}, {Enke}, {Zwitter}, {Guiglion}, {McMillan}, {Kordopatis}, {Valentini}, {Chiappini}, {Casagrande}, {Wojno}, {Anguiano}, {Bienaym{\'e}}, {Bijaoui}, {Binney}, {Burton}, {Cass}, {de Laverny}, {Fiegert}, {Freeman}, {Fulbright}, {Gibson}, {Gilmore}, {Grebel}, {Helmi}, {Kunder}, {Munari}, {Navarro}, {Parker}, {Ruchti}, {Recio-Blanco}, {Reid}, {Seabroke}, {Siviero}, {Siebert}, {Stupar}, {Watson}, {Williams}, {Wyse}, {Anders}, {Antoja}, {Birko}, {Bland-Hawthorn}, {Bossini}, {Garc{\'\i}a}, {Carrillo}, {Chaplin}, {Elsworth}, {Famaey}, {Gerhard}, {Jofre}, {Just}, {Mathur}, {Miglio}, {Minchev}, {Monari}, {Mosser}, {Ritter}, {Rodrigues}, {Scholz}, {Sharma}, {Sysoliatina}, \& {RAVE Collaboration}}]{2020AJ....160...82S}
{Steinmetz}, M., {Matijevi{\v{c}}}, G., {Enke}, H., {et~al.} 2020, \href{http://dx.doi.org/10.3847/1538-3881/ab9ab9}{\color{magenta}\aj}, \href{https://ui.adsabs.harvard.edu/abs/2020AJ....160...82S}{160, 82}

\bibitem[{{Stone-Martinez} {et~al.}(2024){Stone-Martinez}, {Holtzman}, {Imig}, {Nitschelm}, {Stassun}, \& {Brownstein}}]{2024AJ....167...73S}
{Stone-Martinez}, A., {Holtzman}, J.~A., {Imig}, J., {et~al.} 2024, \href{http://dx.doi.org/10.3847/1538-3881/ad12a6}{\color{magenta}\aj}, \href{https://ui.adsabs.harvard.edu/abs/2024AJ....167...73S}{167, 73}

\bibitem[{{Sysoliatina} \& {Just}(2022)}]{2022A&A...666A.130S}
{Sysoliatina}, K. \& {Just}, A. 2022, \href{http://dx.doi.org/10.1051/0004-6361/202243780}{\color{magenta}\aap}, \href{https://ui.adsabs.harvard.edu/abs/2022A&A...666A.130S}{666, A130}

\bibitem[{{Tuntipong} {et~al.}(2024){Tuntipong}, {van de Sande}, {Croom}, {Barsanti}, {Bland-Hawthorn}, {Brough}, {Bryant}, {Casura}, {Fraser-McKelvie}, {Lawrence}, {Ristea}, {Sweet}, \& {Zafar}}]{2024MNRAS.533.4334T}
{Tuntipong}, S., {van de Sande}, J., {Croom}, S.~M., {et~al.} 2024, \href{http://dx.doi.org/10.1093/mnras/stae2042}{\color{magenta}\mnras}, \href{https://ui.adsabs.harvard.edu/abs/2024MNRAS.533.4334T}{533, 4334}

\bibitem[{{van de Sande} {et~al.}(2024){van de Sande}, {Fraser-McKelvie}, {Fisher}, {Martig}, {Hayden}, \& {Geckos Survey Collaboration}}]{2024IAUS..377...27V}
{van de Sande}, J., {Fraser-McKelvie}, A., {Fisher}, D.~B., {et~al.} 2024, in IAU Symposium, Vol. 377, Early Disk-Galaxy Formation from JWST to the Milky Way, ed. {Tabatabaei}, F., {Barbuy}, B., \& {Ting}, Y.-S., \href{https://ui.adsabs.harvard.edu/abs/2024IAUS..377...27V}{27--33}

\bibitem[{{van de Sande} \& {Scott}(2021)}]{2021NatAs...5..879V}
{van de Sande}, J. \& {Scott}, N. 2021, \href{http://dx.doi.org/10.1038/s41550-021-01433-w}{\color{magenta}Nature Astronomy}, \href{https://ui.adsabs.harvard.edu/abs/2021NatAs...5..879V}{5, 879}

\bibitem[{{Vasiliev}(2019)}]{2019MNRAS.482.1525V}
{Vasiliev}, E. 2019, \href{http://dx.doi.org/10.1093/mnras/sty2672}{\color{magenta}\mnras}, \href{https://ui.adsabs.harvard.edu/abs/2019MNRAS.482.1525V}{482, 1525}

\bibitem[{{Vasiliev} \& {Valluri}(2020)}]{2020ApJ...889...39V}
{Vasiliev}, E. \& {Valluri}, M. 2020, \href{http://dx.doi.org/10.3847/1538-4357/ab5fe0}{\color{magenta}\apj}, \href{https://ui.adsabs.harvard.edu/abs/2020ApJ...889...39V}{889, 39}

\bibitem[{{Vazdekis} {et~al.}(2010){Vazdekis}, {S{\'a}nchez-Bl{\'a}zquez}, {Falc{\'o}n-Barroso}, {Cenarro}, {Beasley}, {Cardiel}, {Gorgas}, \& {Peletier}}]{2010MNRAS.404.1639V}
{Vazdekis}, A., {S{\'a}nchez-Bl{\'a}zquez}, P., {Falc{\'o}n-Barroso}, J., {et~al.} 2010, \href{http://dx.doi.org/10.1111/j.1365-2966.2010.16407.x}{\color{magenta}\mnras}, \href{https://ui.adsabs.harvard.edu/abs/2010MNRAS.404.1639V}{404, 1639}

\bibitem[{{Verro} {et~al.}(2022){Verro}, {Trager}, {Peletier}, {Lan{\c{c}}on}, {Gonneau}, {Vazdekis}, {Prugniel}, {Chen}, {Coelho}, {S{\'a}nchez-Bl{\'a}zquez}, {Martins}, {Arentsen}, {Lyubenova}, {Falc{\'o}n-Barroso}, \& {Dries}}]{2022A&A...660A..34V}
{Verro}, K., {Trager}, S.~C., {Peletier}, R.~F., {et~al.} 2022, \href{http://dx.doi.org/10.1051/0004-6361/202142388}{\color{magenta}\aap}, \href{https://ui.adsabs.harvard.edu/abs/2022A&A...660A..34V}{660, A34}

\bibitem[{{Wang} {et~al.}(2024){Wang}, {Sharma}, {Hayden}, {van de Sande}, {Bland-Hawthorn}, {Vaughan}, {Martig}, \& {Pinna}}]{2024MNRAS.534.1175W}
{Wang}, Z., {Sharma}, S., {Hayden}, M.~R., {et~al.} 2024, \href{http://dx.doi.org/10.1093/mnras/stae2148}{\color{magenta}\mnras}, \href{https://ui.adsabs.harvard.edu/abs/2024MNRAS.534.1175W}{534, 1175}

\bibitem[{{Wegg} \& {Gerhard}(2013)}]{2013MNRAS.435.1874W}
{Wegg}, C. \& {Gerhard}, O. 2013, \href{http://dx.doi.org/10.1093/mnras/stt1376}{\color{magenta}\mnras}, \href{https://ui.adsabs.harvard.edu/abs/2013MNRAS.435.1874W}{435, 1874}

\bibitem[{{Wegg} {et~al.}(2015){Wegg}, {Gerhard}, \& {Portail}}]{2015MNRAS.450.4050W}
{Wegg}, C., {Gerhard}, O., \& {Portail}, M. 2015, \href{http://dx.doi.org/10.1093/mnras/stv745}{\color{magenta}\mnras}, \href{https://ui.adsabs.harvard.edu/abs/2015MNRAS.450.4050W}{450, 4050}

\bibitem[{{Williams} {et~al.}(2024){Williams}, {Lee}, {Larson}, {Leroy}, {Sandstrom}, {Schinnerer}, {Thilker}, {Belfiore}, {Egorov}, {Rosolowsky}, {Sutter}, {DePasquale}, {Pagan}, {Berger}, {Anand}, {Barnes}, {Bigiel}, {Boquien}, {Cao}, {Chastenet}, {Chevance}, {Chown}, {Dale}, {Deger}, {Eibensteiner}, {Emsellem}, {Faesi}, {Glover}, {Grasha}, {Hannon}, {Hassani}, {Henshaw}, {Jim{\'e}nez-Donaire}, {Kim}, {Klessen}, {Koch}, {Li}, {Liu}, {Meidt}, {M{\'e}ndez-Delgado}, {Murphy}, {Neumann}, {Neumann}, {Neumayer}, {Oakes}, {Pathak}, {Pety}, {Pinna}, {Querejeta}, {Ramambason}, {Romanelli}, {Sormani}, {Stuber}, {Sun}, {Teng}, {Usero}, {Watkins}, \& {Weinbeck}}]{2024ApJS..273...13W}
{Williams}, T.~G., {Lee}, J.~C., {Larson}, K.~L., {et~al.} 2024, \href{http://dx.doi.org/10.3847/1538-4365/ad4be5}{\color{magenta}\apjs}, \href{https://ui.adsabs.harvard.edu/abs/2024ApJS..273...13W}{273, 13}

\bibitem[{{Xiang} \& {Rix}(2022)}]{2022Natur.603..599X}
{Xiang}, M. \& {Rix}, H.-W. 2022, \href{http://dx.doi.org/10.1038/s41586-022-04496-5}{\color{magenta}\nat}, \href{https://ui.adsabs.harvard.edu/abs/2022Natur.603..599X}{603, 599}

\bibitem[{{Zakharova} {et~al.}(2024){Zakharova}, {Tikhonenko}, {Sotnikova}, \& {Smirnov}}]{2024MNRAS.527.3038Z}
{Zakharova}, D., {Tikhonenko}, I.~S., {Sotnikova}, N.~Y., \& {Smirnov}, A.~A. 2024, \href{http://dx.doi.org/10.1093/mnras/stad3468}{\color{magenta}\mnras}, \href{https://ui.adsabs.harvard.edu/abs/2024MNRAS.527.3038Z}{527, 3038}

\bibitem[{{Zasowski} {et~al.}(2025){Zasowski}, {Imig}, \& {Coluccio}}]{2025ApJ...991...36Z}
{Zasowski}, G., {Imig}, J., \& {Coluccio}, H. 2025, \href{http://dx.doi.org/10.3847/1538-4357/adef53}{\color{magenta}\apj}, \href{https://ui.adsabs.harvard.edu/abs/2025ApJ...991...36Z}{991, 36}

\bibitem[{{Zhao} {et~al.}(2012){Zhao}, {Zhao}, {Chu}, {Jing}, \& {Deng}}]{2012RAA....12..723Z}
{Zhao}, G., {Zhao}, Y.-H., {Chu}, Y.-Q., {Jing}, Y.-P., \& {Deng}, L.-C. 2012, \href{http://dx.doi.org/10.1088/1674-4527/12/7/002}{\color{magenta}Research in Astronomy and Astrophysics}, \href{https://ui.adsabs.harvard.edu/abs/2012RAA....12..723Z}{12, 723}

\bibitem[{{Zhou} {et~al.}(2023){Zhou}, {Arag{\'o}n-Salamanca}, {Merrifield}, {Andrews}, {Drory}, \& {Lane}}]{2023MNRAS.521.5810Z}
{Zhou}, S., {Arag{\'o}n-Salamanca}, A., {Merrifield}, M., {et~al.} 2023, \href{http://dx.doi.org/10.1093/mnras/stad853}{\color{magenta}\mnras}, \href{https://ui.adsabs.harvard.edu/abs/2023MNRAS.521.5810Z}{521, 5810}

\bibitem[{{Zibetti} {et~al.}(2024){Zibetti}, {Rossi}, \& {Gallazzi}}]{2024MNRAS.528.2790Z}
{Zibetti}, S., {Rossi}, E., \& {Gallazzi}, A.~R. 2024, \href{http://dx.doi.org/10.1093/mnras/stae178}{\color{magenta}\mnras}, \href{https://ui.adsabs.harvard.edu/abs/2024MNRAS.528.2790Z}{528, 2790}

\end{thebibliography}

\begin{appendix}\label{sec4::appendix}

\section{Simulation of the MW}\label{sec4::simulations_appendix}
\begin{figure*}
\centering  
\adjincludegraphics[height=0.3\textheight, clip, trim=0 0 {0.14\textwidth} 0]{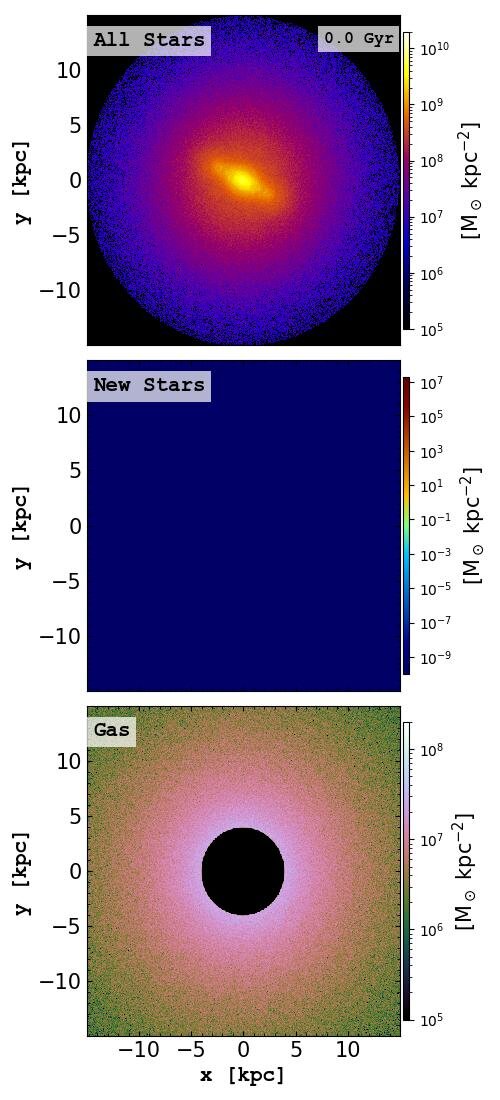} 
\adjincludegraphics[height=0.3\textheight, clip, trim={0.12\textwidth} 0 {0.14\textwidth} 0]{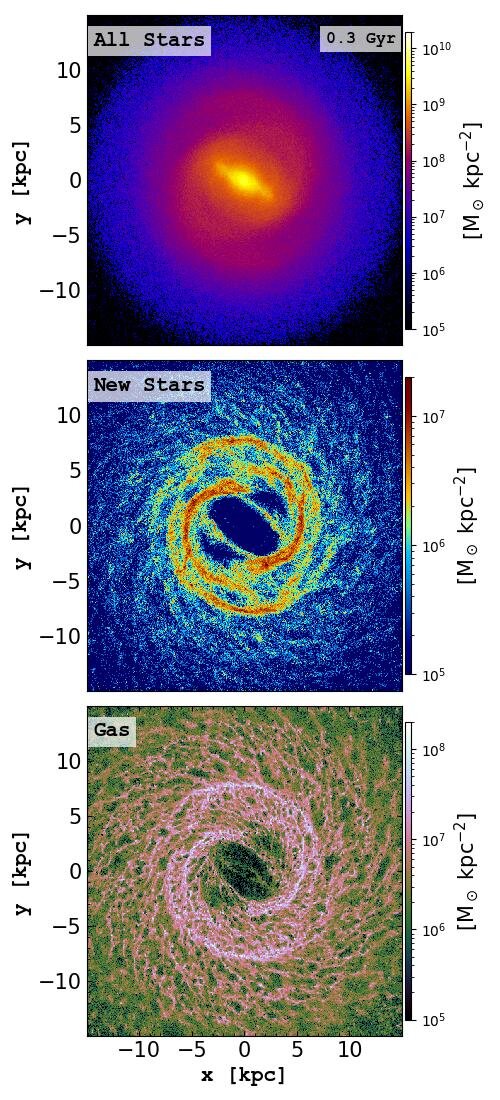}
\adjincludegraphics[height=0.3\textheight, clip, trim={0.12\textwidth} 0 {0.14\textwidth} 0]{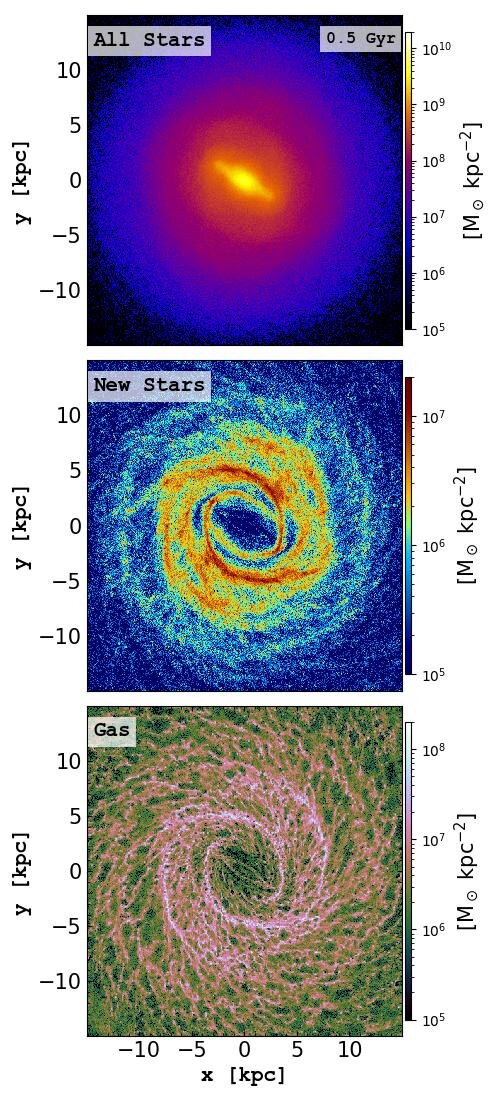}
\adjincludegraphics[height=0.3\textheight, clip, trim={0.12\textwidth} 0 {0.14\textwidth} 0]{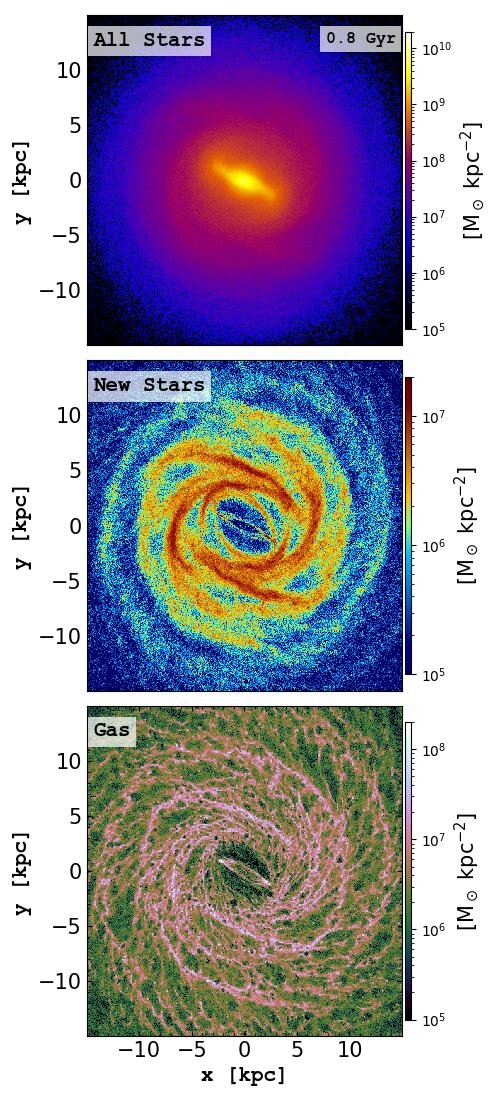}
\adjincludegraphics[height=0.3\textheight, clip, trim={0.12\textwidth} 0 {0.14\textwidth} 0]{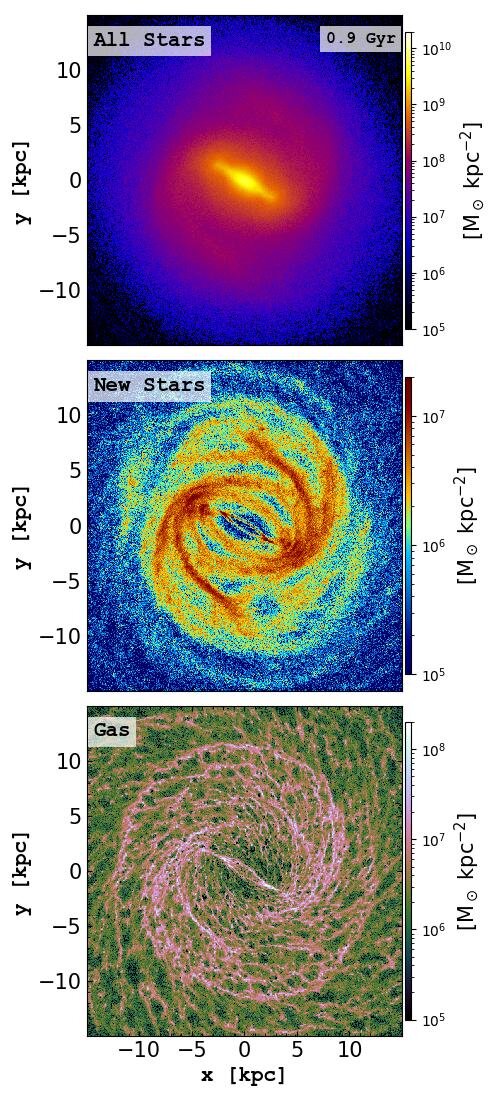}
\adjincludegraphics[height=0.3\textheight, clip, trim={0.12\textwidth} 0 {0.14\textwidth} 0]{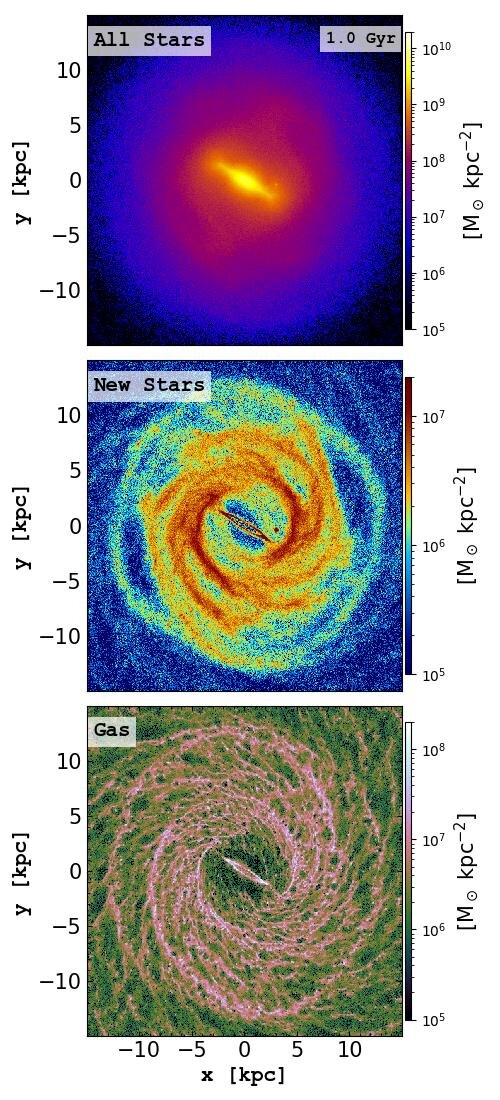}
\adjincludegraphics[height=0.3\textheight, clip, trim={0.12\textwidth} 0 0 0]{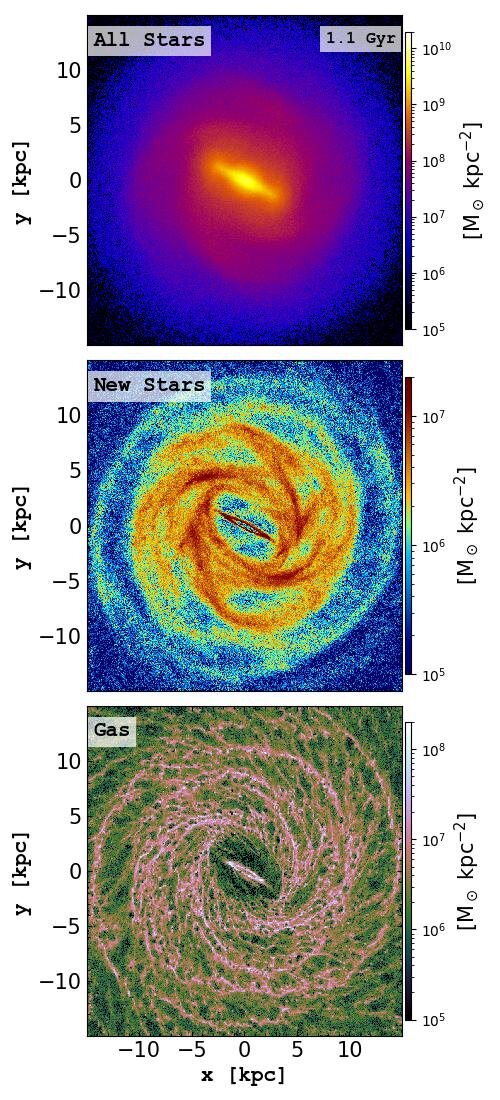}
\caption{Simulated evolution of the MW over $1.1$~Gyr, using initial conditions sampled from the orbit-superposition solution~(see Section \ref{sec4::simulations}). From top to bottom, the panels show the face-on stellar surface density, the distribution of newly formed stars, and the gas density. The bar is realigned in all snapshots so that the solar position corresponds to (-8.12, 0) kpc. The snapshot at 1 Gyr was used in the RT calculations and in the generation of IFU datacubes for models \texttt{eMW.0170} and \texttt{eMW.0174}.}
\label{fig04::emw_simulation}
\end{figure*}

In this section, we illustrate the time evolution of the MW over 1.1~Gyr with the ICs sampled from the orbit superposition solution, as described in Section~\ref{sec4::simulations}. In Figure~\ref{fig04::emw_simulation} we show the stellar density, gas and newly formed stars in the face-on projection. Although the bar is always oriented in the same way, the spiral arms appear to rotate at different pattern speeds, producing a somewhat non-steady appearance. 

\section{Intrinsic LOSVD versus Gauss-Hermite analytic fit}\label{sec04::losvd_error_appendix}

\begin{figure*}
    \centering
    \includegraphics[width=0.49\linewidth]{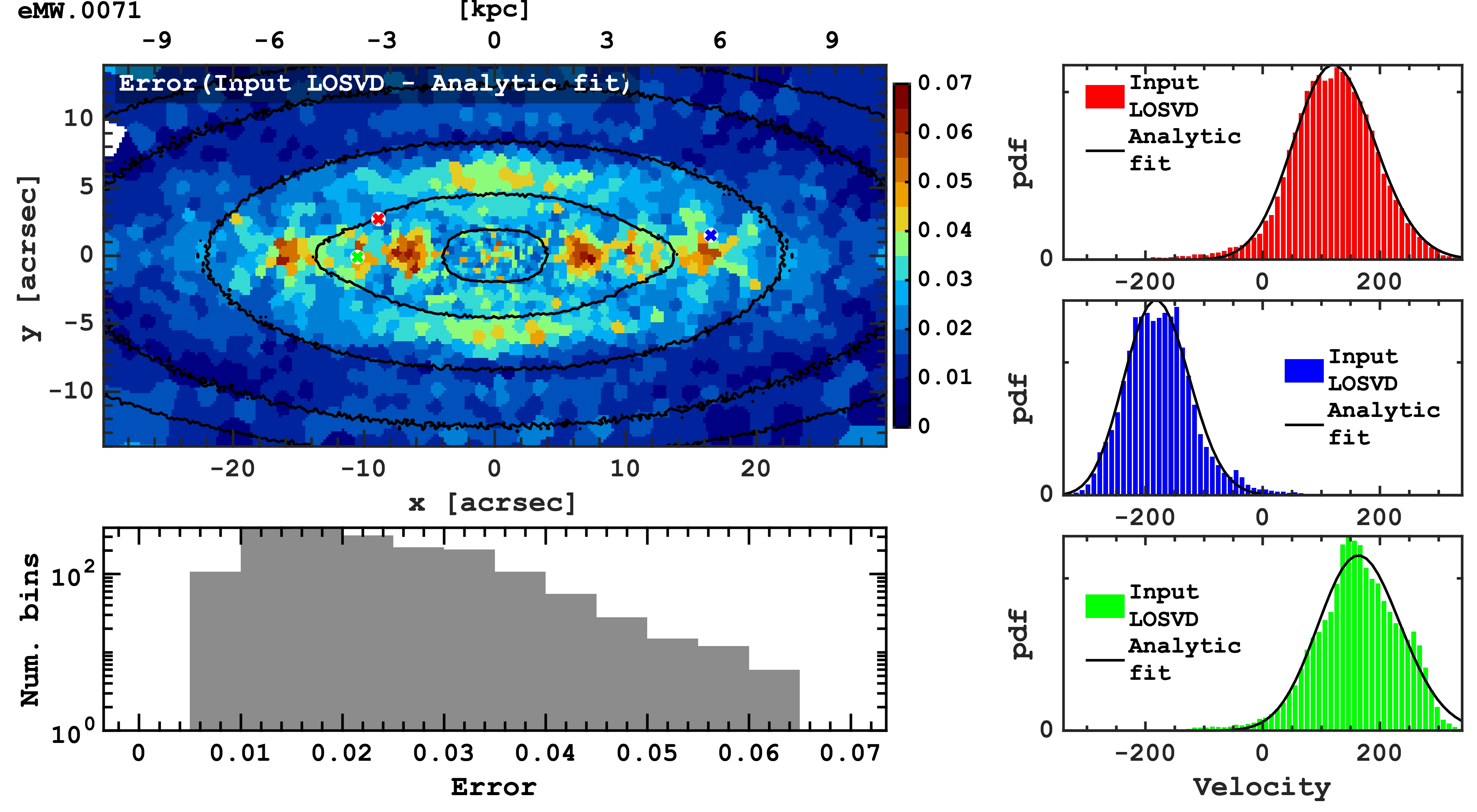} 
    \includegraphics[width=0.49\linewidth]{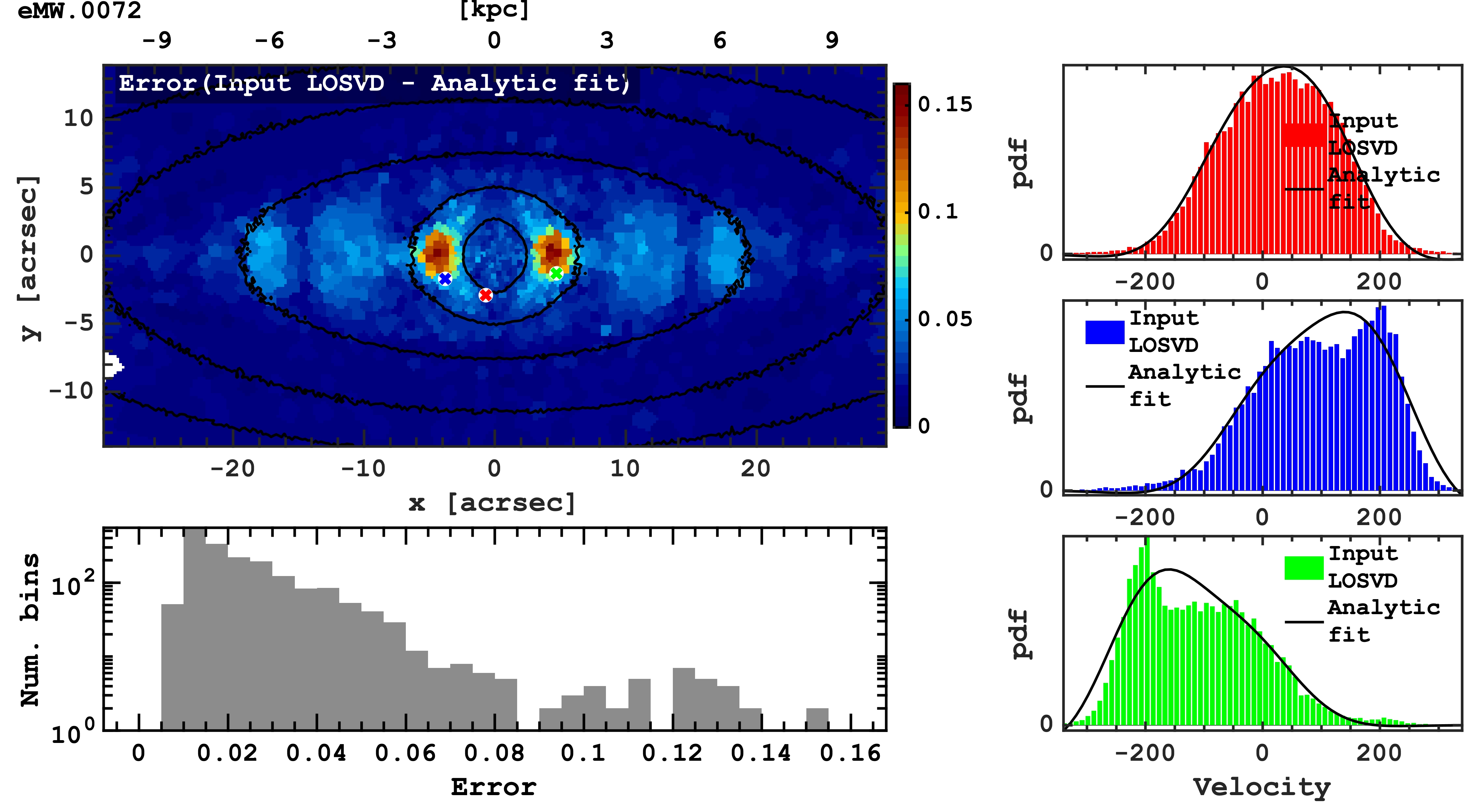}    
    \includegraphics[width=0.49\linewidth]{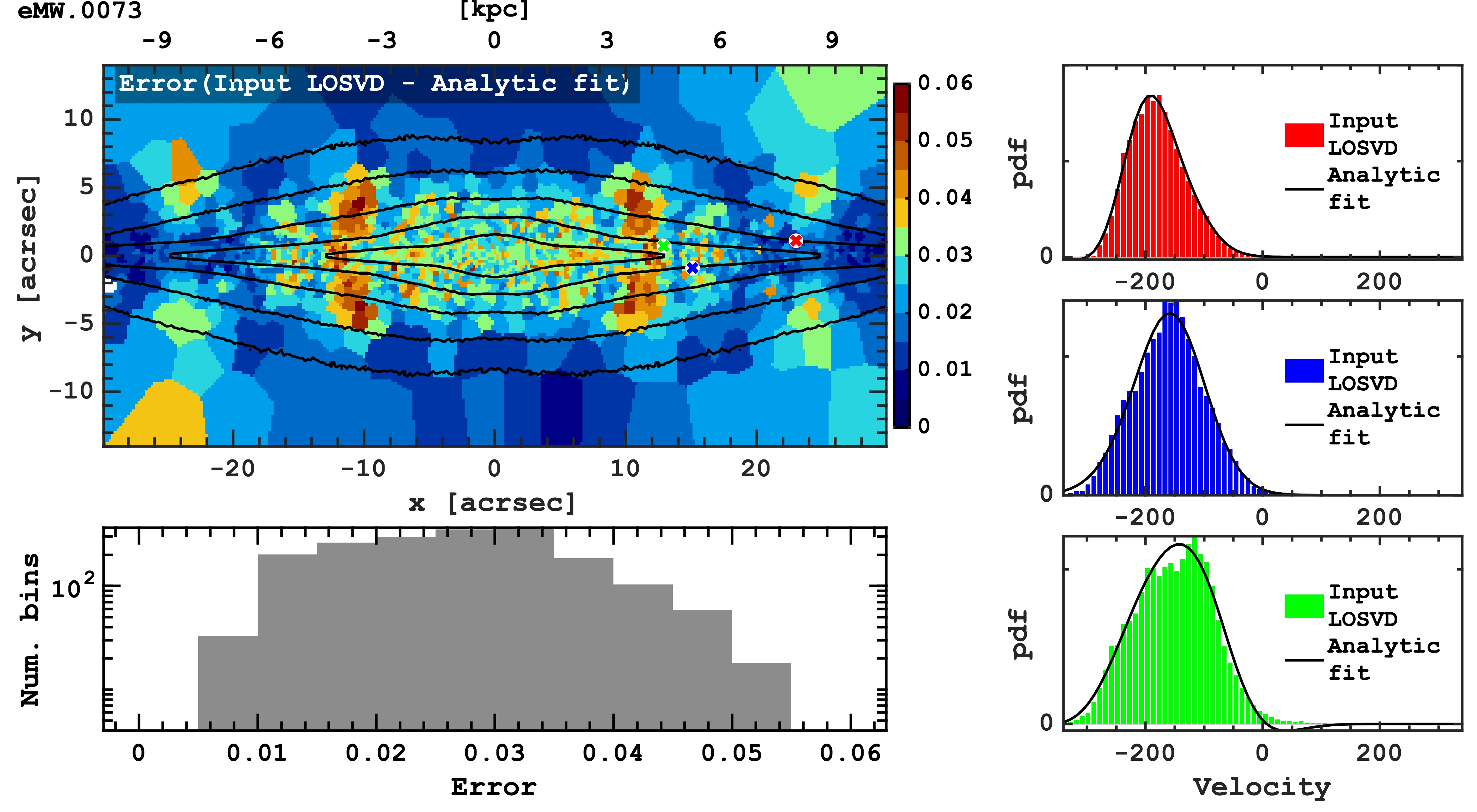}    
    \includegraphics[width=0.49\linewidth]{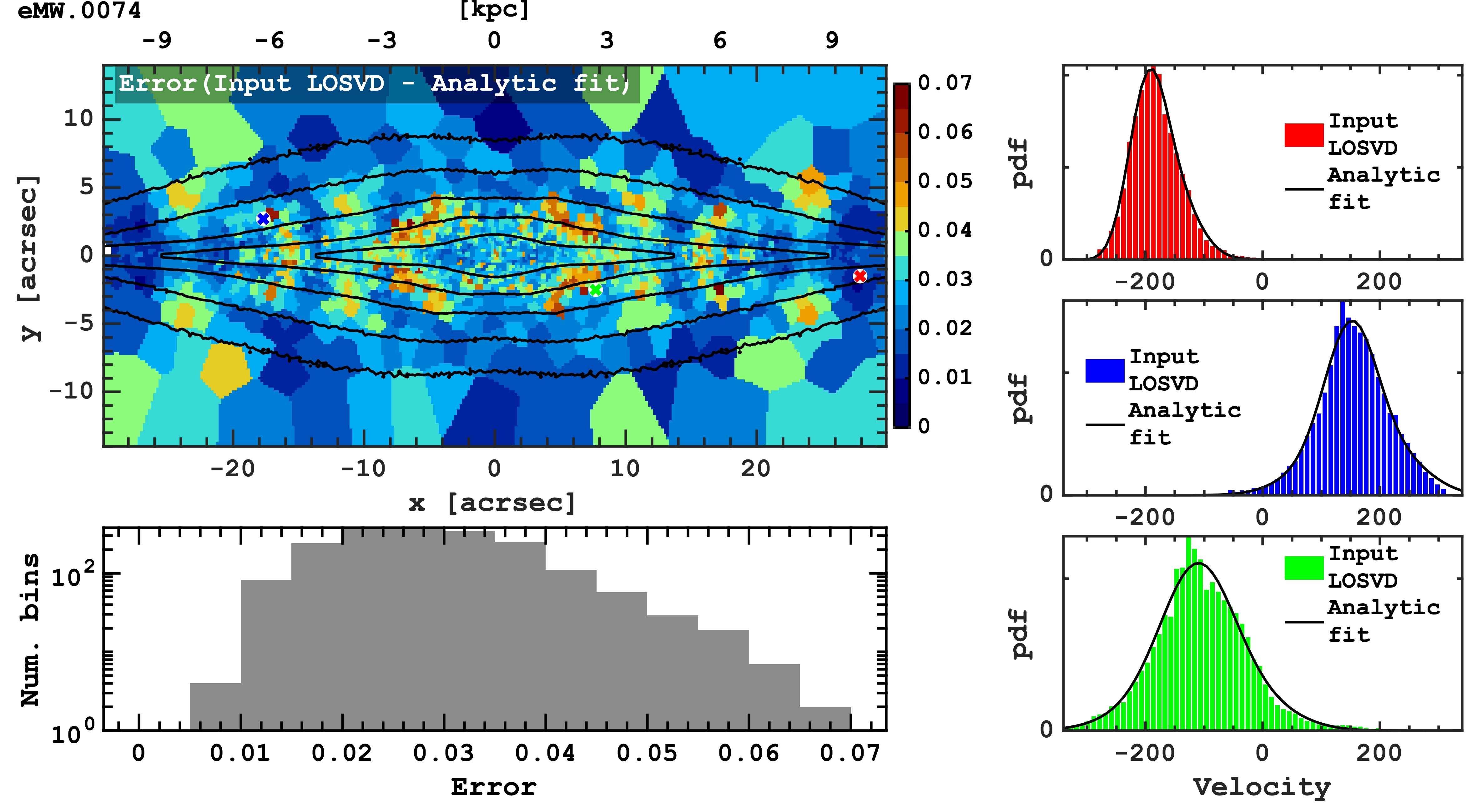}    
    \includegraphics[width=0.499\linewidth]{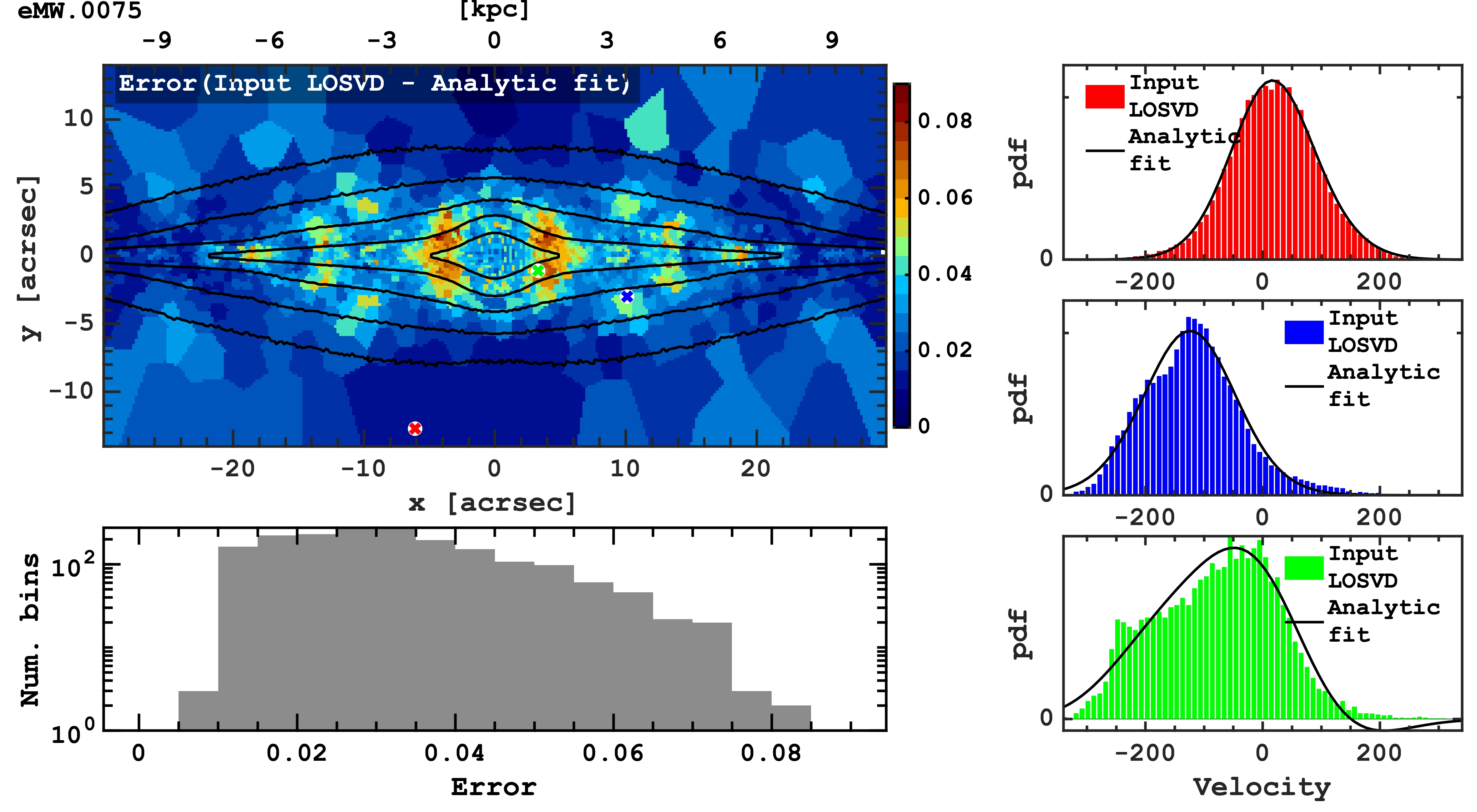}
    \caption{Deviation of the input LOSVD from the analytic G-H fit in different CSP eMW realisations, assuming different projections and bar orientation.}
    \label{fig4::losvd_error_appendix}
\end{figure*}

This section complements the analysis presented in Section~\ref{sec4::results_kinematics_csp}, where we quantified the differences between the intrinsic LOSVDs of the eMW model in various projections and their analytic representations using a G–H parametrisation. Figure~\ref{fig4::losvd_error_appendix} presents the corresponding error maps, together with illustrative examples of the LOSVDs and their G-H fits for three randomly selected Voronoi bins. These examples highlight the typical magnitude and spatial structure of the residuals, and demonstrate how deviations from the analytic approximation depend on both viewing geometry and local kinematic complexity.

\section{Results of the composite stellar populations IFU mock}\label{sec4::kinematics_appendix}
In this section, we illustrate the kinematic and stellar populations recovered from mock IFU data for various CSP models with different bar and disc orientations~(see Table~\ref{tab4:models}), while the detailed analysis of \texttt{eMW.0070} and \texttt{eMW.0074} models are provided in the main body of the paper, in Sect.~\ref{sec4::CSP_results}.
    
\begin{figure*}
  \centering  
  \includegraphics[width=0.49\linewidth]{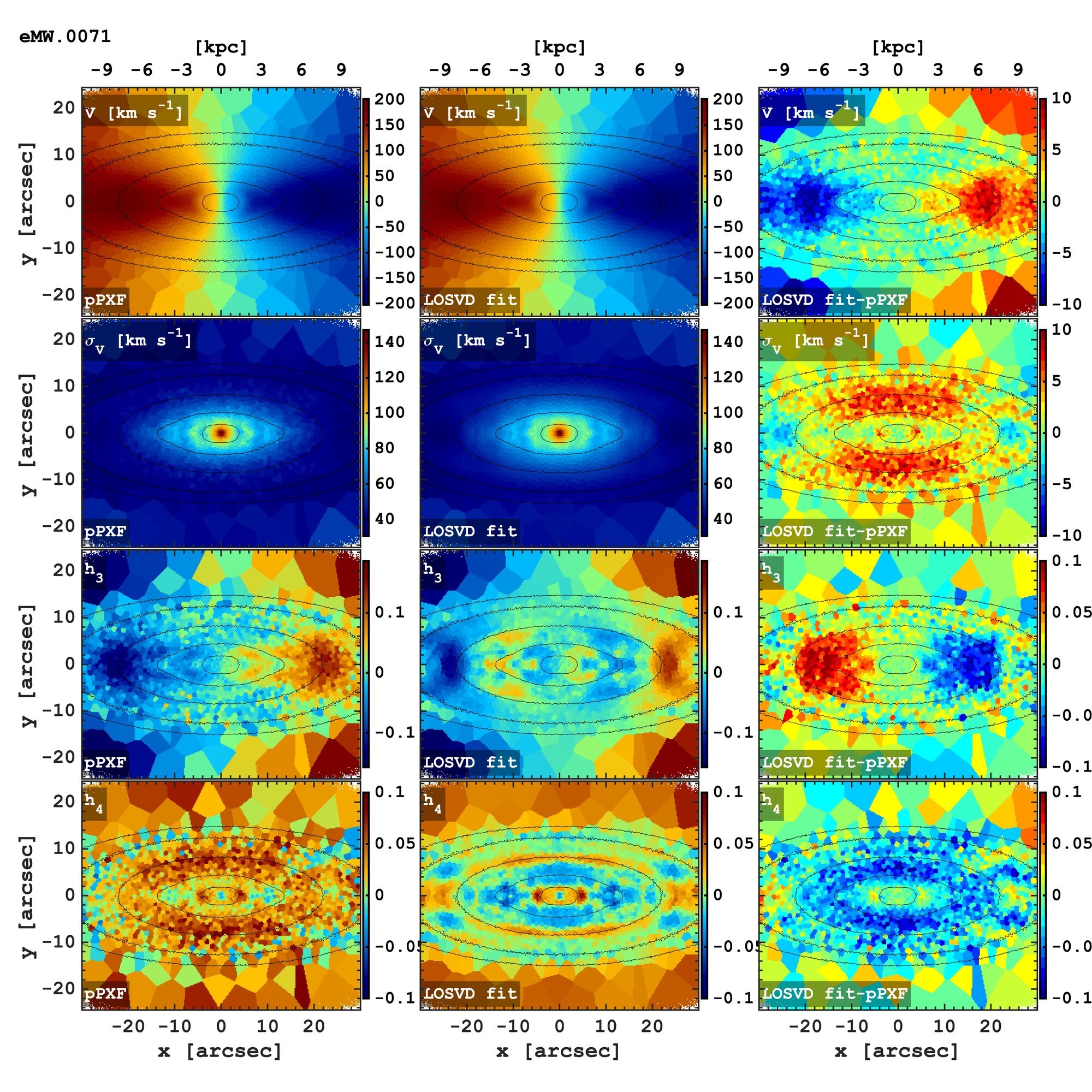} 
  \includegraphics[width=0.49\linewidth]{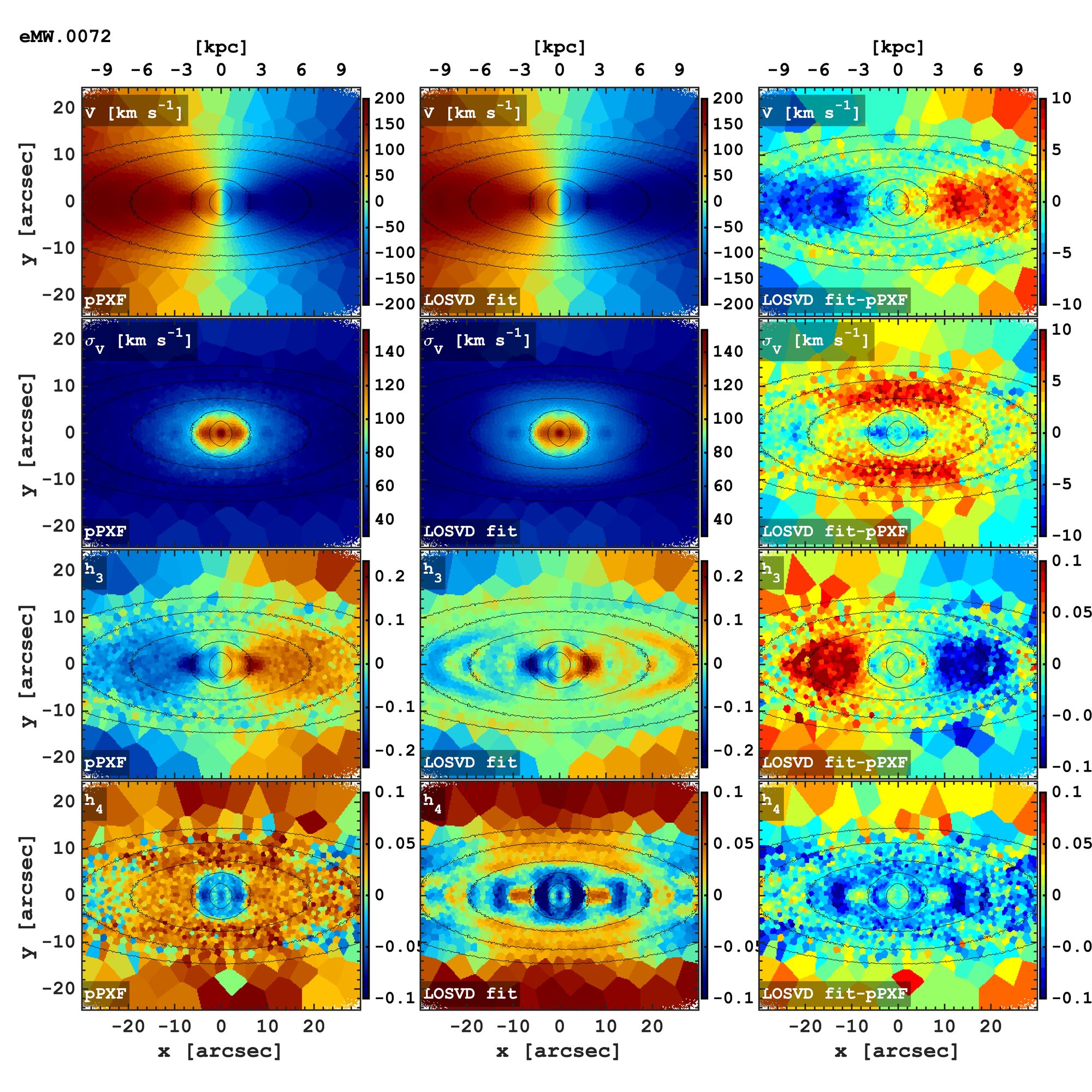} 
  \includegraphics[width=0.49\linewidth]{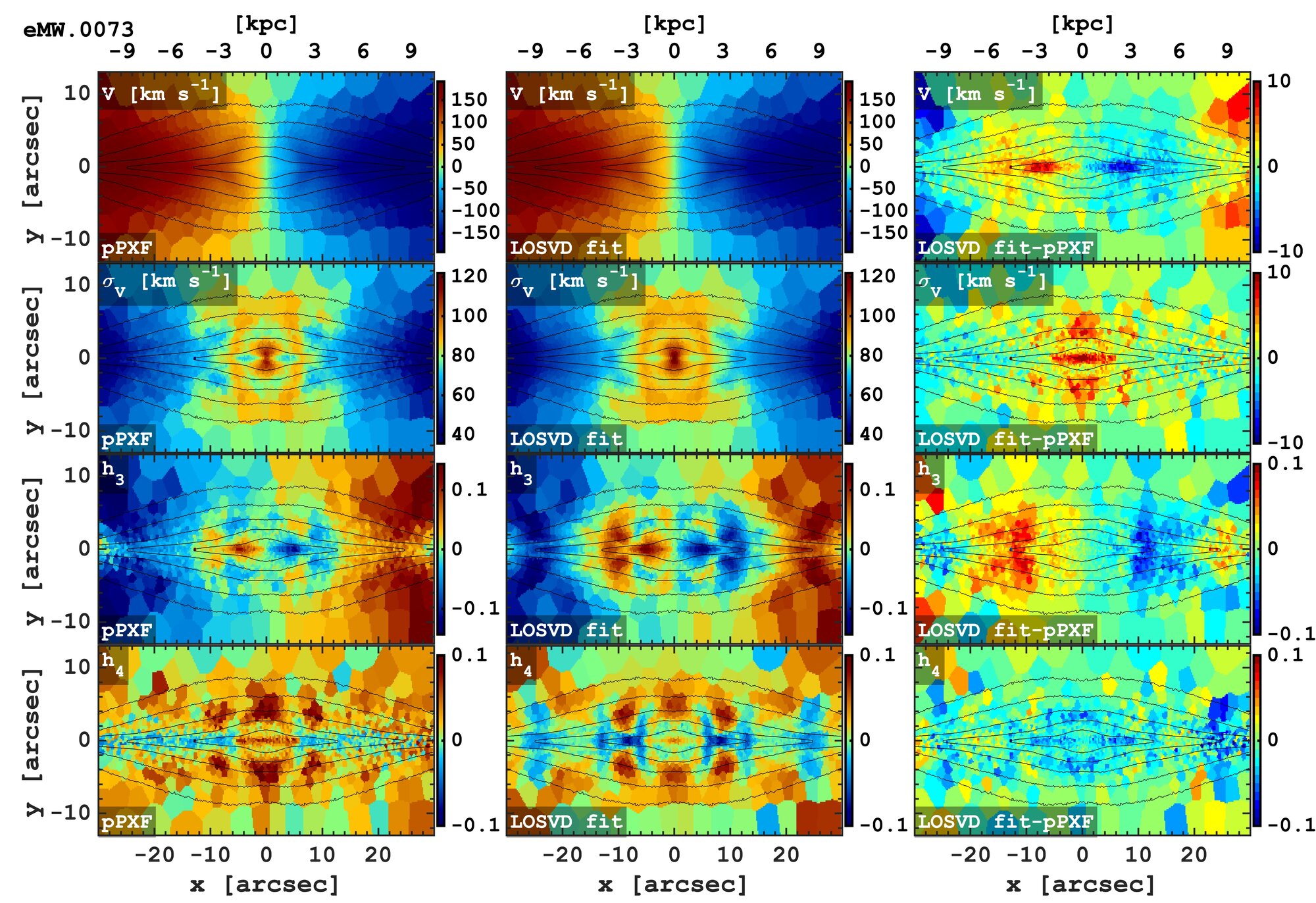} 
  \includegraphics[width=0.49\linewidth]{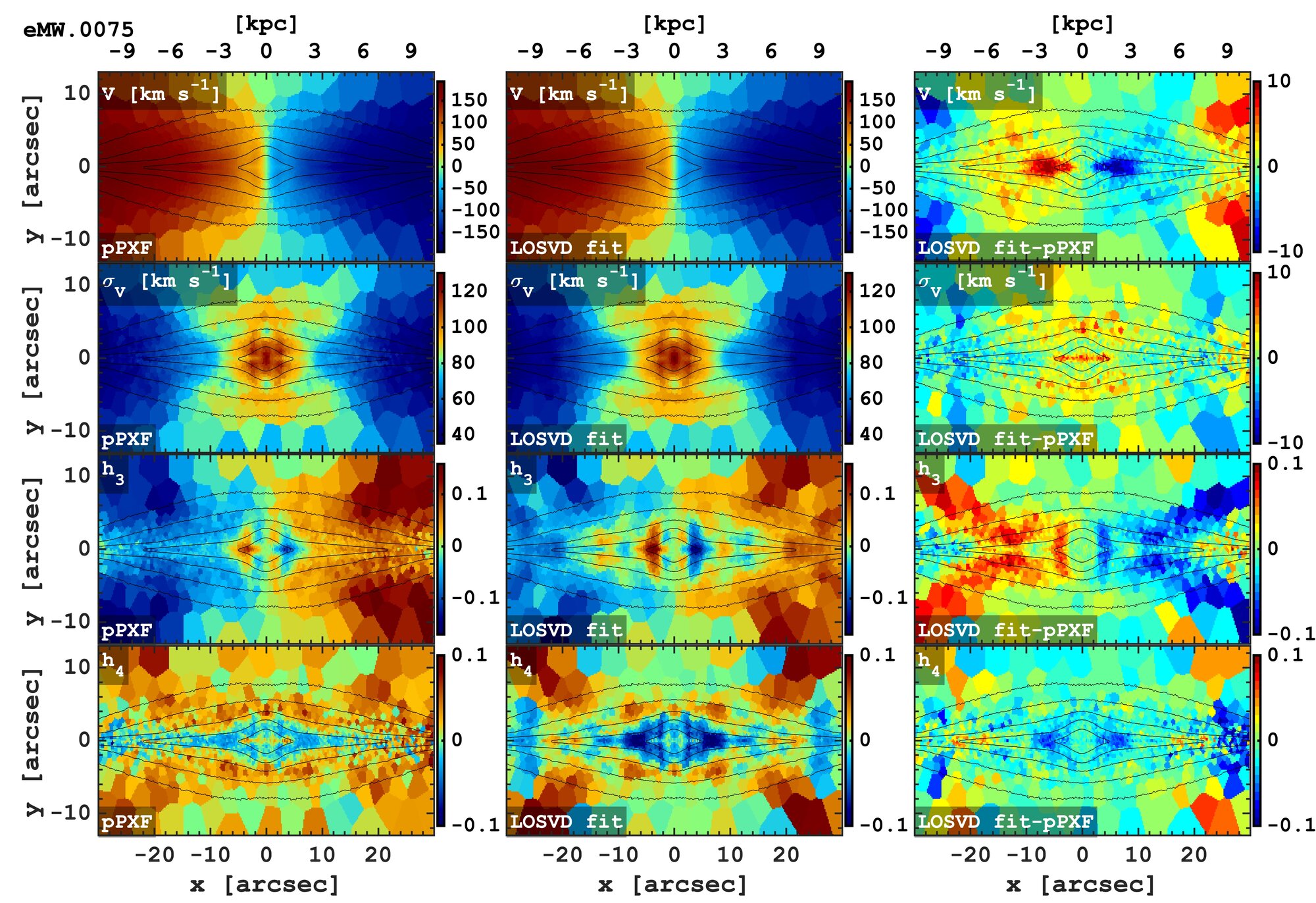} 
  \caption{Same as in Fig.~\ref{fig04:kinematics_70} and Fig.~\ref{fig04:kinematics_74} but for a single \ppxf\ fit of CSP models without regularisation; see parameters in Table~\ref{tab4:models}.}\label{fig04:emw.all_kin_continue0}
\end{figure*}

\begin{figure*}
  \centering  
  \includegraphics[width=0.49\linewidth]{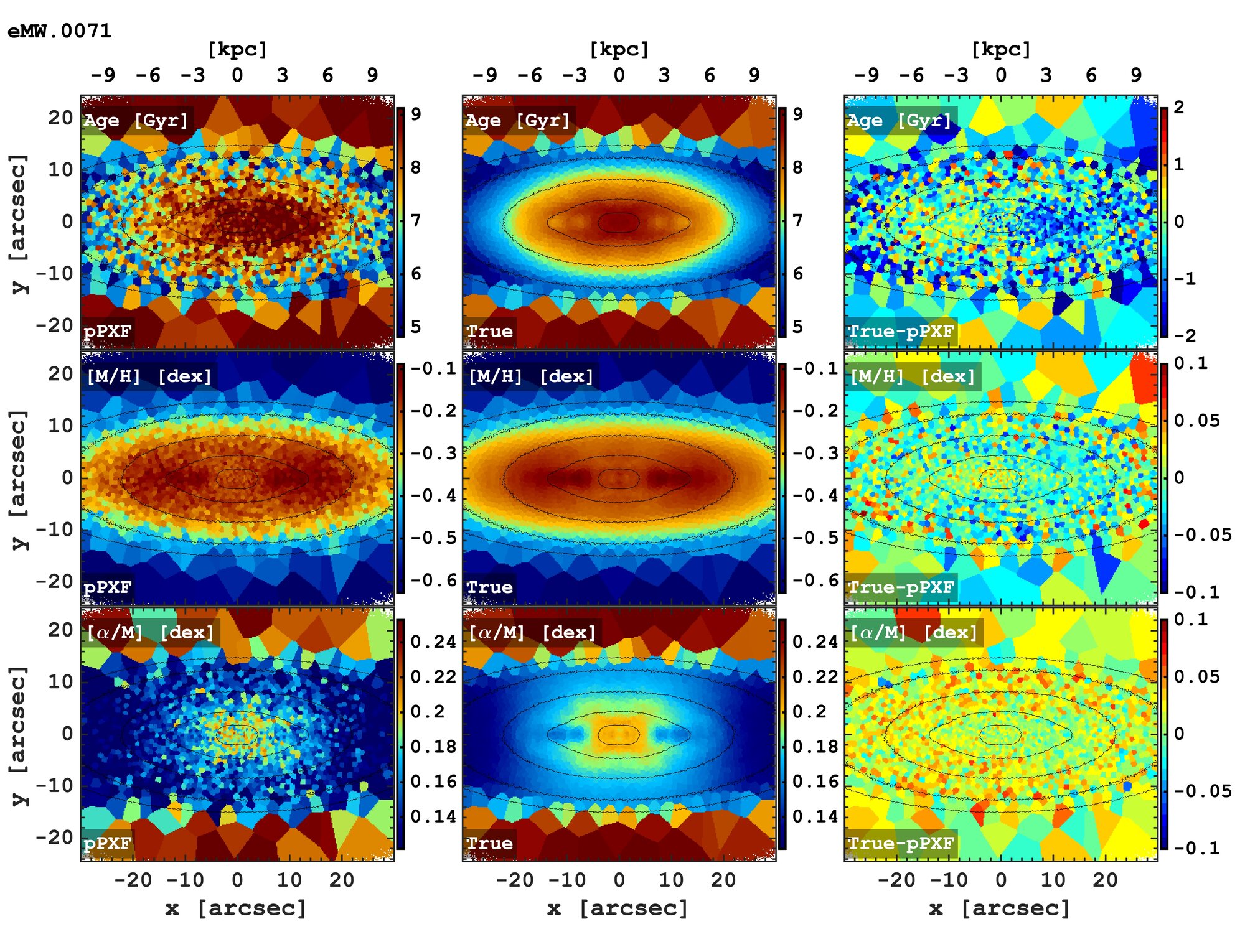} 
  \includegraphics[width=0.49\linewidth]{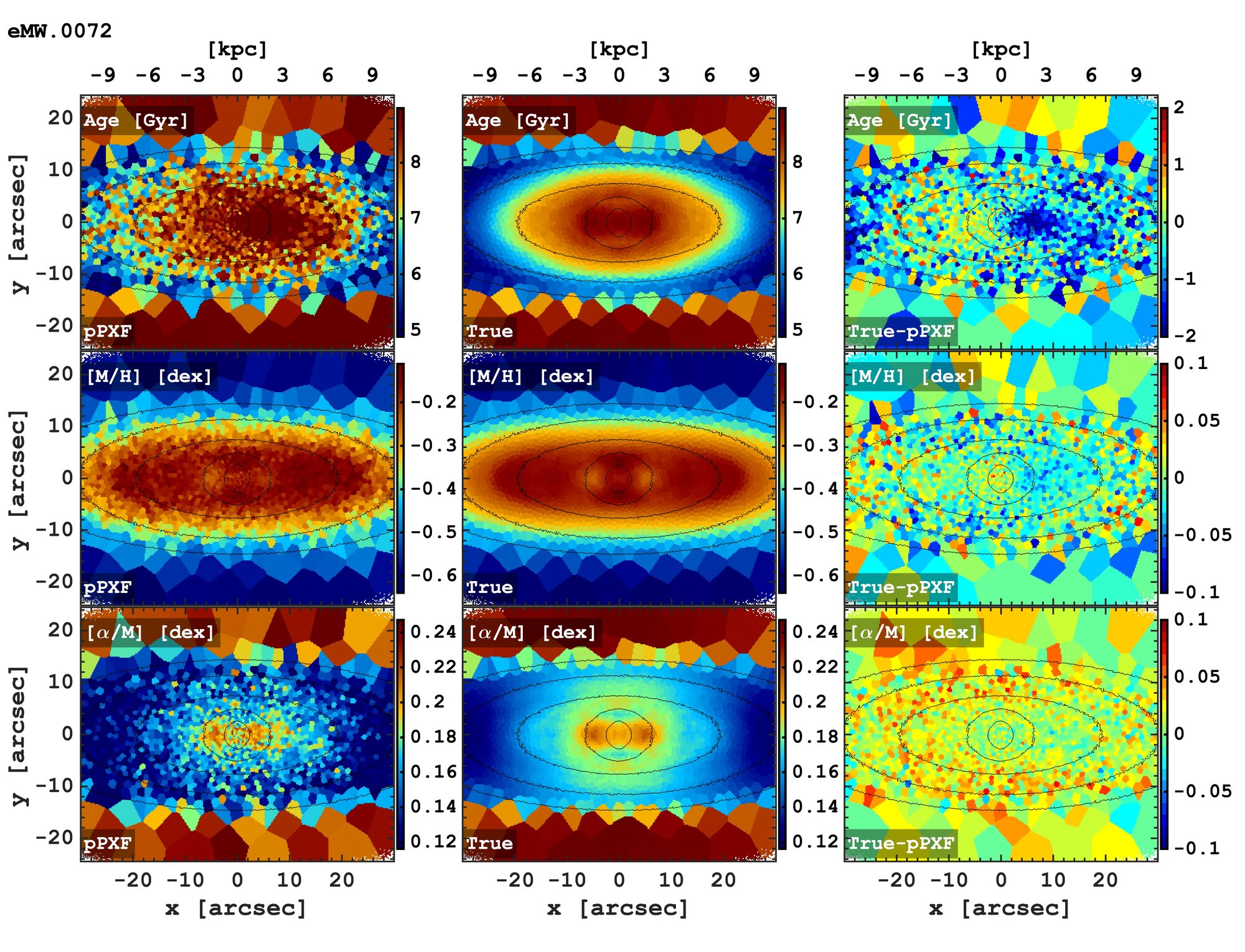} 
  \includegraphics[width=0.49\linewidth]{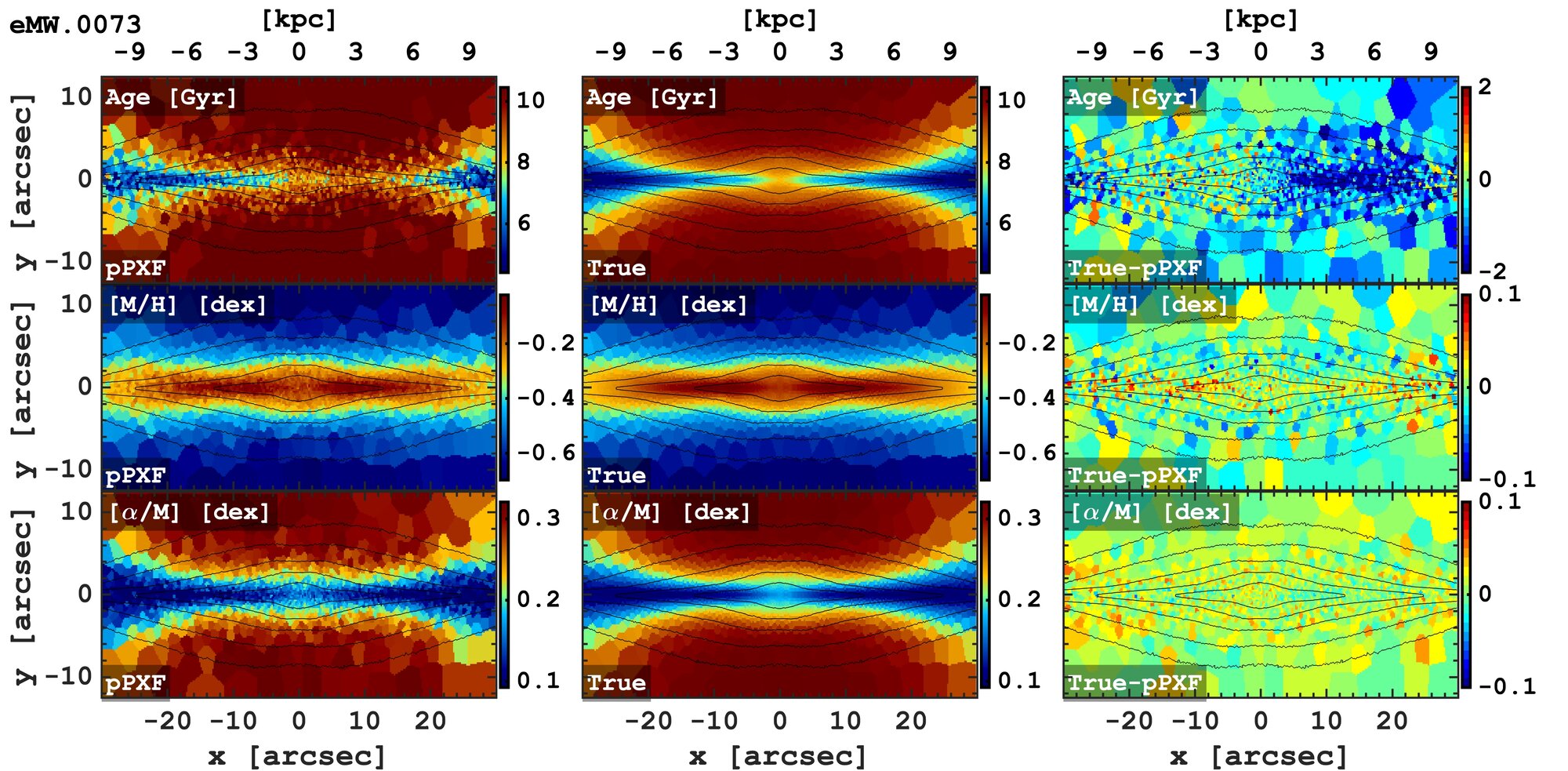} 
  \includegraphics[width=0.49\linewidth]{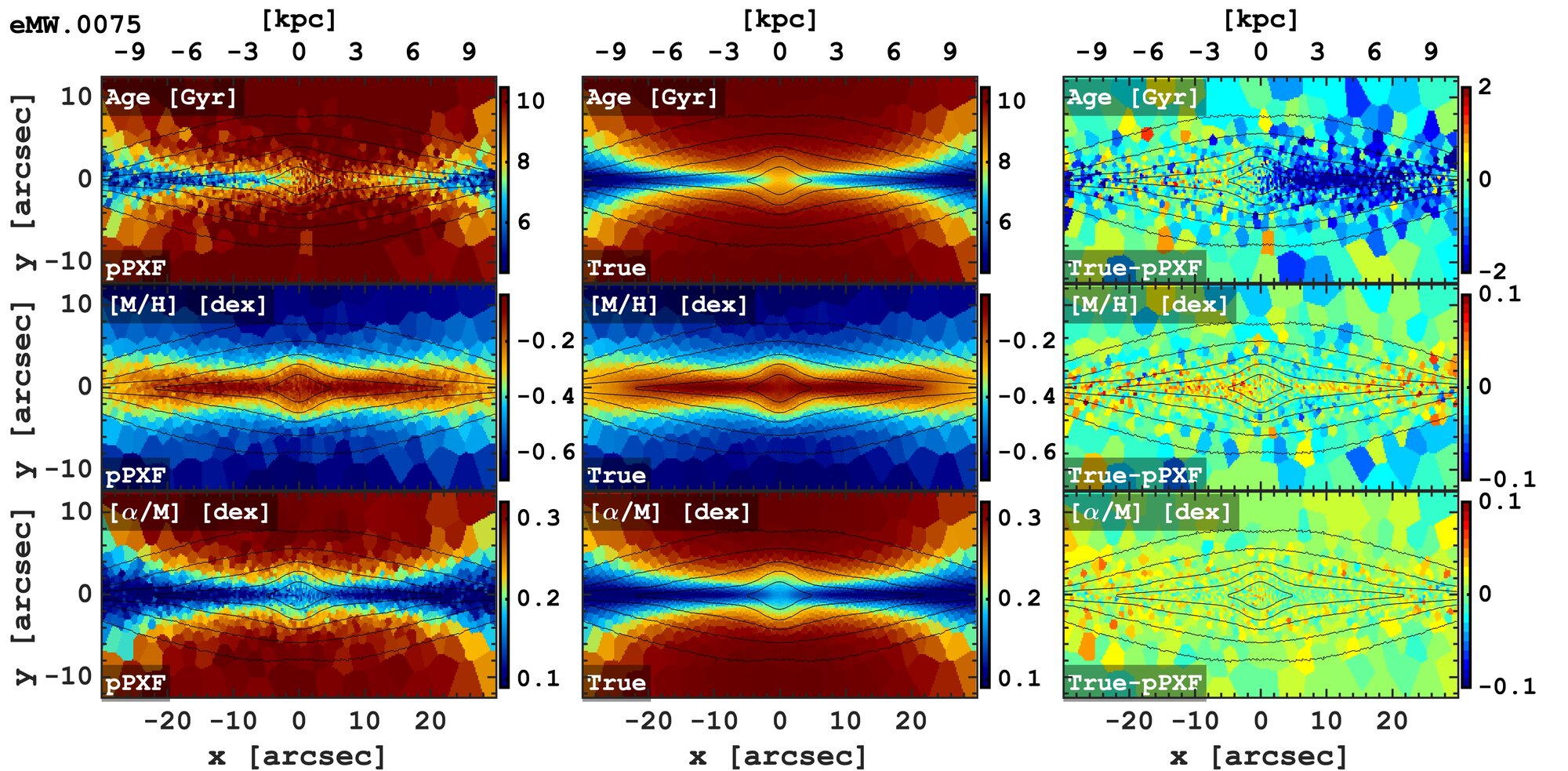} 
  \caption{Same as in Fig.~\ref{fig04:sfh_70} and Fig.~\ref{fig04:sfh_74} but for the remaining  CSP models without regularisation; see parameters in Table~\ref{tab4:models}.}\label{fig04:emw.all_sfh_continue0}
\end{figure*}


\section{Attenuation in extragalactic MW}\label{sec4::dust_appendix}

\begin{figure*}
    \centering
    \includegraphics[width=1\linewidth]{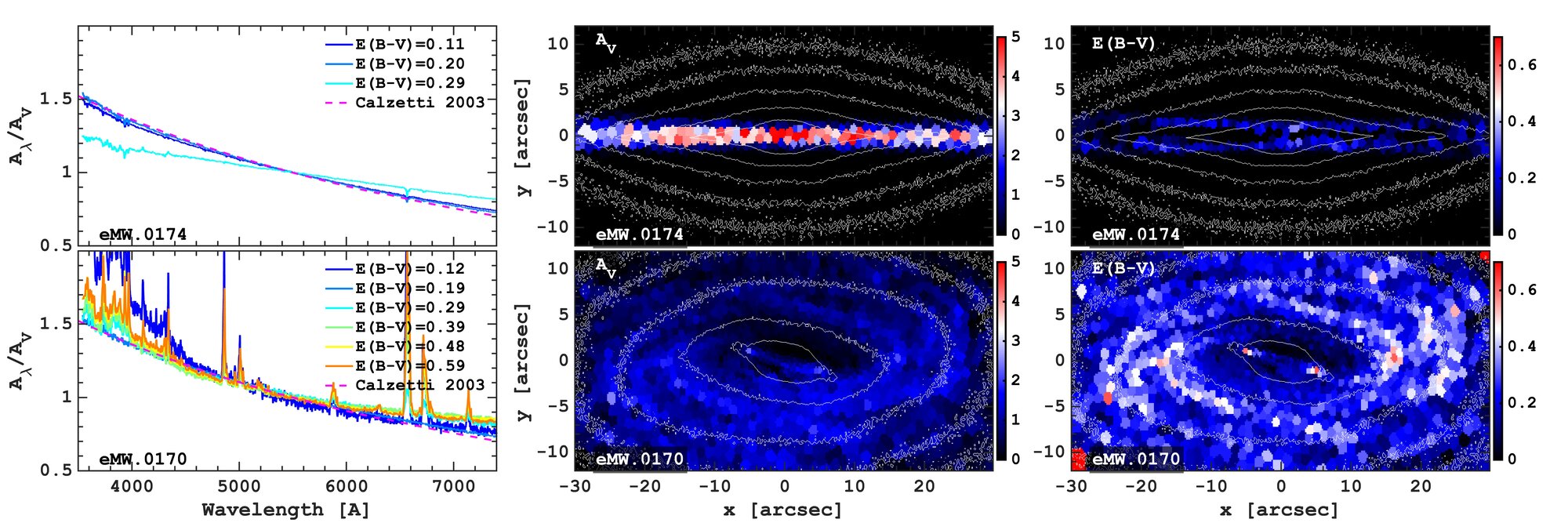}
    \caption{
    Left: Examples of normalised attenuation curves, $A_\lambda/A_V$, measured at different spatial locations in the two models \texttt{eMW.0170} (top) and \texttt{eMW.0174} (bottom). Middle and Right: Corresponding maps of the $A_V$ extinction and E(B-V), with contours indicating the underlying stellar surface density. Note that the contours trace the intrinsic stellar mass distribution and are therefore unaffected by dust.}
    \label{fig04::dust}
\end{figure*}

In this section, we quantify the impact of dust attenuation in the RT-based IFU models. In Fig.~\ref{fig04::dust}, we present attenuation curves for individual Voronoi bins spanning a range of $E(B!-!V)$ values. To isolate the energy attenuated by dust, each model/projection is rerun without diffuse dust and without subgrid dust associated with photodissociation regions around young stars.

The resulting attenuation curves show broad agreement with the Calzetti law~\citep{2000ApJ...533..682C}, while exhibiting numerous narrow features associated with emission from star-forming regions. The corresponding maps of visual extinction, $A_V$, and colour excess, $E(B-V)$, are shown in the middle and right panels, respectively. Although $A_V$ and $E(B-V)$ trace similar large-scale structures, they do not scale linearly on resolved spatial scales; variations in geometry, scattering, and the underlying stellar populations introduce significant local differences between their distributions.

In particular, the spatial distributions of the extinction amplitude, $A_V$, and the colour excess, $E(B-V)$, exhibit markedly different behaviour in highly inclined projections of the galaxy. While $A_V$ peaks strongly along the disc midplane, reflecting the large dust column density, the corresponding colour excess, $E(B-V) = A_B - A_V$, remains low in these regions and instead reaches its maximum at moderate heights above and below the plane. This behaviour arises because, in optically thick regions ($A_V \gtrsim 2$--$3$), both the $B$- and $V$-band fluxes are heavily attenuated, such that $A_B \simeq A_V$ and the reddening saturates. In addition, the mixed geometry of stars and dust in the disc midplane, together with scattering, leads to a flattening of the effective attenuation curve, further reducing the wavelength dependence of the attenuation. As a result, high values of $A_V$ do not necessarily correspond to large $E(B-V)$. In contrast, at intermediate optical depths, the attenuation retains a stronger wavelength dependence, producing elevated $E(B-V)$ despite lower total extinction. This decoupling of $A_V$ and $E(B-V)$ is a natural outcome of radiative transfer in dusty, inclined discs and highlights that $A_V$ primarily traces dust column density, whereas $E(B-V)$ probes the shape of the effective attenuation curve.

\end{appendix}

\end{document}